\documentstyle[hep93,epsfig]{article}
\def\tmt{\times 10^{-2}}
\def\tmth{\times 10^{-3}}
\def\tmf{\times 10^{-4}}
\def\tmfv{\times 10^{-5}}
\def\dag{\hspace{-3mm}\not}
\newcommand{\elevenbf}{\bf}
\newcommand{\elevenit}{\it}
\newcommand{\beq}{\begin{equation}}
\newcommand{\eeq}{\end{equation}}
\newcommand{\bea}{\begin{eqnarray}}
\newcommand{\eea}{\end{eqnarray}}
\newcommand{\barr}{\begin{array}}
\newcommand{\earr}{\end{array}}
\newcommand{\bc}{\begin{center}}
\newcommand{\ec}{\end{center}}
\newcommand{\btab}{\begin{tabular}}
\newcommand{\etab}{\end{tabular}}
\newcommand{\gv}{\mbox{GeV}}
\newcommand{\tv}{\mbox{TeV}}
\newcommand{\mv}{\mbox{MeV}}
\newcommand{\nn}{\nonumber}
\newcommand{\ra}{\rightarrow}
\newcommand{\lra}{\leftrightarrow}
\newcommand{\ipi}{\frac{i}{16\pi^2}}
\newcommand{\dkm}{\mu^{4-D} \int \frac{d^Dk}{(2\pi)^D} }

\newcommand{\sz}{\Sigma^{Z}}
\newcommand{\sw}{\Sigma^{W}}
\newcommand{\sgz}{\Sigma^{\gamma Z}}

\newcommand{\dmz}{\frac{\delta M_Z^2}{M_Z^2}}
\newcommand{\dmw}{\frac{\delta M_W^2}{M_W^2}}
\newcommand{\dmmz}{\delta M_Z^2}
\newcommand{\dmmw}{\delta M_W^2}
\newcommand{\mwb}{M_W^{0\,2}}
\newcommand{\mzb}{M_Z^{0\,2}}
\newcommand{\mzoo}{M_{Z_0}^2}
\newcommand{\mzop}{M_{Z'}^2}
\newcommand{\dro}{\Delta\rho}
\newcommand{\droz}{\Delta\rho_{Z'}}
\newcommand{\drb}{\Delta\overline{\rho}}

\newcommand{\al}{\alpha}

\newcommand{\alspi}{\frac{\alpha_s}{\pi}}

\newcommand{\m}{\mu}
\newcommand{\g}{\gamma}
\newcommand{\G}{\Gamma}
\newcommand{\Gmu}{G_{\mu}}
\newcommand{\gamu}{\gamma_{\mu}}
\newcommand{\gimu}{\gamma^{\mu}}
\newcommand{\ganu}{\gamma_{\nu}}

\newcommand{\giro}{\gamma^{\rho}}

\newcommand{\garo}{\gamma_{\rho}}
\newcommand{\gasi}{\gamma_{\sigma}}
\newcommand{\gafi}{\gamma_5}

\newcommand{\Pig}{\Pi^{\gamma}}
\newcommand{\Pigr}{\hat{\Pi}^{\gamma}}
\newcommand{\Pizr}{\hat{\Pi}^Z}
\newcommand{\Pir}{\hat{\Pi}}

\newcommand{\Pgzrz}{\hat{\Pi}^{\gamma Z}(M_Z^2)}

\newcommand{\noi}{\noindent}
\newcommand{\epmf}{e^+e^- \rightarrow f\bar{f}}

\newcommand{\epm}{e^+e^-}

\newcommand{\gve}{(g_V^{e\,2}+g_A^{e\,2})}
\newcommand{\gvf}{(g_V^{f\,2}+g_A^{f\,2})}

\newcommand{\rc}{radiative corrections }
\newcommand{\sm}{Standard Model }
\newcommand{\su}{SU(2)$\times$U(1) }
\newcommand{\siw}{\sin^2\theta_W}
\newcommand{\cow}{\cos^2\theta_W}
\newcommand{\sinw}{\sin\theta_W}
\newcommand{\cosw}{\cos\theta_W}
\newcommand{\sinm}{\sin\theta_M}
\newcommand{\sinmm}{\sin^2\theta_M}
\newcommand{\cosm}{\cos\theta_M}
\newcommand{\cosmm}{\cos^2\theta_M}
\newcommand{\swo}{s_W^{0\,2}}
\newcommand{\cwo}{c_W^{0\,2}}

\newcommand{\sefl}{\sin^2\theta_e}
\newcommand{\snue}{(1-4\sin^2\!\theta_{\!\nu e})}
\newcommand{\dal}{\Delta\alpha}
\newcommand{\dalms}{\Delta\hat{\alpha}}

\newcommand{\mz}{M_Z^2}
\newcommand{\mw}{M_W^2}

\newcommand{\mm}{\hat{M}_Z^2}
\newcommand{\gm}{\hat{\Gamma}_Z^2}

\newcommand{\real}{\mbox{Re}}
\newcommand{\imag}{\mbox{Im}}
\newcommand{\Dr}{\Delta r}

\newcommand{\eps}{\epsilon}
\newcommand{\veps}{\varepsilon}

\newcommand{\oal}{O(\alpha^2)}
\newcommand{\alr}{A_{LR}}
\newcommand{\afb}{A_{FB}}

\newcommand{\as}{asymmetry }
\newcommand{\ass}{asymmetries }

\newcommand{\plus}{\phi^+}
\newcommand{\mis}{\phi^-}
\newcommand{\nul}{\phi^0}
\newcommand{\adnu}{\overline{\nu}}
\newcommand{\adl}{\overline{l}}

\newcommand{\adu}{\overline{u}}
\newcommand{\add}{\overline{d}}
\newcommand{\lyu}{{\cal L}_{Y\!ukawa}}

\newcommand{\est}{e_*^2}
\newcommand{\sst}{s_*^2}
\newcommand{\Gst}{G_{\mu *}}

\newcommand{\pr}{{\it Phys.\ Rev.\ }}
\newcommand{\zp}{{\it Z.\ Phys.\ }}
\newcommand{\pl}{{\it Phys.\ Lett.\ }}
\newcommand{\prl}{{\it Phys.\ Rev.\ Lett.\ }}
\newcommand{\np}{{\it Nucl.\ Phys.\ }}
\newcommand{\plb}{{\it Phys.\ Lett.\ }}
\newcommand{\prd}{{\it Phys.\ Rev.\  }}
\newcommand{\ctanw}{\frac{c_W^2}{s_W^2}}

\newcommand{\deltaz}{\Delta-\log\frac{M_Z^2}{\mu^2}
                     +\frac{5}{3} }
\newcommand{\deltat}{\Delta-\log\frac{m_t^2}{\mu^2} }
\newcommand{\deltak}{\Delta-\log\frac{k^2}{\mu^2} }
\newcommand{\fvge}{F_{V}^{\g e}}
\newcommand{\fage}{F_{A}^{\g e}}

\newcommand{\fvgf}{F_{V}^{\g f}}
\newcommand{\fagf}{F_{A}^{\g f}}
\newcommand{\fvzf}{F_{V}^{Z  f}}
\newcommand{\fazf}{F_{A}^{Z  f}}
\newcommand{\dkap}{\Delta\kappa}

\newcommand{\sth}{s_{\theta}}
\newcommand{\cth}{c_{\theta}}
\newcommand{\stt}{s_{\theta}^2}
\newcommand{\Rnu}{R_{\nu}}

\newcommand{\sms}{\hat{s}^2}
\newcommand{\cms}{\hat{c}^2}
\newcommand{\smsb}{\hat{s}}
\newcommand{\cmsb}{\hat{c}}
\newcommand{\roms}{\hat{\rho}}
\newcommand{\Drms}{\Delta\hat{r}}
\newcommand{\ems}{\hat{e}}
\newcommand{\ms}{\overline{MS}}
\newcommand{\xms}{X_{\overline{MS}}}
\newcommand{\dvb}{\hat{\delta}_{VB}}
\newcommand{\ams}{\hat{\alpha}}
\newcommand{\mvms}{\hat{M}_V^2}
\newcommand{\mwms}{\hat{M}_W^2}
\newcommand{\mzms}{\hat{M}_Z^2}
\newcommand{\dk}{\frac{d^4k}{(2\pi)^4}}
\newcommand{\Dkk}{[k^2-m_1^2]}
\newcommand{\Dkq}{[(k+q)^2-m_2^2]}

\newcommand{\km}{k_{\mu}}
\newcommand{\kmn}{k_{\mu}k_{\nu}}

\begin{document}
%----------------------------------------------------------------------
\title{Electroweak Theory}
%----------------------------------------------------------------------
\author{Wolfgang Hollik \\
        {\em Institut f\"ur Theoretische Physik}\\
        {\em Universit\"at Karlsruhe}\\
        {\em D-76128 Karlsruhe, Germany}
       }
%----------------------------------------------------------------------
\abstract{
In these lectures we give a discussion of the structure of
the electroweak 
 \sm and its quantum corrections for tests of the electroweak theory.
 The predictions for  the
vector boson masses, neutrino scattering
cross sections and the $Z$ resonance observables
 are presented in some detail.   
 We show
comparisons with the recent experimental data and
 their implications for the present status of the Standard Model.
 Finally we address
the question how virtual New Physics can influence the predictions
for the precision observables and discuss
 the minimal supersymmetric standard model
 as a special example of
particular theoretical interest.
}
%%%%%%%%%%%%%%%%%%%%%%%%%%%%%%%%%%%%%%%%%%%%%%%%%%%%%%%%%%%%%%%%%%%%%%%%
%\psdraft
%%%%%%%%%%%%%%%%%%%%%%%%%%%%%%%%%%%%%%%%%%%%%%%%%%%%%%%%%%%%%%%%%%%%%%%%
\maketitle
%%%%%%%%%%%%%%%%%%%%%%%%%%%%%%%%%%%%%%%%%%%%%%%%%%%%%%%%%%%%%%%%%%%%%%%%
\section{Introduction}\label{sec:intro}
The present theory of the electroweak interaction, known as the
``Standard Model'' [1-4],
 is a gauge invariant quantum field theory with
the symmetry group \su spontaneously broken by the Higgs mechanism.
It contains three free parameters to describe the gauge bosons
$\g,W^{\pm},Z$
and their interactions with the fermions.
For a
comparison between theory and experiment three independent experimental
input data are required. The most natural choice is
given by the electromagnetic fine structure constant $\al$, the
muon decay constant (Fermi constant) $\Gmu$, and the mass of the $Z$
boson which has meanwhile been measured with high accuracy.
              Other measurable quantities
are predicted in terms of the input data. Each additional precision
experiment which allows the detection of small deviations from the
lowest order predictions can be considered a test of the electroweak
theory at the quantum level.
In the Feynman graph expansion of the scattering amplitude
for a given process the   higher order terms
show up as diagrams containing closed loops.
The lowest order amplitudes could also be derived
from  a corresponding classical field theory whereas the loop
contributions can only be obtained from the  quantized version.
 The renormalizability of the \sm \cite{thooft} ensures that
it retains its predictive power also  in higher orders.
The higher order terms, commonly called     radiative corrections,
are the quantum effects of the electroweak theory. They  
are  complicated in their concrete form, but
they are finally the consequence of the basic Lagrangian with a
simple structure.
The quantum corrections 
 contain the self-coupling of the vector bosons as well as their
interactions with the Higgs field and the top quark,
and  provide the theoretical basis for electroweak precision
tests. Assuming the validity of the Standard model, the presence of the
top quark and the Higgs boson in the loop contributions to electroweak
observables allows to obtain significant bounds on their masses from
precision measurements of these observables. 

\smallskip
The present generation of high  precision experiments hence imposes
stringent tests on the Standard Model.
Besides the impressive achievements in the determination of the
$Z$ boson parameters \cite{lep} and the $W$ mass
\cite{wmass}, the most important step has been the discovery 
of the top quark at the Tevatron \cite{top} with the mass determination
$           m_t = 180 \pm 12$ GeV,
which coincides perfectly with the indirectly obtained mass range via
the radiative corrections.

\smallskip
The high experimental
sensitivity in the electroweak observables, at the
level of the quantum effects,  requires the highest standards
on the theoretical side as well. A sizeable amount of work has
contributed over the last few years to a steadily rising
improvement of the standard model predictions
pinning down the theoretical uncertainties to a level sufficiently
small for the current interpretation of the precision data, but
still sizeable enough to provoke conflict with a further increase
in the experimental accuracy.

\smallskip
 The lack
of direct signals from ``New Physics'' makes the high
precision experiments
 also a unique tool in the search for {\it indirect} effects:
through definite deviations of the experimental results
from the theoretical predictions of the minimal Standard Model.
Since such deviations are expected to be small, of the typical
size of the Standard Model radiative corrections,
it is inevitable to have the 
 the standard loop effects in the precision observables under control.

\smallskip
In these lectures we give a brief discussion of the structure of
the \sm and its quantum corrections for testing  the electroweak theory
at present and future colliders. 
 The predictions for  the
vector boson masses, neutrino scattering
cross sections, and the $Z$ resonance observables like the width
of the $Z$ resonance, partial widths, effective neutral current
coupling constants and mixing angles at the $Z$ peak, are presented in some
detail.   
 We show
comparisons with the recent experimental data and
 their implications for the present status of the Standard Model.
 Finally we address
the question how virtual New Physics can influence the predictions
for the precision observables and discuss
 the minimal supersymmetric standard model
 as a special example of
particular theoretical interest.
%%%%%%%%%%%%%%%%%%%%%%%%%%%%%%%%%%%%%%%%%%%% end of section
\section{The electroweak Standard Model}
 \subsection{The Standard Model Lagrangian}
%text
The phenomenological basis for the formulation of the Standard Model
is given by the following empirical facts:
\begin{itemize}
\item
The \su family structure of the fermions: \\
The fermions appear as families with left-handed doublets and
right-handed singlets:
$$ \left( \barr{l} \nu_e \\ e  \earr \right)_L \, , \;\;\;
   \left( \barr{l} \nu_{\m} \\ \m   \earr \right)_L \, , \;\;\;
   \left( \barr{l} \nu_{\tau} \\ \tau  \earr \right)_L \, , \;\;\;
   e_R, \;\;\; \m_R, \;\;\; \tau_R   $$
$$ \left( \barr{l} u \\ d \earr \right)_L, \;\;\;
   \left( \barr{l} c \\ s \earr \right)_L, \;\;\;
   \left( \barr{l} t \\ b \earr \right)_L, \;\;\;
    u_R, \;\;\; d_R, \;\;\; c_R, \cdots   $$
They can be characterized by the quantum numbers of the weak
isospin $I$, $I_3$, and the weak hypercharge $Y$.
\item
The Gell-Mann-Nishijima relation: \\
Between the quantum numbers classifying the fermions with respect to
the group \su and their electric charges $Q$ the relation
\beq
   Q\, = \, I_3 \, + \, \frac{Y}{2}
\eeq
is valid.
\item
The existence of vector bosons: \\
There are 4 vector bosons as carriers of the electroweak force
$$ \gamma, \;\;\; W^+, \;\;\; W^-, \;\;\; Z    $$
where the photon is massless and the $W^{\pm}$, $Z$ have masses
$M_W \neq 0$, $M_Z \neq 0$.
\end{itemize}
This empirical structure can be embedded in a gauge invariant
field theory
of the unified electromagnetic and weak interactions by interpreting
\su as the group of gauge transformations under which the Lagrangian
is invariant. This full symmetry has to be broken by the Higgs
mechanism down to the electromagnetic gauge symmetry; otherwise the
$W^{\pm},\, Z$ bosons would also be massless. The minimal formulation,
the Standard Model,
requires a single scalar field (Higgs field) which is a doublet under
SU(2).

\bigskip
According to the general principles of constructing a gauge invariant
field theory with spontaneous symmetry breaking, the gauge,
Higgs, and fermion parts of the electroweak Lagrangian
\begin{equation}
{\cal L}_{cl}={\cal L}_G+{\cal L}_H+{\cal L}_F
\end{equation}
are specified in the following way:

\bigskip       \noi
           {\bf Gauge fields} \\
\su is a non-Abelian  group which is generated by the isospin
operators $I_1,\; I_2,\; I_3$ and the hypercharge Y (the
elements of the corresponding Lie algebra). Each of these
generalized charges is associated with a vector field: a triplet
of vector fields
$W_{\m}^{1,2,3}$ with $I_{1,2,3}$ and a singlet field $B_{\m}$
with $Y$.
The isotriplet $W_{\mu}^a$, $a=1,2,3$, and the
isosinglet  $B_{\mu}$
                                               lead to the
field strength tensors
$$ W_{\mu\nu}^a= \partial _{\mu}W_{\nu}^a- \partial_{\nu}W_{\mu}^a
   +g_2 \, \epsilon_{abc} \, W_{\mu}^bW_{\nu}^c ,$$
\begin{equation}
   B_{\mu\nu}=\partial_{\mu}B_{\nu}-\partial_{\nu}B_{\mu}.
\end{equation}
$g_2$ denotes the non-Abelian SU(2) gauge coupling constant and
$g_1$  the Abelian U(1) coupling. From the field
tensors (3) the pure gauge field Lagrangian
\begin{equation}
 {\cal L}_G=-\frac{1}{4} \, W_{\mu\nu}^aW^{\mu\nu,a}-
     \frac{1}{4}\, B_{\mu\nu}B^{\mu\nu}
\end{equation}
is formed according to the rules for the non-Abelian
case.

\bigskip   \noi
{\bf Fermion fields and  fermion-gauge interaction}  \\
The left-handed fermion fields of each lepton and quark family
(colour index is suppressed)
$$ \psi_j^L= \left ( \begin{array}{c}
              \psi_{j+}^L \\
              \psi_{j-}^L
              \end{array} \right )  $$
with family index $j$ are grouped into SU(2) doublets with component
index $\sigma = \pm$,   and   the right-handed fields into singlets
$$ \psi_j^R=\psi^R_{j\sigma}.$$
Each left- and right-handed multiplet is an eigenstate of the weak
hypercharge Y such that the relation (1) is fulfilled. The covariant
derivative
\begin{equation}
 D_{\mu}=\partial_{\mu}\, -\, i\, g_2\, I_aW_{\mu}^a\,
         +\, i\,g_1\, \frac{Y}{2}\, B_{\mu}
\end{equation}
induces the fermion-gauge field interaction via the minimal
substitution rule:
\begin{equation}
{\cal L}_F=\sum_j\,  \bar{\psi}_j^Li\gamma^{\mu}D_{\mu}\psi_j^L\,
     +\, \sum_{j,\sigma}  \,
    \bar{\psi}_{j\sigma}^R i \gamma^{\mu}D_{\mu} \psi_{j\sigma}^R
\end{equation}

\bigskip   \noi
{\bf  Higgs field,  Higgs - gauge field and Yukawa
                interaction} \\
For spontaneous breaking of the SU(2)$\times$U(1) symmetry leaving
the electromagnetic gauge subgroup $U(1)_{em}$ unbroken, a single
complex scalar doublet field with hypercharge $Y=1$
\begin{equation}
 \Phi(x)= \left ( \begin{array}{c}
              \phi^+(x) \\ \phi^0(x)
          \end{array} \right )
\end{equation}
is coupled to the gauge fields
\begin{equation}
 {\cal L}_H=(D_{\mu}\Phi)^+(D^{\mu}\Phi) - V(\Phi)
\end{equation}
with the covariant derivative
$$
 D_{\mu}=\partial_{\mu}\, -\, i\, g_2\, I_aW_{\mu}^a\,
         +\,i\,  \frac{g_1}{2}\, B_{\mu}  \, .
$$
The Higgs field self-interaction
\begin{equation}
 V(\Phi)=-\mu^2\, \Phi^+\Phi + \frac{\lambda}{4}\, (\Phi^+\Phi)^2
\end{equation}
is constructed in such a way that it
has  a
non-vanishing vacuum expectation value $v$,
related to the coefficients of the potential $V$ by
\begin{equation}
 v=\frac{2\mu}{\sqrt{\lambda}} \; .
\end{equation}
The field  (7) can be written in the following way:
\begin{equation}
 \Phi(x)= \left ( \begin{array}{c}
          \phi^+(x) \\
          (v+H(x)+i\chi(x))/\sqrt{2}
          \end{array} \right )
\end{equation}
where the components $\phi^+$, $H$, $\chi$ now have vacuum
expectation values zero.
Exploiting the invariance of the Lagrangian one notices that
the components $\phi^+,\,\chi$ can be gauged away which means that they
are unphysical (Higgs ghosts or would-be Goldstone bosons). In this
particular gauge, the unitary gauge, the Higgs field has
the simple form
$$ \Phi(x) = \frac{1}{\sqrt{2}} \left(
\begin{array}{c}
  0 \\
  v+H(x)
\end{array} \right) \, .
$$
The real part of $\phi^0$, $H(x)$, describes physical neutral scalar
particles with mass
\begin{equation}
 M_H=\mu\sqrt{2}.
\end{equation}
The Higgs field components have triple and quartic
self couplings following from $V$, and couplings to the gauge fields
via the kinetic term of Eq.\ (8).

\medskip
In addition, Yukawa
couplings to fermions are introduced in order to make the
charged fermions massive.
The Yukawa term is conveniently expressed in the doublet field
components (7). We write it down for one family of leptons and quarks:
\bea
  \lyu      & = & -g_l\,(\adnu_L\,\plus\,l_R\,+\,\adl_R\,\mis\,\nu_L
 \,+\, \adl_L\,\nul\,  l_R\, +\, \adl_R\, \phi^{0*}\,  l_L ) \nn \\
            & = & -\, g_d \,
 (\adu_L\,\plus\,d_R\,+\,\add_R\,\mis\,u_L\,+\,\add_L\,\nul\,d_R
                    \,+\,\add_R\,\phi^{0*}\,d_L )     \nn  \\
                &   & -\,g_u \,
 (-\adu_R\,\phi^+\,d_L\,-\,\add_L\,\phi^-\,u_R\,+\,
  \adu_R\,\phi^0\,u_L\,+\,\adu_L\,\phi^{0*}\,u_R )  \; .
\eea
$\phi^-$ denotes the adjoint of $\phi^+$.

\smallskip\noi
By $v\neq 0$ fermion mass terms are induced.
The  Yukawa coupling constants $g_{l,d,u}$  are related to the
masses of the charged fermions by Eq.\ (23).
In the unitary gauge the Yukawa Lagrangian is particularly simple:
\beq
\lyu\, = \,
-\sum_f \,m_f\,\bar{\psi}_f \psi_f - \sum_f \frac{m_f}{v}\,
 \bar{\psi}_f\psi_f \, H \, .
\eeq
As a remnant of this mechanism for generating fermion masses in a
gauge invariant way, Yukawa interactions between the massive
fermions and the physical Higgs field occur with coupling
constants proportional to the fermion masses.

\bigskip  \noi
         {\bf  Physical fields and parameters}   \\
The gauge invariant Higgs-gauge field interaction in the kinetic part
of Eq.\ (8) gives rise to mass terms for the vector bosons in the
non-diagonal form
\begin{equation}
 \frac{1}{2}\, \left ( \frac{g_2}{2}v \right )^2\,
                     (W_1^2+W_2^2)
 +\frac{v^2}{4} \left ( W_{\mu}^3,B_{\mu} \right )
  \left ( \begin{array}{cc}
       g_2^2  & g_1g_2 \\
       g_1g_2 & g_1^2
       \end{array} \right )
  \left ( \begin{array}{c}
          W_{\mu}^3 \\
          B_{\mu}
          \end{array} \right ) \; .
\end{equation}
The physical content becomes transparent by performing a
transformation from the fields $W_{\mu}^a$, $B_{\mu}$ (in terms of
which the symmetry is manifest) to the ``physical'' fields
\begin{equation}
 W_{\mu}^{\pm}=\frac{1}{\sqrt{2}}\, (W_{\mu}^1\pm i W_{\mu}^2)
\end{equation}
and
\begin{eqnarray}
 Z_{\mu} & = & +\cos\theta_W\, W_{\mu}^3\, +\sin\theta_W\, B_{\mu} \\
 A_{\mu} & = & -\sin\theta_W\, W_{\mu}^3\, +\cos\theta_W\, B_{\mu}
\nonumber
\end{eqnarray}
In these fields the mass term (15) is diagonal and has the form
\begin{equation}
 M_W^2\, W_{\mu}^+W^{- \mu}\, +\,
  \frac{1}{2}\, (A_{\mu},Z_{\mu})
  \left ( \begin{array}{cc}
        0 & 0 \\
        0 & M_Z^2
        \end{array} \right )
  \left ( \begin{array}{c}
         A^{\mu} \\
         Z^{\mu}
         \end{array} \right )
\end{equation}
with
\begin{eqnarray}
 M_W & = & \frac{1}{2}\, g_2 v   \\
 M_Z & = & \frac{1}{2}\sqrt{g_1^2+g_2^2}\, v  \nonumber
\end{eqnarray}
The  mixing angle in the rotation (17)
is given by
\begin{equation}
 \cos\theta_W =
 \frac{g_2}{\sqrt{g_1^2+g_2^2}} =\frac{M_W}{M_Z} \, .
\end{equation}
Identifying $A_{\mu}$ with the photon field which couples via  the
electric charge $e=\sqrt{4\pi\alpha}$ to the electron, $e$ can be
expressed in terms of the gauge couplings in the following way
\begin{equation}
 e= \frac{g_1g_2}{\sqrt{g_1^2+g_2^2}}
\end{equation}
or
\begin{equation}
 g_2 = \frac{e}{\sin\theta_W},\;\;\;\;
 g_1 = \frac{e}{\cos\theta_W} .
\end{equation}
Finally, from the Yukawa coupling terms in Eq.\ (13) the fermion
masses are obtained:
\begin{equation}
 m_f = g_f\, \frac{v}{\sqrt{2}} = \sqrt{2}\,\frac{g_f}{g_2}\,
       M_W \, .
\end{equation}

\medskip
\noi
The relations above
allow one to replace the original set of parameters
\begin{equation}
 g_2,\;     g_1,\;     \lambda,\;     \mu^2,\;     g_f
\end{equation}
by the equivalent set of more physical parameters
\begin{equation}
 e,\;     M_W,\;     M_Z,\;     M_H,\;     m_f
\end{equation}   where each of them can
(in principle) directly  be measured  in a suitable experiment.

\smallskip
An additional very precisely measured parameter is the Fermi constant
$\Gmu$ which is the effective 4-fermion coupling constant in the
the Fermi model, measured by the
muon lifetime:
              $$\Gmu=1.16639(2)\cdot 10^{-5}\, \gv^{-2}$$
Consistency of the \sm at $q^2 \ll M_W^2$ with the Fermi model
requires the identification
(see section 5)
\beq
\frac{\Gmu}{\sqrt{2}} \, =\, \frac{e^2}{8\siw M_W^2} \, ,
\eeq
which allows us to relate  the vector
boson masses to the parameters $\al,\, \Gmu$, and $\siw$
as follows:
\bea
 M_W^2 & = & \frac{\pi\al}{\sqrt{2}\Gmu}\cdot \frac{1}{\siw}  \nn \\
 M_Z^2 & = & \frac{\pi\al}{\sqrt{2}\Gmu}\cdot \frac{1}{\siw\cow}
\eea
and thus to establish also the $M_W-M_Z$ interdependence:
\beq
\mw\left(1-\frac{\mw}{\mz}\right)
= \frac{\pi\al}{\sqrt{2}\Gmu} \, .
\eeq
\subsection{Gauge fixing and ghost fields}
%text
Since the S matrix element for any physical process is a gauge
invariant quantity it is possible to work in the unitary gauge with
no unphysical particles in internal lines. For a systematic treatment
of the quantization of ${\cal L}_{cl}$ and for higher order
calculations, however, one better refers to a renormalizable
gauge. This can be done by adding to ${\cal L}_{cl}$ a gauge fixing
Lagrangian, for example
\begin{equation}
 {\cal L}_{fix} = - \frac{1}{2}\, \left (
 F_{\gamma}^2 + F_Z^2 + 2 F_+F_- \right )
\end{equation}
with linear gauge fixings of the 't Hooft type:
\begin{eqnarray}
 F_{\pm} & = & \frac{1}{\sqrt{\xi^W}} \,
   \left ( \partial^{\mu}W_{\mu}^{\pm} \mp i M_W \xi^W \phi^{\pm}
   \right ) \nn \\
 F_Z & = & \frac{1}{\sqrt{\xi^Z}} \,
 \left ( \partial^{\mu}Z_{\mu} - M_Z \xi^Z \chi \right ) \nonumber \\
 F_{\gamma} & = & \frac{1}{\sqrt{\xi^{\gamma}}} \,
     \partial^{\mu} A_{\mu}
\end{eqnarray}
with arbitrary parameters $\xi^{W,Z,\g}$.
In this class of 't Hooft gauges,
the vector boson propagators have the form
$$
 \frac{i}{k^2-M_V^2} \, \left ( -g^{\mu\nu} \, + \,
 \frac{(1-\xi^V) k^{\mu} k^{\nu}}{k^2-\xi^V M_V^2} \right )
$$
\beq  = \,
 \frac{i}{k^2-M_V^2} \, \left ( -g^{\mu\nu} \, + \,
 \frac{k^{\mu} k^{\nu}}{k^2} \right ) \,+\,
  \frac{i\xi^V}
 {k^2-\xi^V M_V^2}   \,
 \frac{ k^{\mu} k^{\nu}}{k^2} \, ,
\eeq
the propagators for the unphysical Higgs fields are given by
\bea
 \frac{i}{k^2-\xi^W M_W^2} & \mbox{for} & \phi^{\pm}    \\
 \frac{i}{k^2-\xi^Z M_Z^2} & \mbox{for} & \chi^0  \, ,
\eea
and Higgs-vector boson transitions do not occur.

\medskip
\noi
For completion of the renormalizable Lagrangian the Faddeev-Popov
ghost term ${\cal L}_{gh}$
has to be added \cite{fp}
 in order to balance the undesired effects
in the unphysical components introduced by ${\cal L}_{fix}$ :
\begin{equation}
 {\cal L} = {\cal L}_{cl}\,+\,{\cal L}_{fix}\,+\,{\cal L}_{gh}
\end{equation}
where
\begin{equation}
 {\cal L}_{gh} = \bar{u}^{\alpha}(x) \, \frac{\delta F^{\alpha}}
         {\delta \theta^{\beta}(x)} \, u^{\beta}(x)
\end{equation}
with ghost fields $u^{\gamma}$, $u^Z$, $u^{\pm}$, and
$\frac{\delta F^{\alpha}}{\delta \theta^{\beta}} $ being the
change of the gauge fixing operators (30) under infinitesimal
gauge transformations characterized by
$ \theta^{\alpha}(x) = \{ \theta^a(x), \, \theta^Y(x) \} $.

\bigskip
In the 't Hooft-Feynman gauge $(\xi = 1)$ the vector boson
propagators (31) become particularly simple: the transverse and
longitudinal components, as well as the propagators for the unphysical
Higgs fields $\phi^{\pm}$, $\chi$ and the ghost fields
$u^{\pm}$, $u^Z$ have poles which coincide with the masses of
the corresponding physical particles $W^{\pm}$ and $Z$.
\subsection{Feynman rules}
Expressed in terms of the physical parameters we can write down the
Lagrangian
$$ {\cal L} (A_{\m},W_{\m}^{\pm},Z_{\m},H,\phi^{\pm},\chi,
    u^{\pm},u^Z,u^{\g};\, M_W,
   M_Z,e,\dots) $$
 in a way which allows us to read off the propagators and
the vertices most directly. We specify them in the $R_{\xi=1}$ gauge
where the vector boson propagators have the simple algebraic
form $\sim g_{\m\nu}$.
\bea
 {\cal L}_G + {\cal L}_H \, =     \nn  \\
              &   &  \nn \\
 \frac{1}{2}
 A_{\m}\,\Box\, A^{\m}  & \;\;\; \lra \;\;\; &
 \frac{-i\, g_{\m\nu}}{q^2}   \nn \\
              &   &  \nn \\
 +\,W^-_{\m}\,(\Box+M_W^2)\,W^{+\m}  & \;\;\; \lra \;\;\; &
 \frac{-i\, g_{\m\nu}}{q^2-M_W^2}   \nn \\
              &   &  \nn \\
 +\, \frac{1}{2}
  Z_{\m}\,(\Box+M_Z^2)\,Z^{\m}  & \;\;\; \lra \;\;\; &
 \frac{-i\, g_{\m\nu}}{q^2-M_Z^2}  \nn     \\
              &   &  \nn \\
 +\, \frac{1}{2}
  H\,(\Box+M_H^2)\,H  & \;\;\; \lra \;\;\; &
 \frac{i}{q^2-M_H^2}   \nn \\
    &  & \nn \\
+ \, \mbox{interaction terms}    \nn \\
  VV,\; VH, \; HH   \nn \\
    &  &  \nn \\
+ \, (\mbox{unphysical degrees of freedom} )    \nn
\eea

\medskip
\bea
{\cal L}_F + {\cal L}_{Y\!ukawa} \, =  \nn \\
   &  &  \nn \\
\sum_f \, \bar{f}\,(i\partial - m_f) \, f
     & \;\;\; \lra \;\;\;  &
\frac{i}{q\dag -m_f}  \nn \\
   &  &  \nn \\
   &  &  \nn \\
+ \, J_{em}^{\m}\, A_{\m}
      & \;\;\; \lra \;\;\; &
 -i\,e\,Q_f \gamu \nn \\
   &  &  \nn \\
   &  &  \nn \\
+ \, J_{NC}^{\m}\, Z_{\m}
      & \;\;\; \lra \;\;\; &
 i\,\frac{e}{2\sinw\cosw}\,\gamu(v_f-a_f\gafi) \nn \\
   &  &  \nn \\
   &  &  \nn \\
+ \, J_{CC}^{\m}\, W_{\m}
      & \;\;\; \lra \;\;\; &
 i\,\frac{e}{2\sqrt{2}\sinw}\,\gamu(1-\gafi) \, V_{jk} \nn \\
   &  &  \nn \\
   &  &  \nn \\
- \, \frac{g_f}{\sqrt{2}}\, \bar{f} f\, H
      & \;\;\; \lra \;\;\; &  -
 i\,\frac{g_f}{\sqrt{2}} \,=\, i\,\frac{e}{2\sinw}\,\frac{m_f}{M_W}
  \nn \\
+ \, (\mbox{unphysical degrees of freedom} )          \\
   &  &  \nn
\eea
These Feynman rules
provide the ingredients to
calculate the lowest order amplitudes for fermionic processes.
For the complete list of all interaction vertices we refer
to the literature \cite{bhs}.

\bigskip
In order to describe scattering processes between light fermions in
lowest order we can, in most cases,
neglect the exchange of Higgs bosons because of their
small Yukawa couplings to the known fermions. The standard
processes
accessible by the experimental facilities are basically
4-fermion processes. These are mediated by the gauge bosons and,
sufficient in lowest order, defined by the vertices for the
fermions interacting with the vector bosons.               They
are given in the Lagrangian above for the electromagnetic,
neutral and charged current interactions.
The neutral current coupling constants in (36) read
 \begin{eqnarray}
  v_f & = &       I_3^f-2Q_f\,\sin^2\theta_W   \nn \\
  a_f & = &       I_3^f      \, .
 \end{eqnarray}
$Q_f$ and $I_3^f$ denote the charge and the third isospin component
of $f_L$.

\smallskip
The quantities $V_{jk}$ in the charged current  vertex
are the elements of the unitary 3$\times$3 matrix
\beq
 U_{KM} \,= \, \left(
\begin{array}{l l l}
 V_{ud} & V_{us} & V_{ub}  \\
 V_{cd} & V_{cs} & V_{cb}  \\
 V_{td} & V_{ts} & V_{tb}
\end{array} \right)
\eeq
which describes family mixing in the quark sector \cite{cabibbo}.
Its origin is the diagonalization of the quark mass
matrices from the Yukawa coupling which appears since
quarks of the same charge have
different masses. For massless neutrinos no mixing in the
leptonic sector is present. Due to the unitarity of $U_{KM}$
the mixing is absent in the neutral current.

\smallskip
For a proper treatment of the charged current vertex at the one-loop
level, the matrix $U_{KM}$ has to be renormalized as well.
As it was shown in \cite{ds},
 where the renormalization procedure was
extended to $U_{KM}$, the resulting effects
 are completely negligible for the
known light fermions. We therefore skip the renormalization of
$U_{KM}$ in our discussion of radiative corrections.
%%%%%%%%%%%%%%%%%%%%%%%%%%%%%%%%%%%%%%%%%%% end of section
\section{Renormalization}
 \subsection{General remarks}
The tree level Lagrangian (2) of the minimal SU(2)$\times$U(1)
model involves a certain number of free parameters which are not
fixed by the theory. The definition of these parameters and their
relation to measurable quantities is the content of a
renormalization scheme.   The parameters (or appropriate combinations)
can be determined from specific experiments with help of the
theoretical results for cross sections and lifetimes.
After this procedure of defining the physical input, other
observables can be predicted allowing verification or
 falsification
of the theory
by comparison with the corresponding experimental results.

\medskip
In higher order perturbation theory the
relations between the formal parameters and measureable quantities
are different from the tree level relations in general.
Moreover, the procedure is obscured by the appearence of divergences
from the loop integrations.
For a mathematically
 consistent treatment one has to regularize the theory, e.g.\
by dimensional regularization (performing the calculations in $D$
dimensions). But then the relations between physical quantities
and the parameters become cutoff dependent. Hence, the parameters
of the basic Lagrangian, the ``bare'' parameters, have no
physical meaning. On the other hand, relations between
measureable physical quantities, where the parameters
drop out, are finite and independent of the cutoff. It is therefore
in principle possible to perform tests of the theory in terms of
such relations by eliminating the bare parameters \cite{pa,pass}.

\smallskip
Alternatively, one may replace the bare parameters by renormalized ones
by multiplicative renormalization for each bare parameter  $g_0$
\beq
   g_0 = Z_g\, g = g +\delta g
\eeq
with renormalization constants $Z_g$ different from 1 by a
1-loop term.
The renormalized parameters $g$  are finite and fixed by a set of
renormalization conditions. The decomposition (39) is to a large
extent arbitrary. Only the divergent parts are determined directly
by the structure of the divergences of the one-loop amplitudes.
The finite parts depend on the choice of the explicit
renormalization conditions.

\smallskip
This procedure of parameter renormalization is sufficient to obtain
finite S-matrix elements when wave function renormalization for
external on-shell particles is included.
Off-shell Green functions, however, are not
finite by themselves. In order
obtain finite propagators and vertices, also the bare fields in
$\cal L$ have to be redefined in terms of renormalized fields
by multiplicative renormalization
\beq
   \phi_0 = Z_{\phi}^{1/2}\, \phi \, .
\eeq

\smallskip \noi
Expanding the renormalization constants according to
$$ Z_i = 1 + \delta Z_i $$
the  Lagrangian is split into a
``renormalized'' Lagrangian and a counter term Lagrangian
\beq
 {\cal L}(\phi_0,g_0) =
 {\cal L}(Z_{\phi}^{1/2}\phi, Z_g g) =
 {\cal L}(\phi,g)
  + \delta {\cal L}(\phi,g,\delta Z_{\phi},\delta g)
\eeq
 which
renders the results for all Green functions in a given order
finite.

\medskip
The simplest way to obtain a set of finite Green functions is
the ``minimal subtraction scheme'' \cite{minsub}
 where (in dimensional
regularization) the singular part of each divergent diagram is
subtracted and the parameters are defined at an arbitrary
mass scale $\mu$. This scheme, with slight modifications,  has
 been applied in QCD where due to
the confinement of quarks and gluons  there is no
distinguished  mass scale in the renormalization
procedure.

\medskip
The  situation is different in QED and in the electroweak theory.
There the classical Thomson scattering and the particle masses
set natural scales where the parameters can be defined. In QED
the favoured renormalization scheme is the on-shell scheme
where $e=\sqrt{4\pi\alpha}$  and the electron, muon, \dots masses
are used as input parameters.
The finite parts of the counter terms are fixed by the
renormalization conditions that the fermion propagators have poles
at their physical masses, and $e$ becomes the $ee\gamma$ coupling
constant in the Thomson limit of Compton scattering. The
extraordinary meaning of the Thomson limit for the definition of the
renormalized
coupling constant is elucidated by the theorem  that the exact
 Compton cross section at low energies becomes equal to the
classical
Thomson cross section. In particular this means that $e$ resp.\
$\alpha$ is free of infrared corrections, and that its numerical
value is independent of the order of perturbation theory,
only determined by the accuracy of the experiment.

\medskip
This feature of $e$ is preserved in the electroweak theory.
In the electroweak Standard Model a distinguished set
for parameter renormalization is given in terms of
$e,M_Z,M_W,M_H,m_f$ with the masses of the corresponding particles.
This electroweak on-shell scheme is the straight-forward
 extension of the
familiar QED renormalization, first proposed by Ross and Taylor
\cite{ross}
and used in many practical applications
\cite{bhs,pav,con,sirmar,dubna,fjeg,aoki,maiani,dz,hollik,LEP}.
For stable particles, the masses are well defined quantities
and can be measured with high accuracy.
The masses of the $W$ and $Z$ bosons are related to the
resonance peaks in cross sections where they are produced and
hence can also be accurately determined. The mass of the Higgs boson,
as long as it is  experimentally unknown,
is treated as a free input parameter.
The light quark masses can only be considered as effective
parameters. In the cases of practical interest
 they can be replaced in terms of directly
measured quantities like the cross section for
$\epm \ra \mbox{hadrons}$.

\bigskip
The electroweak mixing angle is related to the
 vector boson masses in general
by
\beq
\siw \, =\, 1 - \frac{M_W^2}{\rho_0\,M_Z^2}
\eeq
where $\rho_0\neq 1$ at the tree level in case of a
Higgs system
 more complicated
 than with doublets only. We
want to restrict
our discussion of \rc primarily
 to the minimal model with $\rho_0=1$.
For $\rho\neq 1$ see section 9.2.

\medskip \noi
Instead of the set  $e,\,M_W,\,M_Z$ as  basic free
parameters one may
alternatively use as basic parameters $\al$, $\Gmu$,
$M_Z$ \cite{pittau}
 or $\al$, $\Gmu$, $\siw$ with the mixing angle deduced
from neutrino-electron scattering \cite{green}
 or perform the loop
calculations in the $\overline{MS}$ scheme
 \cite{pavelt,msbar,msbar1,msbar2}.
The so-called $*$-scheme \cite{star,star1}
 is a different way of book-keeping
in terms of effective running couplings.
Here we follow the line of the on-shell scheme as specified in
detail in \cite{bhs,hollik}, but skip field renormalization.
\subsection{Mass renormalization}
We have now to discuss the 1-loop contributions to the
on-shell
parameters and their renormalization. Since the boson masses are part
of the propagators we have to investigate the effects of the $W$ and
$Z$ self-energies.

\bigskip
We restrict our discussion to the transverse parts $\sim g_{\m\nu}$.
In the electroweak theory, differently from QED,
the longitudinal components $\sim q_\mu q_\nu$ of the
vector boson propagators
do not give zero results in physical
matrix elements. But for light external fermions the contributions
are suppressed by $(m_f/M_Z)^2$ and we are allowed to neglect them.
Writing the self-energies as
\beq
\Sigma^{W,Z}_{\m\nu} = g_{\m\nu} \Sigma^{W,Z} + \cdots
\eeq
with scalar functions $\Sigma^{W,Z}(q^2)$
we have for the
1-loop propagators ($V=W,Z$)
\beq
 \frac{-ig^{\m\sigma}}{q^2-M_V^2} \left(-i\,\Sigma^V_{\rho\sigma}
 \right) \frac{-ig^{\rho\nu}}{q^2-M_V^2} =
 \frac{-i g^{\m\nu}}{q^2-M_V^2} \left( \frac
 {-\Sigma^V(q^2)}{q^2-M_V^2} \right)
\eeq

\medskip \noi
(the factor $-i$ in the self energy insertion is a convention).
Besides the fermion loop contributions in the electroweak
theory there are also the non-Abelian gauge boson loops and loops
involving the Higgs boson. The Higgs boson and the top quark thus
 enter the 4-fermion amplitudes as experimentally
unknown objects at the level of radiative corrections and have
to be treated as additional free parameters.
  In the graphical representation, the self-energies
for the vector bosons denote the sum of all the diagrams
with  virtual fermions, vector bosons, Higgs and ghost loops. 

\bigskip
Resumming  all self energy-insertions  yields a
geometrical series for the dressed propagators:
\bea   &   &
\frac{-ig_{\m\nu}}{q^2-M_V^2} \left[1 + \left(
\frac{-\Sigma^V}{q^2-M_V^2} \right) + \left(
\frac{-\Sigma^V}{q^2-M_V^2} \right)^2 + \cdots \right] \nn \\
 &  & = \, \frac{-i g_{\m\nu}}{q^2-M_V^2+\Sigma^V(q^2) } \, .
\eea

\medskip \noi
The self-energies have the following properties:
\begin{itemize}
\item
$\imag \Sigma^V(M_V^2) \neq 0$ for both $W$ and $Z$.
This is because $W$ and $Z$ are unstable particels and can decay
into pairs of light fermions. The imaginary parts correspond to the
total decay widths of $W,\,Z$
and remove the poles from the real axis.
\item
$ \real \Sigma^V(M_V^2) \neq 0$ for both $W$ and $Z$ and they are
UV divergent.
\end{itemize}

The second feature tells us that the locations of the poles in
the propagators are shifted by the loop contributions.
Consequently, the principal step in \it mass renormalization
\rm consists in a re-interpretation of the parameters: the masses
in the Lagrangian cannot be the physical masses of $W$ and $Z$
but are the
``bare masses''
 related to the physical masses $M_W,\,M_Z$
by
\bea
 \mwb & = & M_W^2 + \dmmw  \\
 \mzb & = & M_Z^2 + \dmmz  \nn
\eea
with counterterms of 1-loop order. The ``correct'' propagators
according to this prescription are given by
\beq
\frac{-ig_{\m\nu}}{q^2-M_V^{0\,2} + \Sigma^V(q^2) } \, =\,
\frac{-ig_{\m\nu}}{q^2-M_V^2-\delta M_V^2 + \Sigma^V(q^2) } \, .
\eeq              instead of Eq.\ (45).
The renormalization conditions which ensure that $M_{W,Z}$ are the
physical masses fix the mass counterterms to be
\bea
\dmmw & = & \real\, \Sigma^W(M_W^2)   \\
\dmmz & = & \real \, \Sigma^Z(M_Z^2) \, . \nn
\eea
In this way, two of our input parameters and their counterterms have
been defined.
\subsection{Charge renormalization}
Our third input parameter is the electromagnetic charge $e$.
The electroweak \it charge renormalization \rm is very similar to that
in pure QED.
As in QED, we want to maintain the definition of $e$ as the classical
charge in the Thomson cross section
$$
\sigma_{Th} = \frac{e^4}{6\pi\,m_e^2} \, .
$$
 Accordingly, the Lagrangian carries the bare charge
 $ e_0 =  e+\delta e$ with the charge counter term
$\delta e$ of 1-loop order.
The  charge counter term $\delta e$ has to absorb the
electroweak loop contributions to the $ee\g$ vertex in the Thomson
limit.
 This charge renormalization condition is simplified by the validity
of a generalization of the QED Ward identity \cite{ward}
 which implies that those
corrections related to the external particles cancel
each other. Thus for $\delta e$  only two universal
contributions are left:
\beq
 \frac{\delta e}{e} \,=\, \frac{1}{2}\, \Pi^{\g}(0) \, -\,
\frac{s_W}{c_W}\,\frac{\Sigma^{\g Z}(0)}{M_Z^2}  \, .
\eeq
The first one, quite in analogy to QED,
is given by the vacuum polarization of
the photon.
But now, besides the fermion loops, it
contains also bosonic loop diagrams from $W^+W^-$ virtual states and the
corresponding ghosts.
The second term contains the mixing between photon and $Z$, in general
described as a mixing propagator with $\sgz$ normalized as  
$$
 \Delta^{\g Z} =
  \frac{-ig_{\m\nu}}{q^2} \left(
\frac{-\Sigma^{\g Z}(q^2)}{q^2-M_Z^2} \right) \, .  $$
The fermion loop contributions to $\Sigma^{\g Z}$ vanish at
$q^2=0$; only
the non-Abelian bosonic loops yield $\Sigma^{\g Z}(0)\neq 0$.

\smallskip
To be more precise, the charge renormalization as discussed
above, is a condition only for the vector coupling constant of
the photon. The axial coupling vanishes for on-shell photons
as a consequence of the Ward identity.

\bigskip
From the diagonal photon self-energy
 $$ \Sigma^{\g}(q^2) = q^2\,\Pi^{\g}(q^2)  $$
no  mass term arises for the photon since, besides the fermion
loops, also the bosonic loops behave like
$$ \Sigma_{bos}^{\g}(q^2) \simeq q^2\, \Pi^{\g}_{bos}(0) \ra 0 $$
for $q^2\ra 0$ leaving the pole at $q^2=0$ in the propagator.
The absence of mass terms for the photon in all orders
                               is a consequence of the
unbroken electromagnetic gauge invariance.

\bigskip
Concluding this discussion we summarize the principal structure
of electroweak calculations:

\begin{itemize}
\item
The classical Lagrangian ${\cal L}(e,M_W,M_Z,\dots)$
is  sufficent for lowest order
calculations and the parameters can be identified with the physical
parameters.
\item
For higher order calculations, $\cal L$ has to be considered as the
``bare'' Lagrangian of the theory ${\cal L}(e_0,M_W^0,M_Z^0,\dots)$
with  ``bare'' parameters which are
related to the physical ones by
$$ e_0 =e+\delta e,\;\;\; \mwb = M_W^2 +\dmmw,\;\;\;
   \mzb = M_Z^2 + \dmmz \, . $$
The counter terms are fixed
  in terms of a certain subset of 1-loop diagrams
  by specifying the definition of the physical parameters.
\item     For any 4-fermion process
                               we can write down the 1-loop matrix
element with the bare parameters and the
loop diagrams for this process. Together with
the counter terms
the matrix element is finite when expressed in terms of the
physical parameters, i.e.\ all UV
singularities
are removed.
\end{itemize}
\section{One-loop calculations}
In this section we provide technical details for the
calculation of radiative corrections for electroweak precision
observables. The methods used are essentially based on the work
of \cite{pav} and \cite{thooftvelt}.
 \subsection{Dimensional regularization}
The diagrams with closed loops occuring in higher order
perturbation theory involve integrals over the loop momentum.
These integrals are in general divergent for large integration
momenta (UV divergence). For this  reason
we need a regularization, which is a procedure to redefine the integrals
in such a way that they become finite and mathematically well-defined
objects.
The widely used regularization procedure for gauge theories is that
of dimensional regularization \cite{dimreg},
 which is Lorentz and gauge invariant:
replace the dimension 4 by a lower
dimension $D$ where the integrals are convergent:
\beq
 \int \dk \, \ra \,        \dkm
\eeq
An (arbitrary) mass parameter $\m$ has been introduced in order to
keep the dimensions of the coupling constants in front of the
integrals independent of $D$. After renormalization the results for
physical quantities are finite in the limit $D\ra 4$.

\smallskip \noi
The metric tensor in $D$ dimensions has the property
\beq
 g_{\m}^{\m} = g_{\m\nu} g^{\nu\m} = \mbox{Tr}(1) = D \, .
\eeq
The Dirac algebra in $D$ dimensions
 \beq
%\{\gamu,\ganu\}\,= \, 2\, g_{\m\nu}\, \mbox{\boldmath $1$}
 \{\gamu,\ganu\}\,= \, 2\, g_{\m\nu}\, \bf 1
\eeq
has the consequences
\bea
%\gamu\gimu & = & D\, \mbox{\boldmath $1$} \nn \\
 \gamu\gimu & = & D\, \bf 1 \nn \\
 \garo\gamu\giro & = & (2-D)\, \gamu \nn \\
%\garo\gamu\ganu\giro & = & 4 g_{\m\nu}\, \mbox{\boldmath $1$}\,
 \garo\gamu\ganu\giro & = & 4 g_{\m\nu}\, {\bf 1}\,
    -\,(4-D)\,\gamu\ganu \nn \\
 \garo\gamu\ganu\gasi\giro & = & -2\,\gasi\ganu\gamu \,+\,
    (4-D)\,\gamu\ganu\gasi
\eea
A consistent treatment of $\gafi$ in $D$ dimensions is more
subtle \cite{breitmaison}.
 In theories which are anomaly free like the
 Standard Model we can use $\gafi$ as anticommuting with $\gamu$:
    \beq
\{\gamu,\gafi\} \, =\, 0 \, .
\eeq
\subsection{One- and two-point integrals}
In the calculation of self energy diagrams the following types of
one-loop integrals appear:

\medskip \noi
\underline{1-point integral:}
\bea
     &     &   \nn \\
     \dkm\,\frac{1}{k^2-m^2} & =: & \frac{i}{16\pi^2}\,A(m) \\
     &     &   \nn
\eea

 \noi
\underline{2-point integrals:}
\bea
     &     &   \nn \\
     \dkm \,\frac{1}{\Dkk\Dkq} & =: & \frac{i}{16\pi^2} \,
           B_0(q^2,m_1,m_2)      \\
     &     &   \nn \\
\dkm \,\frac{\km;\,\kmn}{\Dkk\Dkq} & =: & \frac{i}{16\pi^2} \,
           B_{\m;\, \m\nu}(q^2,m_1,m_2) \, . \\
     &     &   \nn
\eea
The vector and tensor integrals $B_{\m},\,B_{\m\nu}$ can be
expanded into Lorentz covariants and scalar coefficients:
\bea
 B_{\m}    & = &     q_{\m}\,
          B_1(q^2,m_1,m_2) \nn \\
 B_{\m\nu} & = &
g_{\m\nu} B_{22}(q^2,m_1,m_2) \,+\,
 q_{\m}q_{\nu} B_{21}(q^2,m_1,m_2)  \, .
\eea

\bigskip \noi
The coefficient functions can be obtained algebraically from the
scalar 1- and 2-point integrals $A$ and $B_0$.
Contracting (58) with $q^{\m},\,
 g^{\m\nu}$ and
$ q^{\m}q^{\nu}$
yields:
\bea
\int \frac{kq}{\Dkk\Dkq} & = & \ipi \,  q^2 B_1    \nn \\
         &   &  \nn \\
\int
     \frac{k^2}{\Dkk\Dkq} & = & \ipi \left( D B_{22} + q^2 B_{21}
                               \right)     \nn \\
\int \frac{(kq)^2}{\Dkk\Dkq} & = & \ipi \left( q^2 B_{22}
                                 + q^4 B_{21} \right) \, . \\
   &   & \nn
\eea
Solving these equations and making use of the decompositions
\bea
   &   & \nn \\
\int \frac{k^2}{\Dkk\Dkq} & = & \int \frac{1}{k^2-m_2^2}  +
m_1^2 \int\frac{1}{\Dkk\Dkq}  \nn \\
   &   & \nn \\
% & = & \ipi \left\{ A(m_2) + m_1^2 B_0(q^2,m_1,m_2) \right\}\,,\nn\\
%  &   & \nn \\
 \int\frac{kq}{\Dkk\Dkq} & = & \frac{1}{2} \,
 \int\frac{1}{k^2-m_1^2} - \frac{1}{2} \,
  \int\frac{1}{k^2-m_2^2}  \nn \\
 &   &  +\frac{m_2^2-m_1^2-q^2}{2}\, \int\frac{1}{\Dkk\Dkq}
                             \nn \\
   &   & \nn \\
\int\frac{(kq)^2}{\Dkk\Dkq} & = & \frac{1}{2} \int
 \frac{kq}{k^2-m_1^2} \,- \,\frac{1}{2}\int \frac{kq}
                               {(k+q)^2-m_2^2}         \nn \\
 &  & +\frac{m_2^2-m_1^2-q^2}{2}\int\frac{kq}{\Dkk\Dkq}  \nn \\
 &   &   \nn
\eea
and of the definition (56,57) we obtain:
\bea
  &    & \nn \\
 B_1(q^2,m_1,m_2) & = & \frac{1}{2q^2} \left[
  A(m_1)-A(m_2) +(m_1^2-m_2^2-q^2)
                       B_0(q^2,m_1,m_2)\right] \nn \\
  &    & \nn \\
 B_{22}(q^2,m_1,m_2) & = & \frac{1}{6}\, \left[
       A(m_2) + 2 m_1^2 B_0(q^2,m_1,m_2) \right.  \nn \\
 &   & \left. +  (q^2+m_1^2-m_2^2) B_1(q^2,m_1,m_2)
        +m_1^2+m_2^2 -\frac{q^2}{3} \right]  \nn   \\
   &   &  \nn \\
 B_{21}(q^2,m_1,m_2) & = & \frac{1}{3q^2}\, \left[
       A(m_2) -m_1^2 B_0(q^2,m_1,m_2) \right.  \nn \\
 &   & \left. - 2  (q^2+m_1^2-m_2^2) B_1(q^2,m_1,m_2) -
  \frac{m_1^2+m_2^2}{2} +\frac{q^2}{6} \right]  \, .  \\
   &   &  \nn
\eea

\bigskip \noi
Finally we have to calculate the scalar integrals $A$ and $B_0$.
 With help of the
Feynman parametrization
$$ \frac{1}{ab} = \int^1_0 dx \,
   \frac{1}{[ax+b(1-x)]^2}  $$
and after a shift in the k-variable, $B_0$ can be written in the form
\beq   \ipi\, B_0(q^2,m_1,m_2)
          = \int^1_0 dx \, \frac{\m^{4-D}}{(2\pi)^D} \int
 \frac{d^Dk}{[k^2-x^2q^2+x(q^2+m_1^2-m_2^2)-m_1^2]^2 } \, .
\eeq
The advantage of this parametrization is a simpler $k$-integration
where the integrand is only a function of $k^2=(k^0)^2-\vec{k}^2$.
In order to transform it into a Euclidean integral we perform the
substitution
\footnote{The $i\eps$-prescription in the masses ensures that this is
compatible with  the pole structure of the integrand.}
$$ k^0 = i\,k_E^0,\;\, \vec{k} =\vec{k_E},\;\;
   d^D k = i\,d^D k_E   $$
where the new integration momentum $k_E$ has a definite metric:
$$ k^2 = -k_E^2, \;\; \;
   k_E^2 = (k^0_E)^2 + \cdots + (k_E^{D-1})^2 \, .   $$
This leads us to a Euclidean integral over $k_E$:
\beq
\ipi\, B_0 = i \int^1_0 dx \frac{\m^{4-D}}{(2\pi)^D}
\int \frac{d^Dk_E}{(k_E^2 + Q)^2 }
\eeq
where
\beq
 Q = x^2q^2-x(q^2+m_1^2-m_2^2)+m_1^2 - i\veps
\eeq
is a constant with respect to the $k_E$-integration.

\medskip
Also the 1-point integral $A$ in (55) can be transformed into a
Euclidean integral:
\beq
\ipi \,A(m) = -i\, \frac{\m^{4-D}}{(2\pi)^D}
\int \frac{d^Dk_E}{k_E^2+m^2} \, .
\eeq
Both $k_E$- integrals are of the general type
$$ \int \frac{d^Dk_E}{(k_E^2+L)^n}   $$
of rotational invariant integrals in a $D$-dimensional Euclidean
space. They can be evaluated in $D$-dimensional polar
coordinates ($k_E^2 = R$)
$$ \int\frac{d^Dk_E}{(k_E^2+L)^n}\, =\, \frac{1}{2}
\int d\Omega_D \int^{\infty}_0 dR\, R^{\frac{D}{2}-1} \,
\frac{1}{(R+L)^n}  \, ,  $$
yielding
\beq
\frac{\m^{4-D}}{(2\pi)^D} \int \frac{d^Dk_E}{(k_E^2+L)^n} \, =\,
\frac{\m^{4-D}}{(4\pi)^{D/2}} \cdot
\frac{\Gamma(n-\frac{D}{2})}{\Gamma(n)}\cdot L^{-n+\frac{D}{2}} \, .
\eeq

\medskip \noi
The singularities of our initially 4-dimensional integrals are now
recovered
as poles of the $\Gamma$-function for $D=4$ and  values
$n  \leq 2$.

\medskip
Although the l.h.s. of
Eq.\ (65) as a $D$-dimensional integral is sensible
   only for integer values of $D$, the r.h.s. has
an analytic continuation in the variable $D$: it is well defined for
all complex values $D$ with $n-\frac{D}{2}\neq 0,-1,-2,\dots$,
 in particular for
$$  D = 4 -\eps \;\;\; \mbox{ with } \eps >  0 \, . $$
For physical reasons we are interested in the vicinity of $D=4$.
Hence  we consider the limiting case $\eps \ra 0$
and perform an expansion around $D=4$ in powers of $\eps$.
For this task we need the following properties of the
$\Gamma$-function at  $x\ra 0$:
\bea      &   &
 \Gamma(x)   =   \frac{1}{x}\, - \,\g\, +\,O(x) \, , \nn \\
          &   &
 \Gamma(-1+x)   =   -\,\frac{1}{x} \,+\,\g \,-\, 1 \,+\,O(x)
\eea
with
$$ \g = -\,\Gamma'(1) = 0.577\dots $$
known as Euler's constant.

\bigskip   \noi
$n=1$:

\smallskip \noi
Combining (64) and (65) we obtain the scalar 1-point integral for
$D=4-\eps$:
\bea
A(m) & = & -\,\frac{\m^{\eps}}{(4\pi)^{-\eps/2}} \cdot
    \frac{\Gamma(-1+\frac{\eps}{2})}{\Gamma(1)} \cdot
    \left(m^2 \right)^{1-\eps/2 }  \nn \\
    & = &  m^2 \left(\frac{2}{\eps}-\g +\log 4\pi
           -\log\frac{m^2}{\m^2} + 1 \right)
          +\,O(\eps)  \nn \\
    & \equiv &  m^2 \left( \Delta
           -\log\frac{m^2}{\m^2} + 1 \right)
          +\,O(\eps)
\eea
Here we have introduced the abbreviation for the singular part
\beq
\Delta  = \frac{2}{\eps} -\g + \log 4\pi \, .
\eeq

\bigskip \noi
$n=2 :$

\smallskip \noi
For the scalar 2-point integral $B_0$ we evaluate the integrand of
the $x$-integration in Eq.\ (62) with help of Eq.\ (65) as follows:
\bea
\frac{\m^{\eps}}{(4\pi)^{2-\eps/2}} \cdot
\frac{\Gamma(\frac{\eps}{2})}{\Gamma(2)} \cdot
 Q^{-\eps/2}   & = & \frac{1}{16\pi^2} \left(
\frac{2}{\eps} -\g + \log 4\pi -\log\frac{Q}{\m^2} \right)
+\, O(\eps)  \nn \\
 & = & \frac{1}{16\pi^2} \left( \Delta -\log\frac{Q}{\m^2} \right)
        +\, O(\eps) \, .
\eea
Since the $O(\eps)$ terms vanish in the limit $\eps\ra 0$ we skip
them in the following formulae. Insertion into Eq.\ (62)
 with $Q$ from
Eq.\ (63) yields:
\bea
  &    & \nn \\
B_0(q^2,m_1,m_2) & = & \Delta \, -
\int^1_0 dx\,\log\frac{x^2 q^2-x(q^2+m_1^2-m_2^2)+m_1^2-i\veps}{\m^2} \\
  &    &  \nn
\eea
The explicit analytic formula can be found in \cite{bhs}.

\bigskip
For the calculation of one-loop amplitudes also 3- and 4-point
functions have to be included.
In  low energy processes, like muon decay or neutrino scattering,
where the external momenta can be neglected in view of the internal
gauge boson masses, the 3-point and 4-point integrals
 can immediately be
reduced to 2-point integrals.
The analytic results for the $\g,Zff$ vertices
 can be found in the literature \cite{hollik}.
Massive box diagrams are negligible around the $Z$ resonance.
\subsection{Vector boson self energies}
The diagrams contributing to the self energies of the photon,
$W,\, Z$ and the photon-$Z$ transition contain fermion, vector boson,
Higgs and ghost loops. Here we 
 consider the fermion loops in more detail, since they yield the 
biggest contributions. \\
\underline{Photon self energy}:
\vglue 0.2cm
We give the expression for a single fermion with charge $Q_f$ and
mass $m$. The total contribution is obtained by summing over
all fermions.
 Evaluating the fermion loop diagram we obtain in the notation
of section 4.2:
\bea
\Sigma^{\g}(k^2)
 & = &\frac{\al}{\pi}\,Q_f^2\, \{-A(m)+\frac{k^2}{2}
             B_0(k^2,m,m) + 2\,B_{22}(k^2,m,m) \} \nn \\
  &   &  \nn \\
                 & = & \frac{\al}{3\pi}\,Q_f^2 \left\{
k^2 \left( \Delta-\log\frac{m^2}{\m^2} \right) + \,
(k^2+2 m^2) \, \bar{B}_0(k^2,m,m)\,-\,
\frac{k^2}{3}  \right\}    .  \\
  &    &   \nn
\eea
$\bar{B}_0$ denotes the finite function
\beq
 \bar{B}_0(k^2,m,m) \,=\, - \int^1_0 dx\,\log \left(
 \frac{x^2k^2-x k^2 +m^2}{m^2}-i\veps \right)
\eeq
in the decomposition
\beq
 B_0(k^2,m,m) \,=\, \Delta - \log\frac{m^2}{\m^2}
  \,+\, \bar{B}_0(k^2,m,m) \, .
\eeq

\smallskip \noi
The dimensionless quantity
\beq
 \Pi^{\g}(k^2) \, =\, \frac{\Sigma^{\g}(k^2)}{k^2}
\eeq
is usually denoted as the photon ``vacuum polarization''.
We list two simple expressions arising from Eq.\ (71) for
special situations of practical interest:
 \begin{itemize}
\item
light fermions ($\mid k^2 \mid \gg m^2$):
\beq
 \Pi^{\g}(k^2) = \frac{\al}{3\pi}Q_f^2 \left( \Delta
 -\log\frac{m^2}{\m^2} +\frac{5}{3} -\log\frac{\mid k^2\mid}{m^2}
\, + \, i \pi \,\theta(k^2) \right)
\eeq
\item heavy fermions ($\mid k^2\mid \ll m^2$):
\beq
\Pi^{\g}(k^2) = \frac{\al}{3\pi}Q_f^2 \left( \Delta -
\log\frac{m^2}{\m^2} + \frac{k^2}{5m^2} \right)
\eeq
\end{itemize}
\vglue 0.2cm       \noi
\underline{Photon - $Z$ mixing: }
\vglue 0.2cm   \noi
Each charged fermion yields a
contribution
\bea
\Sigma^{\g Z}(k^2) & = &
    - \frac{\al}{3\pi}\,
        \frac{v_fQ_f}{2s_Wc_W}
       \left\{
k^2 \left( \Delta-\log\frac{m^2}{\m^2} \right) + \,
(k^2+2 m^2) \, \bar{B}_0(k^2,m,m)\,-\,
\frac{k^2}{3}  \right\} \, .  \nn \\
\eea
As in the photon case, the fermion loop contribution
vanishes for $k^2=0$. \\

\bigskip \noi
\underline{$Z$ and $W$ self energies:}
\vglue 0.2cm \noi
We give the formulae for a  single doublet,
leptons or quarks, with $m_{\pm},\, Q_{\pm},\, v_{\pm}, \, a_{\pm}$
denoting mass, charge, vector and axial vector coupling of the
up(+) and the down(-) member. At the end, we have to
 perform the sum over the
various doublets,
 including color summation.
\bea
\sz (k^2)\,= \,\frac{\al}{\pi} \sum_{f=+,-} &\left\{ \right.&
\frac{v_f^2+a_f^2}{4s_W^2c_W^2} \left[
 2B_{22}(k^2,m_f,m_f)+\frac{k^2}{2}B_0(k^2,m_f,m_f)-A(m_f)\right]\nn\\
 &  & - \,\frac{m_f^2}{8s_W^2c_W^2}\, B_0(k^2,m_f^2,m_f^2)
      \left.   \right\}  \nn  \\
 &   &   \nn \\
\sw  (k^2) \, =\, \frac{\al}{\pi} \cdot
   \frac{1}{4s_W^2}
 &\left\{ \right.&
 2 B_{22}(k^2,m_+,m_-)-\frac{A(m_+)+A(m_-)}{2}  \nn \\
 &   &  +\frac{k^2-m_+^2-m_-^2}{2} \, B_0(k^2,m_+,m_-)
         \left. \right\}
\eea
%
%  soweit in egmont
%
%

\medskip \noi
Again, the following two cases are of particular practical
interest:
\begin{itemize}
\item Light fermions:
\end{itemize}
In the light fermion limit
 $ k^2 \gg  m_{\pm}^2$ the$Z$ and $W$
self-energies simplify considerably:
\bea
%\frac{\Sigma^W(0)}{M_W^2} = O\left(\frac{m_{\pm}^2}{M_W^2}\right)
%  \simeq 0  \, ,   \nn \\
&   & \sz (k^2) \, =\,
 \frac{\al}{3\pi}\cdot \frac{v_+^2+a_+^2+v_-^2+a_-^2}
       {4 s_W^2 c_W^2}\, k^2\, \left( \deltak \, +\, i\pi
        \right) \, ,  \nn \\
&   & \sw (k^2) \, =\,
 \frac{\al}{3\pi}\cdot \frac{k^2}{4 s_W^2}
             \left( \deltak \,+\, i\pi \right)  \, .
%- \frac{\al}{16\pi s_W^2} \, \log c_W^2 \, ,\nn  \\
\eea

\begin{itemize}
\item Heavy fermions:
\end{itemize}
Of special interest is the case of a heavy top quark which yields
a large correction $\sim m_t^2$. In order to extract this
part we keep for simplicity only those terms which are either
singular or quadratic in the top mass $m_t\equiv m_+$ ($N_C=3$):
\bea
%&   & \real\, \Pi^{\g}(M_Z^2) = N_C \,\frac{\al}{3\pi}
%     (Q_+^2+Q_-^2) \left(\deltat\right) \, +\cdots   \nn \\
%&    & \nn \\
&   & \sz (k^2) \, = \,
 N_C\,\frac{\al}{3\pi} \left\{
     \frac{v_+^2+a_+^2+v_-^2+a_-^2}{4s_W^2 c_W^2} \,k^2 \, -
     \frac{3m_t^2}{8s_W^2c^2_W} \right \}
       \left( \deltat \right) \, + \cdots   \nn \\
 &  &   \nn \\
&   & \sw (k^2) \, = \,
 N_C\,\frac{\al}{3\pi} \left\{
      \frac{k^2}{4s_W^2} \left(\deltat\right) -
      \frac{3m_t^2}{8s_W^2} \left(\deltat +\frac{1}{2}
       \right) \right\}  \,+ \cdots  \nn \\
\eea

\medskip \noi
The quantity \cite{rho}            \beq \dro =
   \frac{\sz(0)}{M_Z^2}-\frac{\sw(0)}{M_W^2} \eeq
is finite as far as the leading fermion contribution is considered
which yields for the top quark:
\beq
\dro = N_C\,\frac{\al}{16\pi s_W^2 c_W^2} \,
       \frac{m_t^2}{M_Z^2} \, .
\eeq

%%%%%%%%%%%%%%%%%%%%%%%%%%%%%%%%%%%%%%%%%%%% end of section
%
%\vglue 0.6cm
\section{The vector boson  masses}
 \subsection{One-loop corrections to the muon lifetime}
%text
The interdependence between the gauge boson masses is established
through the accurately measured muon lifetime or the Fermi
coupling constant $\Gmu$, respectively.
Originally, the $\m$-lifetime $\tau_{\m}$ has been calculated
within the framework of the effective 4-point Fermi interaction.
If QED corrections are included one
obtains the result \cite{muon}
\beq
\frac{1}{\tau_{\m}} = \frac{\Gmu^2m_{\m}^5}{192\pi^3}
\left( 1-\frac{8m_e^2}{m_{\m}^2} \right)
\left[ 1+\frac{\al}{2\pi} (\frac{25}{4}-\pi^2) \right] \, .
\eeq
The leading $2nd$ order correction  is  obtained by replacing
$$ \al \ra \al \left(1+\frac{2\al}{3\pi} \log
   \frac{m_{\m}}{m_e} \right) \, . $$
This formula is used as the defining equation for $\Gmu$ in terms
of the experimental $\m$-lifetime. In lowest order, the Fermi
constant is  given by the \sm expression (26) for the decay
amplitude. In 1-loop order, $\Gmu/\sqrt{2}$ coincides with the
expression
\beq
\frac{\Gmu}{\sqrt{2}} \, = \,
\frac{e_0^2}{8\swo \mwb} \left[ 1+\frac{\Sigma^W(0)}{M_W^2}
+\,(      vertex, box ) \right] \, .
\eeq
This equation  contains the bare parameters
with  the bare mixing angle
\beq
\swo = 1 -\frac{\mwb}{\mzb} \, .
\eeq
The term $(vertex,box)$ schematically summarizes the vertex
corrections and box diagrams in  the decay
amplitude.
A set of infra-red divergent ``QED correction'' graphs has
been   removed from this class of diagrams. These left-out
diagrams, together with the real bremsstrahlung contributions,
reproduce the QED correction factor of the Fermi
model result in Eq.\ (83) and therefore have no influence on the
relation between $\Gmu$ and the \sm parameters.

 \bigskip                \noi
Next we evaluate Eq.\ (84) to 1-loop order by expanding the bare
parameters
\bea
 e_0^2 & = & (e+\delta e)^2 \; =\; e^2(1+2\frac{\delta e}{e}), \nn \\
\mwb & = & M_W^2 \, (1+\dmw) ,  \nn \\
\swo & = & 1-\frac{M_W^2+\dmmw}{M_Z^2+\dmmz}
%    \; =\; 1-\frac{M_W^2}{M_Z^2} + \left(\dmz-\dmw\right) \nn \\
 \;=\;  s_W^2 + c_W^2 \left(
        \dmz-\dmw \right)
\eea
and keeping only terms of 1-loop order in Eq.\ (26):
\bea
\frac{\Gmu}{\sqrt{2}} & = & \frac{e^2}{8s_W^2 M_W^2} \cdot \nn \\
    &   & \cdot \left[
 1+2\frac{\delta e}{e} -\frac{c_W^2}{s_W^2} \left(\dmz-\dmw\right)
 +\frac{\Sigma^W(0)-\dmmw}{M_W^2} +(vertex,\,box) \right] \nn \\
 & \equiv & \frac{e^2}{8s_W^2 M_W^2}\cdot \left[
        1+ \Dr\right]
\eea
which is the 1-loop corrected version of Eq.\ (26).

 \medskip \noi
The  quantity $\Dr(e,M_W,M_Z,M_H,m_t)$ is the finite combination
of loop diagrams and counterterms in Eq.\ (87).
Since we have already determined the counterterms in the previous
subsection in terms of the boson self-energies, it is now only a
technical problem to evaluate the 1-loop diagrams for the final
explicit expression of $\Dr$.
                        Here we  quote the result:
\beq
 (vertex,\,box)   =   \frac{\al}{\pi s_W^2} \left(
 \Delta -\log\frac{M_W^2}{\m^2} \right)
 +\frac{\al}{4\pi s_W^2} \left( 6+\frac{7-4s_W^2}{2s_W^2}
  \log c_W^2 \right) \, .
\eeq
The singular part of Eq.\ (88), up to a factor, coincides with the
non-Abelian bosonic contribution to the charge counterterm
     in Eq.\ (49):
$$
  \frac{\al}{\pi s_W^2} \left(
 \Delta -\log\frac{M_W^2}{\m^2} \right)  \, = \,
 \frac{2}{c_W s_W}\, \frac{\Sigma^{\g Z}(0)}{M_Z^2} \, .
$$
Together with Eq.\ (49) and (88) we obtain from Eq.\ (87):
\bea
\Dr & = & \Pi^{\g}(0) -\frac{c_W^2}{s_W^2} \left(\dmz-\dmw\right)
          +\frac{\Sigma^W(0)-\dmmw}{M_W^2}   \nn    \\
    &   &+ 2\,\frac{c_W}{s_W}\,\frac{\Sigma^{\g Z}(0)}{M_Z^2} \, +\,
\frac{\al}{4\pi s_W^2} \left( 6+
\frac{7-4 s_W^2}{2 s_W^2} \log c_W^2   \right) \, .
\eea

\medskip \noi
The first line is of particular interest: via $\Sigma^W(0)$
and the mass counter terms $\delta M^2_{W,Z}$ also the experimentally
unknown parameters $M_H,\,m_t$ enter $\Dr$, whereas the residual terms
depend only on the vector boson masses.
 We proceed
with a more explicit discussion of the gauge invariant subset of
fermion loop corrections which involves, among others, the top quark.
This subset is also of primordial practical interest since it
constitutes the numerically dominating part of $\Dr$.

\subsection{Fermion contributions to $\Dr$}
In  the fermionic vacuum polarization of Eq.\ (89)
we split off the subtracted part evaluated at $\mz$:
\bea
 \Pi^{\g}(0) & = & - \real \Pi^{\g}(M_Z^2) +\Pi^{\g}(0)
                   + \real\,\Pi^{\g}(M_Z^2)  \nn \\
 & = & -\real\, \Pigr(M_Z^2) + \real\, \Pi^{\g}(M_Z^2)  \, .
\eea
The subtracted
finite quantity $\Pigr(M^2_Z)$ can be split into a leptonic
and a hadronic part:
$$
 \real\,\Pigr (M_Z^2)   = \real\, \Pigr_{lept}(M_Z^2)
                      + \real\, \Pigr_{had}(M_Z^2)\, .
$$
Heavy top quarks decouple from the subtracted vacuum polarization:
\beq
 \Pigr_{top}(M_Z^2) = \frac{\al}{\pi} \,Q_t^2\,
            \frac{M_Z^2}{5\,m_t^2} \, .
\eeq
 Whereas the leptonic content can easily
be obtained from
\beq
 \real\,\Pigr_{lept}(M_Z^2) = \sum_{l=e,\m,\tau}\,
\frac{\al}{3\pi}\left( \frac{5}{3}-\log\frac{M_Z^2}{m_l^2}
 \right)\, ,
\eeq
no light quark masses are available as
reasonable input parameters for the hadronic content. Instead,
the 5 flavor contribution to
$\Pigr_{had}$ can be derived from experimental data with the help
of a dispersion relation
\beq
 \Pigr_{had}(M_Z^2) = \frac{\al}{3\pi} \, M_Z^2 \,
 \int^{\infty}_{4m_{\pi}^2}  ds' \, \frac{R^{\g}(s')}
 {s'(s'-M_Z^2-i\veps)}
\eeq
with
 $$
 R^{\g}(s) = \frac{\sigma(\epm \ra\g^* \ra hadrons)}
 {\sigma(\epm\ra\g^*\ra \m^+\m^-)}
$$
as an experimental quantity up to a scale $s_1$ and applying
perturbative QCD for the tail region above $s_1$.
Using $\epm$ data for  the energy range below 40 GeV 
and perturbative QCD for the high energy tail, the recent updates
\cite{eidelman,burkhardt} yield 
\beq
\real\, \Pigr_{had}(s_0) = - 0.0280\pm 0.0007
\eeq
for $s_0 =M_Z^2$, 
and thus confirm the the previous value \cite{vacpol} with an improved
accuracy. 
The error is almost completely due to the experimental data.
Other determinations \cite{swartz,martin}
agree within one standard deviation.

Combining this result with the leptonic part one obtains
$$   \real\,\Pigr(M_Z^2) = -0.0593\pm 0.0007 \, . $$

\bigskip \noindent
Besides $\Pi^{\g}(M_Z^2)$ we need the $W$ and $Z$ self-energies.
For simplicity we restrict the further discussion to a single family,
leptons or quarks, with $m_{\pm},\, Q_{\pm},\, v_{\pm}, \, a_{\pm}$
denoting mass, charge, vector and axial vector coupling of the
up(+) and down(-) member. At the end, we perform the sum over the
various families.
We discuss the light and heavy fermions separately:
\begin{itemize}
\item {\bf Light fermions:}
\end{itemize}
                                      In the light
fermion limit, i.e.\ neglecting all terms $\sim m_{\pm}/M_{W,Z}$, the
 various ingredients of $\Dr$
follow from (79) to be
(in case of quark doublets with an additional factor $N_C=3$):
\bea  &   &
\frac{\Sigma^W(0)}{M_W^2} = O\left(\frac{m_{\pm}^2}{M_W^2}\right)
  \simeq 0  \, ,   \nn \\
&   & \dmz = \frac{\al}{3\pi}\cdot \frac{v_+^2+a_+^2+v_-^2+a_-^2}
            {4 s_W^2 c_W^2} \left( \deltaz \right) \, ,  \nn \\
&   & \dmw = \frac{\al}{3\pi}\cdot \frac{1}{4 s_W^2}
             \left( \deltaz \right)
 - \frac{\al}{16\pi s_W^2} \, \log c_W^2 \, ,\nn
\eea       together with
 $$   \real\, \Pi^{\g}(M_Z^2) \, = \,
  \frac{\al}{3\pi} \,(Q_+^2+Q_-^2) \left( \deltaz \right)
 \, .  $$
Inserting everything into Eq.\ (89) yields
\bea
 \Dr & = & - \real\, \Pigr(M_Z^2)   \nn \\
&   & +\frac{\al}{3\pi} \left(\deltaz\right) \cdot  \nn \\
&   & \cdot
     \left[ Q_+^2+Q_-^2-\frac{c_W^2}{s_W^2} \left(
     \frac{v_+^2+a_+^2+v_-^2+a_-^2}{4s_W^2 c_W^2} -
     \frac{1}{4s_W^2} \right) - \frac{1}{4s_W^4} \right]  \nn \\
 &   & -\frac{\al}{3\pi}\,\frac{c_W^2-s_W^2}{4s_W^2}\,\log c_W^2 \nn \\
 &   &   \nn \\
 & = & - \real\, \Pigr(M_Z^2)
       -\frac{\al}{3\pi}\,\frac{c_W^2-s_W^2}{4s_W^4}\,\log c_W^2
\eea

\medskip \noi
The term in brackets $[\dots]$ is zero with the coupling constants
in Eq.\ (37).
Thus, the main effect from the light fermions comes from the
subtracted photon vacuum polarization as the remnant from the
renormalization of the electric charge at $q^2=0$. For this
reason, after summing over all light fermions, we can write
\bea
\Delta\al \equiv  -\,\real\,\Pigr(M_Z^2) & = & 0.0593\pm 0.0007\, .
\eea
\begin{itemize}
\item {\bf Heavy fermions:}
\end{itemize}
Of special interest is the case of a heavy top quark which contributes
a large correction $\sim m_t^2$ to $\Dr$. In order to extract this
piece we keep for simplicity only those terms which are either
singular or quadratic in the top mass $m_t\equiv m_+$ ($N_C=3$):
\bea
&   & \real\, \Pi^{\g}(M_Z^2) = N_C \,\frac{\al}{3\pi}
     (Q_+^2+Q_-^2) \left(\deltat\right) \, +\cdots   \nn \\
&    & \nn \\
&   & \dmz = N_C\,\frac{\al}{3\pi} \left\{
     \frac{v_+^2+a_+^2+v_-^2+a_-^2}{4s_W^2 c_W^2} -
     \frac{3m_t^2}{8s_W^2c^2_WM_Z^2} \right \}
       \left( \deltat \right) \, + \cdots   \nn \\
 &  &   \nn \\
&   & \dmw = N_C\,\frac{\al}{3\pi} \left\{
      \frac{1}{4s_W^2} \left(\deltat\right) -
      \frac{3m_t^2}{8s_W^2 M_W^2} \left(\deltat +\frac{1}{2}
       \right) \right\}  \,+ \cdots  \nn \\
 &   &   \nn \\
%&  &
&   & \frac{\Sigma^W(0)}{M_W^2} = -\,N_C\,\frac{\al}{3\pi}
      \cdot \frac{3m_t^2}{8s_W^2M_W^2} \left(
      \deltat +\frac{1}{2} \right) \, .  \nn
\eea

\medskip \noi
Inserting into Eq.\ (89) we verify that the singular parts cancel and
a finite term $\sim m_t^2$ remains:
\beq
 (\Dr)_{b,t} = -\, \real \Pigr_b(M_Z^2)
               \,-\, \frac{c_W^2}{s_W^2} \,\dro \,+\cdots
\eeq
with $\dro$ from Eq.\ (82),
since the $m_t^2$-term is $q^2$-independent.

\bigskip
As a result of our discussion, we have got a simple form for $\Dr$
in the leading terms which is valid also after including the full
non-fermionic contributions:
\beq
 \Dr = \Delta\al -\frac{c_W^2}{s_W^2}\,\dro
         + (\Dr)_{remainder} \, .
\eeq
 $\Delta\al$  contains the large logarithmic
corrections from the light fermions and $\dro$ the leading quadratic
correction from a large top mass. All other terms are collected in
the $(\Dr)_{remainder}$. It should be noted that the remainder also
contains a term logarithmic in the top mass (for which our
approximation above was too crude) which is not negligible
\beq
 (\Dr)^{top}_{remainder} = -\frac{\al}{4\pi s_W^2} \left(
 \ctanw -\frac{1}{3}\right) \log\frac{m_t}{M_Z}\,+\cdots
\eeq
Also the Higgs boson contribution is part of the remainder. For large
$M_H$, it increases only logarithmically (``screening'' of a heavy
Higgs \cite{screening}):
\beq
 (\Dr)^{Higgs}_{remainder} \simeq  \frac{\al}{16\pi s_W^2} \cdot
 \frac{11}{3} \left( \log\frac{M_H^2}{M_W^2}
                   -\frac{5}{6} \right) \, .
\eeq
The typical size of $(\Dr)_{remainder}$ is of the order $\sim 0.01$.

 \subsection{Higher order contributions}
%text
Since $\Dr$ contains two large entries, $\Delta\al$ and $\Delta\rho$,
a careful investigation of higher order effects is necessary.

\medskip
\begin{itemize}
\item[(i)] Summation of large $\dal$ terms:
\end{itemize}
The replacement of the $\dal$-part
$$ 1+\dal\, \ra \, \frac{1}{1-\dal}    $$
of the 1-loop result
in Eq.\ (87)
 correctly takes into account all orders in the leading
logarithmic corrections $(\Delta\al)^n$, as can be shown by
renormalization group arguments \cite{marciano}
The evolution of the electromagnetic coupling with the scale
$\m$ is described by the renormalization group equation
\beq
 \m \, \frac{d\al}{d\mu} = - \,
\frac{\beta_0}{2\pi}\, \al^2
\eeq
with the coefficient of the 1-loop $\beta$-function in QED
\beq
\beta_0 = -\frac{4}{3} \,\sum_{f\neq t} Q_f^2 \, .
\eeq
The solution of the RGE contains the leading logarithms in
the resummed form.
It corresponds to a resummation of the iterated 1-loop vacuum
polarization  to all orders.
The non-leading QED-terms of next order are numerically not
significant. Thus, in a situation where large corrections are only
due to the evolution of the electromagnetic charge between two very
different scales set by $m_f$ and $M_Z$,
the resummed form
\beq
\Gmu = \frac{\pi\al}{\sqrt{2}M_W^2s_W^2} \, \frac{1}{1-\Dr}
= \frac{\pi\al}{\sqrt{2}M_Z^2c_W^2s_W^2} \, \frac{1}{1-\Dr}
\eeq
with $\Dr$ in Eq.\ (98) represents a good approximation to the
full result.

\medskip
\begin{itemize}
\item[(ii)] Summation of large $\dro$ terms:
\end{itemize}

\noi
For the heavy top quark also $\dro$ is large and the powers
$(\dro)^n$ are not correctly resummed in Eq.\ (103). A result correct
in the leading terms up to $\oal$ is instead given by
the independent resummation \cite{chj}
\beq
\frac{1}{1-\Dr} \, \ra\, \frac{1}{1-\Delta\al}\cdot\frac
{1}{1+\frac{c_W^2}{s_W^2}\drb}       +(\Dr)_{remainder}
\eeq
where
\beq
    \drb  = 3 x_t [ 1+ x_t \rho^{(2)}(\frac{M_H}{m_t}) ], \;\;\;
 x_t = \frac{\Gmu m_t^2}{8\pi^2\sqrt{2}}
\eeq
incorporates the result from 2-loop 1-particle
 irreducible diagrams.
For light Higgs bosons $M_H \ll m_t$, where $M_H$
can be neglected, the coefficient
 \beq
\rho^{(2)} =
  19-2\pi^2  \eeq
was first calculated by Hoogeveen and van der Bij \cite{hoog}.
The general function $\rho^{(2)}$, valid for all Higgs masses,
has been derived in \cite{barbieri}. For large Higgs masses $M_H > 2m_t$,
a good approximation is given by
the asymptotic expression with $r = (m_t/M_H)^2$
\cite{barbieri}
\bea
 \rho^{(2)} & = & \frac{49}{4} + \pi^2 +\frac{27}{2} \log r
                  +\frac{3}{2} \log^2 r \nn \\
  &   &
  +\frac{r}{3} \left( 2-12\pi^2+12\log r-27\log^2 r \right) \nn \\
  &   &
  +\frac{r^2}{48} \left(1613-240\pi^2-1500 \log r -720 \log^2 r
    \right) \, .
\eea
 With the resummed $\rho$-parameter
\beq
   \rho = \frac{1}{1-\drb}
\eeq
Eq.\ (104) is
compatible with the following form of the $M_W-M_Z$ interdependence
$$   \Gmu \,=\, \frac{\pi}{\sqrt{2}}\, \frac{\al(\mz)}
{M_W^2\left(1-\frac{M_W^2}{\rho M_Z^2} \right)}\cdot
                                        \left[
1+ (\Dr)_{remainder} \right] \,    $$
with
\beq
   \al(\mz) = \frac{\al}{1-\dal} \, . \eeq

\bigskip  \noi
It is interesting to compare this result with the corresponding
 lowest order
 $M_W-M_Z$
  correlation  in a more
general model with a tree level $\rho$-parameter $\rho_0\neq 1$:
the tree-level $\rho_0$ enters in the same way as the $\rho$ from a
heavy top in the minimal model. The same applies for the quadratic
mass terms from other particles
like scalars or additional heavy fermions in isodoublets
with large mass splittings.
        Hence, up to the small quantity
$(\Dr)_{remainder}$, they are  indistinguishable from an
experimental point of view ($\Delta\al$ is universal).
                 In the minimal model, however, $\rho$
is calculable in terms of $m_t, M_H$
 whereas $\rho_0$ is an {\it additional}
free parameter.

\medskip
\begin{itemize}
\item[(iii)] QCD corrections: 
\end{itemize}
Virtual gluons contribute to the quark loops in the vector
boson self-energies at the 2-loop level. For the light quarks
this QCD correction is already contained in the result for
the hadronic vacuum polarization from the dispersion integral,
Eq.\ (93).
Fermion loops involving the top quark get additional $O(\al\al_s)$
corrections which have been calculated perturbatively
\cite{qcd}.
The dominating term represents the QCD correction
to the leading $m_t^2$ term of the $\rho$-parameter and can be built in
by writing instead of Eq.\ (105): 
 \beq
 \drb= 3 x_t \cdot [ 1+ x_t \,  \rho^{(2)} + \delta\rho_{QCD} ]
\eeq
The QCD term  \cite{adjouadi,tarasov} reads:
\beq
    \delta\rho_{QCD} = -\,
\frac{\al_s(\mu)}{\pi}\, c_1
 +\left(\frac{\al_s(\mu)}{\pi}\right)^2 c_2(\mu)
\eeq
with
\beq
 c_1 =  \frac{2}{3} \left(
\frac{\pi^2}{3}+1\right) = 2.8599
\eeq
and the recently calculated
3-loop coefficent
 \cite{tarasov}
\beq
 c_2=-14.59 \;
 \mbox{  for } \mu =m_t \mbox{ and 6 flavors}
\eeq
with the on-shell top mass $m_t$.
It reduces the scale
dependence of $\rho$ significantly and hence is an important
entry to decrease the theoretical uncertainty of the standard model
predictions for precision observables.
As part of the higher order irreducible contributions to $\rho$,
the QCD correction is
resummed together with the electroweak 2-loop irreducible term
as indicated in Eq.\ (104).

\smallskip 
Beyond the $\Gmu m_t^2\al_s$ approximation through the $\rho$-parameter,
the complete
 $O(\al\al_s)$ corrections to the self energies
 are available from  perturbative  calculations
\cite{qcd} and by means of dispersion relations \cite{dispersion1}.
All the higher order terms contribute with the same positive sign
to $\Dr$,  thus
making the top mass dependence of $\Dr$ significantly flatter.
This is of high importance for the indirect determination
of $m_t$ from $M_W$ measurements, which is affected by the order
of 10 GeV.
Quite recently, also non-leading terms to $\Delta r$ of
the type
$$ \Dr_{(bt)} = 3x_t \left(\frac{\al_s}{\pi}\right)^2 \left(
 a_1 \frac{\mz}{m_t^2} + a_2 \frac{M_Z^4}{m_t^4} \right)
$$
have been computed \cite{cks}. 
 For $m_t = 180$ GeV they contribute an extra
term of
$+0.0001$ to $\Dr$ and thus are within the uncertainty from
$\dal$.

 \medskip
\begin{itemize}
\item[(iv)] Non-leading higher order terms:
\end{itemize}
The modification of Eq.\ (104) by placing $(\Dr)_{remainder}$
into the denominator
\beq
\frac{1}{1-\Dr} \, \ra\, \frac{1}{(1-\Delta\al)\cdot
(1+\frac{c_W^2}{s_W^2}\drb) \, -\,(\Dr)_{remainder}}
\eeq
correctly incorporates the non-leading higher order terms
containing mass singularities of the type $\al^2\log(M_Z/m_f)$
\cite{nonleading}

\smallskip
The treatment of the higher order reducible terms in Eq.\ (114)
can be further refined by performing in
    $(\Dr)_{remainder}$ the following substitution
\beq
\frac{\al}{s_W^2} \ra \frac{\sqrt{2}}{\pi} \Gmu M_W^2 (1-\dal)
\eeq
in the expansion parameter of the combination
$$ \left( \dmz-\dmw \right) -\, \dro $$
after cancellation of the UV singularity in the combination (89)
or in the $\ms$ scheme with $\m = M_Z$.
 This is discussed in \cite{beg}
and is equivalent to the method described in \cite{hollik}
 as well as
to the recipe given at the end of ref.\ \cite{chj}.
Numerically this modification
is not of significance for a top quark not heavier than 200 GeV.

\smallskip
The refined treatment of the non-leading reducible higher order terms
can be considered as an improvement only in case that the
2-loop irreducible non-leading terms are essentially smaller
in size.  Irreducible contributions of the type
$ \Gmu^2 m_t^2\mz$ are unknown, and one has to rely on the
assumption that the suppression by $1/N_C$ relative to the
2-loop reducible term is not compensated by a large coefficient.
For bosonic 2-loop terms reducible and irreducible
contributions are a priori of the same size and one does not gain
from resumming 1-loop terms. In order to be on the
safe side, the differences caused by the summation of non-leading
reducible terms should be considered as a theoretical uncertainty
at the level of
1-loop calculations improved by higher order leading terms.
(see section 7.5).

 \subsection{Predictions and experimental data}
%text
The correlation of the electroweak parameters, complete at the
one-loop level and with the proper incorporation of the leading
higher order  effects,
is given by the following equation:
\bea
 \mw \left(1-\frac{\mw}{\mz}\right) &  = &
\frac{ \pi\al}{\sqrt{2}\Gmu}\cdot \frac{1}{(1-\Delta\al)\cdot
(1+\frac{c_W^2}{s_W^2}\drb) \, -\,(\Dr)_{remainder}} \nn \\
 & \equiv &
\frac{ \pi\al}{\sqrt{2}\Gmu}\cdot
\frac{1}{1-\Dr}  \, .
\eea
The term $\Dr$ in Eq.\ (116) is an effective quantity beyond the 1-loop
order, introduced to obtain  the formal analogy to the naively
 resummed first order result in Eq.\ (103).
 $\drb$ includes the 2-loop irreducible electroweak
and QCD corrections to the $\rho$-parameter 
according to Eq.\ (110).
The correlation (113) allows us to predict a value for the
$W$ mass after the other parameters have been specified.
These predicted values for $M_W$ are put together in table 1
for various Higgs and top masses.
The present experimental value for the $W$ mass
from the combined UA2,  CDF and D0 results \cite{wmass} is
 \beq
  M_W^{exp}\, = \, 80.26 \pm 0.16 \, \gv \, .
\label{mw}
\eeq

\bigskip
\begin{table}
\bc \btab{| l | c  c  c  c |}
\hline
  $m_t$  &  $M_H= 65$  &  100  &  300  & 1000  \\
\hline
     150   &     80.265  &  80.242 &   80.168  &     80.073  \\
     160   &     80.324  &  80.300 &   80.226  &     80.129  \\
     170   &     80.385  &  80.361 &   80.285  &     80.188  \\
     180   &     80.449  &  80.424 &   80.347  &     80.249  \\
     190   &     80.515  &  80.491 &   80.412  &     80.312  \\
     200   &     80.585  &  80.559 &   80.479  &     80.376  \\
\hline
\etab
\caption{The $W$ mass $M_W$  as predicted by the Standard Model
         for $M_Z=91.1884$ GeV and various top and Higgs masses,
         based on Eq.\ (113). $\al_s=0.123$.
         All masses are in GeV. }
\ec
\end{table}

\bigskip \noi
 We can define the quantity $\Dr$ also as a physical
observable by
\beq
\Dr \,=\, 1\,-\,\frac{\pi\al}{\sqrt{2}\Gmu} \, \frac{1}
     {M_W^2 \left(1-\frac{M_W^2}{M_Z^2} \right) } \, .
\eeq
Experimentally, it is determined by $M_Z$ and the ratio
$M_W/M_Z$.
Theoretically, it can be computed from $M_Z,\Gmu,\al$
after specifying the masses $M_H,m_t$ by solving Eq.\ (113).
Both electroweak and QCD higher order effects yield a positive shift
to $\Dr$ and thus diminish the slope of the
first order dependence on
$m_t$ for large top masses.
The effect on $\Dr$ coming from the modified $\rho^{(2)}$ in
Eq.\ (107) is
an additional weakening of the sensitivity to $m_t$ for large
Higgs masses.

\smallskip
The theoretical prediction for $\Dr$ for various Higgs and top masses
is displayed in Figure 1.
%
%
%figure deltar
%
%

%\begin{figure}[htb]
%\vspace{-1cm}
%\centerline{
%\epsfig{figure=deltar.ps,height=15cm,angle=0}}
%\vspace{-1.5cm}
%\caption{$\Dr$ for $M_H=65$ GeV (dotted) and $M_H=1$ TeV (solid).}
%\label{deltar}
%\end{figure}
%%\clearpage

\setlength{\unitlength}{1cm}
\begin{figure}[htb]
\begin{center}
\begin{picture}(15,13)
%\put(0,0){\rule{.2cm}{.2cm}}
%\put(0,13){\rule{.2cm}{.2cm}}
%\put(15,0){\rule{.2cm}{.2cm}}
%\put(15,13){\rule{.2cm}{.2cm}}
\put(-4.0,-11.2){\includegraphics{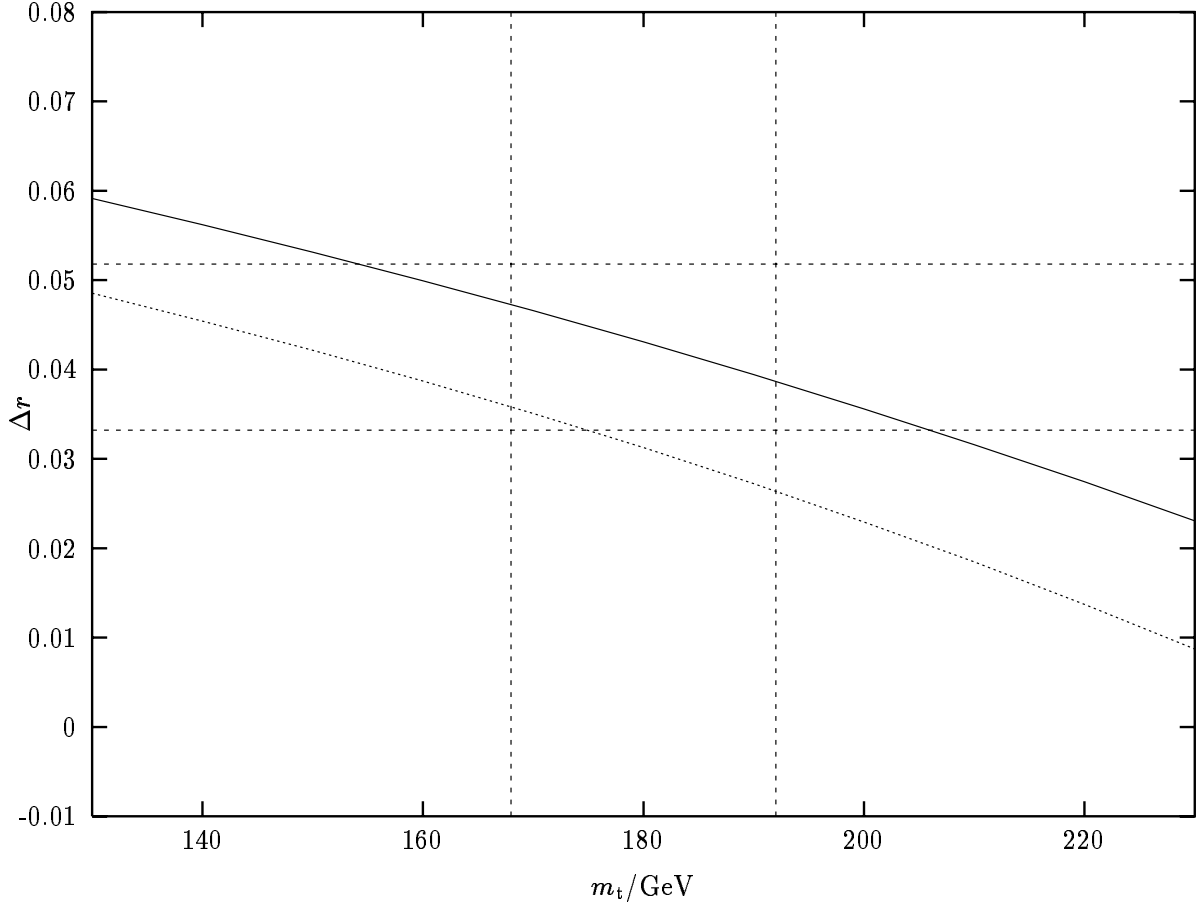}}
\end{picture}
\end{center}
\caption{$\Dr$ for $M_H=65$ GeV (dotted) and $M_H=1$ TeV (solid), with 
          experimental $\pm 1\sigma$ ranges. }
\label{deltar}
\end{figure}
\setlength{\unitlength}{0.7mm}

For comparison with data, the experimental $1\sigma$ limits from the
 direct measurements of $M_Z=91.1884$ at LEP \cite{lep}  and
$M_W$ in $p\bar{p}$, Eq.~(\ref{mw}),  are indicated.
The  constraints on  the top mass obtained from these results
completely coincide with the direct $m_t$ measurement at the
Tevatron of $m_t= 180\pm 12$ GeV \cite{top}.
The present experimental error does not allow a sensitivity to the
Higgs mass. Precision measurements of $M_W$ at LEP 200 will pin
down the error
to $\delta\Dr = 0.0024$
The expected precision in the determination of $\Dr$
matches the
size of $(\Dr)_{remainder}$
and thus will provide some sensitivity also to the Higgs mass.
For virtual Higgs effects, however, the observables from the
$Z$ resonance are more suitable.

\subsection{Input from neutrino scattering}
The quantity $s_W^2$ resp.\  the ratio $M_W/M_Z$
can indirectly be measured in  the class of low enery experiments
comprising neutrino-quark, neutrino-electron, and electron-quark
scattering. The two most precise informations come from the
NC/CC neutrino-nucleon cross section ratios \cite{neutrino}.
For an isoscalar target these ratios do not
depend on the nucleon structure \cite{llewellyn}:
\bea
 & R_{\nu} & = \;\frac{\sigma^{\nu}_{NC}}{\sigma^{\nu}_{CC}}\, =\,
  \left(\frac{M_W}{M_Z}\right)^4
  \frac{1-2s_W^2+\frac{10}{9}(1+r) s_W^4}{2(1-s_W^2)^2}  \nn \\
 & R_{\bar{\nu}} & =\;
  \frac{\sigma^{\bar{\nu}}_{NC}}{\sigma^{\bar{\nu}}_{CC}} \,=\,
  \left(\frac{M_W}{M_Z}\right)^4
  \frac{1-2s_W^2+\frac{10}{9}(1+\frac{1}{r})s_W^4}{2(1-s_W^2)^2}
\eea
with $r=\sigma^{\nu}_{CC}/\sigma^{\bar{\nu}}_{CC} \simeq 0.4$.

\smallskip
The second factor in $R_{\nu}$ has a very weak dependence on $s_W^2$.
Hence, measurements of $\Rnu$ can directly be converted into
values for $M_W/M_Z$. This principal feature remains valid also
after the incorporation of radiative corrections in Eq.\ (119).
Besides the QED corrections, vertex corrections and box diagrams
which do not depend on $m_t$, $M_H$, the dominant effect
$\sim m_t^2$ 
can simply be embedded in Eq.\ (119) by replacing
$$ s_W^2 \ra s_W^2 + c_W^2 \dro  $$
with $\dro$ from Eq.\ (82). This is obvious from the expansion (86)
together with Eq.\ (82). Since an increase in $m_t$ is equivalent to
a slight shift in $s_W^2$, the relation between $\Rnu$ and
$(M_W/M_Z)^4$ is affected only marginally. This explains
qualitatively the stability of $M_W/M_Z$ against variations
of $m_t$ when extracted from $\Rnu$.

\smallskip
The present world average on $s_W^2$ from the experiments
CCFR, CDHS and CHARM \cite{neutrino} 
\beq
s_W^2 = 1 - M_W^2/M_Z^2 = 0.2253 \pm 0.0047  
\eeq
is fully consistent with the direct vector boson mass measurements
and with the standard theory. 

\smallskip
The mixing angle which is measured in (anti)neutrino - electron
scattering has a meaning different from  the quantity $s_W^2$ in
Eq.\ (120). It is much closer to the effective mixing angle
determined at the $Z$ peak and will be dealt with in the context
of the $Z$ boson observables in section 8.3.  
%%%%%%%%%%%%%%%%%%%%%%%%%%%%%%%%%%%%%%%%%%%% end of section
%
\section{Renormalization schemes}
Before one can make predictions from the theory,
a set of independent parameters has to be determined from experiment.
This can either be done for the bare quantities or
for renormalized parameters which have a simple physical
interpretation. In a more restrictive sense, a renormalization scheme
characterizes a specific choice of experimental data points to be
used as input defining the basic parameters of the Lagrangian in
terms of which the perturbative calculation of physical amplitudes
is performed.

Predictions for the relations between physical quantities do not
depend on the choice of a specific renormalization scheme if we
perform the calculation to all orders in the perturbative
expansion. Practical calculations, however, are obtained from
truncated perturbation series, making the predictions depend on
the chosen set of basic parameters  and thus leading to a
scheme dependence.

\smallskip
 Differences between various schemes are formally
of higher order than the one under consideration.
 The study of the
scheme dependence of the perturbative results, after improvement by
resumming the leading terms, allows us to estimate the missing
higher order contributions.

\bigskip \noi
Parametrizations or `renormalization schemes' frequently used in
electroweak calculations are:
\begin{enumerate}
\item
 the on-shell (OS) scheme with
 $$ \al, \; M_W,\; M_Z,\; m_f,\; M_H $$
\item
the $\Gmu$ scheme with the basic parameters
 $$ \al, \; \Gmu,\; M_Z,\; m_f,\; M_H $$
\item
the low energy scheme with the mixing angle as a basic parameter
defined in neutrino-electron scattering:
 $$ \al, \; \Gmu,\; \sin^2\theta_{\nu e},\; m_f,\; M_H $$
\item
the $*$ scheme where the bare parameters $e_0,\, \Gmu^0,\, s_0^2$
 are eliminated and replaced in terms
of dressed running ($k^2$-dependent) parameters
$$ \est(k^2),\; \Gst(k^2),\; \sst(k^2);\; m_f,\; M_H$$
\item
the $\ms$-scheme.
\end{enumerate}
In the following we give some details on
the $\ms$ scheme.

\bigskip
The modified minimal subtraction scheme ($\ms$-scheme)
 \cite{pavelt,msbar,msbar1,msbar2}
is one of the
simplest ways to obtain finite 1-loop expressions by
performing the substitution
$$
\frac{2}{\eps}-\gamma+\log 4\pi +\log \m^2 \, \ra \,
 \log \m^2_{\ms}   $$
in the divergent  parts of the loop integrals,
Eq.\ (68).
Formally, the $\ms$ self energies and vertex corrections
 are obtained by splitting the bare masses and couplings
into $\ms$ parameters and counter terms
\beq
  M_0^2 =  \hat{M}^2 + \delta\hat{M}^2 , \;\;\;
  e_0 = \ems + \delta \ems           ,
\eeq
where the counter terms
together with field renormalization constants
 $$ 1 + \delta \hat{Z}_i $$
are defined in such a way that they absorb the singular parts
proportional to
 $$ \Delta  \, =\,
\frac{2}{\eps}-\gamma+\log 4\pi                 \, .
$$
As a consequence, self energies and vertex corrections in the
$\ms$-scheme depend on the arbitrary scale $\m$.

\bigskip \noi
Perturbative calculations start from the Lagrangian with the
formal $\ms$ parameters
$${\cal L}(\ems,\hat{M}_W,\hat{M}_Z,\dots) .$$
The $\ms$ parameters fulfill the same relations as the corresponding
bare parameters. In particular, the mixing angle in the $\ms$-scheme,
denoted by $\sms$,   can be expressed in terms of the $\ms$ masses
of $W$ and $Z$
in the following way:
 \beq
 \sms \, = \, 1 -\frac{\mwms}{\mzms} \, .
 \eeq

\bigskip \noi
The relation of the $\ms$ parameters to the conventional
OS-parameters is obtained by calculating the dressed vector boson
propagators and the dressed electron-photon vertex in
the Thomson limit in the $\ms$-scheme and identifying the poles
with the OS masses and the electromagnetic coupling with the
classical charge.

%\newpage
\begin{itemize}
\item \underline{The $\ms$ charge}:
\end{itemize}
The $\ms$ analogon of the OS charge renormalization condition
Eq. (49) reads:
\beq
\ems\left[ 1-\frac{1}{2} \, \Pig_{\ms}(0) +
\frac{\smsb}{\cmsb}\,\frac{\sgz_{\ms}(0)}{\mzms} \right] \, = \, e .
\eeq
 The l.h.s.\  is the  coupling constant of the dressed
electromagnetic vertex in the Thomson limit which has to be
identified with the classical charge.

\medskip \noi
The $\ms$ self energies in Eq.\ (123) read explicititly:
\bea
 \Pig_{\ms}(0) &  = &  \frac{\ems^2}{16\pi^2}\, A^{\g}(0), \nn \\
 A^{\g}(0) & = &       \frac{4}{3}\,
 \sum_f\, Q_f^2\,\log\frac{\m^2}{m_f^2} + 3\, \log
 \frac{\mw}{\m^2} -\frac{2}{3}         , \nn \\
\frac{\smsb}{\cmsb}\,\frac{\sgz_{\ms}(0)}{\mzms} &  = &
- \, \frac{\ems^2}{8\pi^2} \,\log \frac{\mw}{\m^2} \, .
\eea
A natural scale for electroweak physics is given by $\m = M_Z$.
 Hence, the correlation between $e$ and $\ems$ involves
large logarithms from the light fermions
 which can be resummed according to the
RGE (101). The bosonic terms are small.
  Resummation
leads to the relation
\beq
 e^2 \,=\, \frac{\ems^2}{1+
 \frac{\ems^2}{16\pi^2} \left[A^{\g}(0)
 +4\,\log\frac{\mw}{\m^2} \right]} .
\eeq
Inverting this equation yields the $\ms$ charge
expressed in terms of the OS charge
\beq
 \ems^2 \,=\, \frac{e^2}{1-\frac{e^2}{16\pi^2} \left[A^{\g}(0)
 +     4\,\log\frac{\mw}{\m^2} \right]} .
\eeq
Choosing $\m=M_Z$ we can evaluate the expression
in (126) to obtain the $\ms$ fine structure constant
at the $Z$ mass scale
\beq
 \hat{\al} = \frac{\al}{1-\Delta\hat{\al}}
\eeq
with  the value
\beq
\dalms  = 0.0682\pm 0.0007
   -\frac{8\al}{9\pi}\log \frac{m_t}{M_Z}
   +\frac{\al}{2\pi} \left(\frac{7}{2}\log c_W^2 -
     \frac{1}{3}\right) \, .
\eeq
The first term is due to the light fermions. It can be obtained
from the quantity in Eq.\ (96) by adding the constant term
$$ \frac{\al}{\pi} \left( \frac{5}{3} +\frac{55}{27}\,
   (1+\frac{\al_s}{\pi})
     \right) \, . $$
The uncertainty in Eq.\ (128) is the hadronic uncertainty of $\dal$
in Eq.\ (96).

\bigskip
$\hat{\al}$ has to be distinguished from the effective charge at
the $Z$ scale
introduced in Eq.\ (109)
which contains
only the light fermion contributions. A heavy top quark decouples
in $\dal$
according to Eq.\ (91), but does not decouple in $\dalms$.
Numerically one finds
\beq
(\hat{\al})^{-1}  =  128.08\, \pm 0.02\pm 0.09
\eeq
for  $m_t=180\pm 12$ GeV.

\bigskip \noi
\begin{itemize}
\item \underline{The $\ms$ mixing angle}:
\end{itemize}
The $\ms$ mass parameters $\mwms,\mzms$ enter the corresponding
transverse propagators together with the self energies as follows
($V=W,Z$):
\beq
 D_V \, = \,\frac{1}{k^2-\mvms+\Sigma_{\ms}^{V}(k^2)}
\eeq
The OS-masses fulfill the pole conditions
\beq
 M_V^2 -\mvms +\real\, \Sigma_{\ms}^{V}(M_V^2) \, = \, 0
\eeq
yielding $\mvms$ expressed in terms of the OS-masses:
\beq
 \mvms = M_V^2+\real\, \Sigma_{\ms}^{V}(M_V^2) .
\eeq
The mass parameters $\mvms$ are $\m$-dependent.
We can choose $\m=M_Z$ as the natural scale
for electroweak calaculations, as done also for $\hat{\al}$.

\medskip
The self energies $\Sigma_{\ms} $  are obtained from the expressions
given in section 4.3 by dropping everywhere the singular term
$\Delta$ and substituting
$$ e\ra \ems,\;\; s_W\ra \smsb,\;\; c_W\ra \cmsb $$
 in the couplings,
with $ \cms=1-\sms$.
It is convenient to remove the overall normalization factors
and to write for the real parts:
\bea
 \real\, \sw_{\ms} & = & \frac{\ems^2}{\sms}\, A_W(k^2), \nn\\
 \real\, \sz_{\ms} & = & \frac{\ems^2}{\sms\cms}\, A_Z(k^2).
\eea

\bigskip \noi
The mixing angle $\sms$ in the $\ms$-scheme, defined in Eq.\ (122),
can be related to the OS mixing angle
$s_W^2 = 1-\mw/\mz$
by substituting $\hat{M}_{W,Z}^2$ from Eq.\ (132), yielding
\beq
 \sms = s_W^2 + c_W^2\, \xms, \;\;\;
 \cms= c_W^2\,(1-\xms)
\eeq
with
\beq
\xms = \frac{\ems^2}{\sms}\left(
 \frac{A_W(\mw)}{\mw}-\frac{A_Z(\mz)}{\cms\mz}\right)
 \left( 1-\frac{\ems^2}{\sms}\,
           \frac{A_Z(\mz)}{\cms\mz}\right)^{-1} \, .
\eeq
Making use of the property
$$
 \xms =
       \frac{\ems^2}{\sms}\,
       \frac{A_W(\mw)}{\mw} \, -\, (1-\xms)\,
          \frac{\ems^2}{\sms}\,
           \frac{A_Z(\mz)}{\cms\mz}
$$
the relation (132) can simplified:
\beq
 \sms \, = \, s_W^2 \,+ \, \frac{\ems^2}{\sms}
           \frac{A_Z(\mz)
                - A_W(\mw)}{\mz}          \, .
\eeq
The leading 2-loop irreducible contributions are incorporated
by adding in (136) the extra term
 $ c_W^2\,\dro^{(2)}$ with
         $\dro^{(2)}$  from Eq.\ (110).

\bigskip
Eq.\ (136) determines $\sms$ in terms of the OS parameters.
$\ems^2$ has to be taken from Eq.\ (126) or (127),
respectively, for $\m= M_Z$. Numerically it is very close to the 
effective leptonic mixing angle at the $Z$ peak.

\bigskip
One can obtain $\sms$ also in a more direct way from the experimental
data points $\al, \Gmu, M_Z$, without passing first through the
OS-calculation, by deriving the effective Fermi constant
in the $\ms$-scheme
\beq
\frac{\Gmu}{\sqrt{2}} \,=\, \frac{\ems^2}{8\,\sms\cms\roms\,\mz}\,
 \cdot \frac{1}{1-\Drms}
\eeq
where
\bea
\Drms &  = &  \frac{\ems^2}{\sms}\,\frac{A_W(0)-A_W(\mw)}{\mw}
           +\dvb , \nn \\
 \dvb     & = & \frac{\hat{\al}}{4\pi\sms} \left[
  6 + \frac{7-5s_W^2+\sms(3 c_W^2/\cms -10)}{2s_W^2} \,
      \log c_W^2 \right],
\eea
together with
\bea
\mw & = & \cms \roms \mz , \nn \\
\roms & = & \frac{1}{1-\xms} \, .
\eea
For given parameters $\al,\Gmu,M_Z,m_t,M_H$ the solution of this set
of equations yields the quantities $\sms,\roms$ together with $M_W$.
$\Drms$ is a small correction and has only a mild dependence
on the top and Higgs masses.

\smallskip
The term $\dvb$ in $\Drms$ is the vertex and box correction
to the muon decay amplitude in the $\ms$-scheme \cite{msbar1}.
          The given expression  refers to a mixed
          $\ms$ - on-shell calculation of the loop diagrams
           where
          $\ms$-couplings are used but on-shell masses in the
          propagators. Numerically  the differences to the
          corresponding expression exclusively with $\ms$
          parameters is insignificant $(< 3\cdot 10^{-4})$.
The main difference to the on-shell quantity $\delta_{VB}$
in Eq.\ (137) (besides the parametrization)
 is the extra additive term
  $$ - \,\frac{\hat{\al}}{\pi} \, \log c_W^2
    \,\equiv \, -\,\frac{\hat{\al}}{\pi}\, \log
     \frac{\mw}{\m^2} \;\; \mbox{ for } \m = M_Z $$
arising from the UV singularity in the sum of the diagrams.

\bigskip
The $\ms$ quantities $\hat{\al},\, \sms$ are formal parameters
which have no simple relation to physical quantities. The interest
in these parameters is based on two important features:
\begin{itemize}
\item
They are universal, i.e.\ process independent, and take
 into account
the universal large effects from fermion loops. Expressing the
NC coupling constants (see section 7.1)
 for the $Zff$ vertices in terms of
$\ams ,\, \sms$  yields  a good approximation to the
complete results (148):
\bea
 \sqrt{2}\Gmu\mz\,\rho_f & = &
 \frac{\ems^2}{4\,\sms\cms} \, (1+ \delta\hat{\rho}_f),\nn \\[0.2cm]
 s_f^2 & = & \sms \, + \, \delta \sms_f \,.
\eea
The flavor dependent residual corrections $\delta\hat{\rho}_f$
and $\delta\sms_f$ are small and practically independent of $m_t$
and $M_H$. An exception is the $Zbb$ vertex, where also
 non-universal
large top terms are present \cite{vertex}.
\item
The knowledge of the values for $\ams$ and $\sms$ at the $Z$ scale
allows the extrapolation of the SU(2) and U(1) couplings
 \beq
 \ams_1(\m^2) = \frac{\ams(\m^2)}{\cms(\m^2)}, \;\;\;
 \ams_2(\m^2) = \frac{\ams(\m^2)}{\sms(\m^2)}
\eeq
to large mass scales and, together with the strong coupling constant
 $\al_s(\m^2)$ in the $\ms$-scheme, to test scenarios
 of Grand Unification. In particular the minimal SU(5) model of
Grand Unfication predicts with $\al$ and $\al_s$ as input
\cite{luo}:
$$
 \sms_{SU(5)}(\mz) \, = \, 0.2102^{+0.0037}_{-0.0031}
$$
which is in disagreement with the experimental  result (table 7).
Supersymmetric models of Grand Unification, however, are in favor
\cite{luo,susy}.
\end{itemize}
%

%%%%%%%%%%%%%%%%%%%%%%%%%%%%
%%%%%%%%%%%%%%%%%%%%%%%%%%%%%%%%%%%%%%%%%%%% end of section
\section{$Z$ physics in electron-positron annihilation}
The measurement of the $Z$ mass from the $Z$ line shape at LEP
provides us with an additional precise input parameter besides
 $\al$ and $\Gmu$. Other observable quantities from the $Z$ peak,
like total and partial decay widths, asymmetries,
$\tau$-polarization,       allow us to perform precision tests
of the theory by comparison with the theoretical predictions.

\medskip
In lowest order, the $Z$ observables are completely fixed in terms
of $\al,\,\Gmu,\,M_Z$ applying the rules and relations of section
2 to compute the Born $\g$ and $Z$ exchange diagrams.
Since 1-loop terms are of the order $\al/\pi$ and typically enhanced
by factors $\sim \log \mz/m_f^2$ or $\sim m_t^2/\mz$, the size
$\al/\pi=0.0023$ in view of experimental precisions of a few
$10^{-3}$ immediately signals the need for dressing the Born
amplitudes by next order contributions.

\subsection{Amplitudes and effective couplings}
A gauge invariant subset of the 1-loop diagrams to $\epmf$ are the
QED corrections: 
The sum of the virtual photon loop graphs is UV finite but IR
(= infra-red) divergent because of the massless photon.
The IR-divergence is cancelled by adding the cross section with
real photon bremsstrahlung  (after integrating over the phase
space for experimentally invisible photons) which always
accompanies a realistic scattering process. Since the phase space
for invisible photons is a detector dependent
quantity the QED
corrections cannot in general be separated from the experimental
device.

\bigskip
Our discussion will concentrate on the residual set of 1-loop
diagrams, the non-QED  or weak  corrections. This  class is free
of IR-singularities but sensitive to the details  beyond
the lowest order amplitudes.
The UV-singular terms associated with the loop diagrams are
cancelled by our counterterms of section 3  as a consequence
of renormalizability. The 1-loop  amplitude for
$\epmf$  contains the sum of the
individual contributions to the self-energies and vertex corrections
(including the external fermion self-energies via wave function
renormalization). The essential steps
for getting the total amplitude finite are: expressing the tree
diagrams in terms of the bare parameters
$e_0,\, \mzb,\, \swo$
expanding the bare quantities according to Eq.\ (46,86),
 and inserting the
counterterms given by Eq.\ (48) and (49). After some lengthy
calculations the total amplitude around the $Z$ pole can be cast into 
a form close to the lowest order amplitude
$$
 A(\epmf \ra f\bar{f}) = A_{\g} + A_Z + (box)
$$
 as
the sum of a dressed photon and a dressed $Z$ exchange amplitude
plus the contribution  from the box diagrams which are numerically not
significant around the peak (relative contribution $< 10^{-4}$).
For theoretical consistency (gauge invariance) they have to be
retained; for practical purposes they can be neglected in $Z$ physics.
Resummation of the iterated self-energy insertions in the photon and
$Z$ propagators brings the finite $Z$ decay width into the
denominator, and treats the higher order leading terms in the
proper way. Since the leading terms arise from fermion loops only, we
do not have problems with gauge invariance; the bosonic loop terms
have to be understood as expanded to strict 1-loop order. Numerically
their resummation does not yield significant differences but allows a
simple and compact notation.

\bigskip \noi
{\bf Dressed photon amplitude: }

\smallskip \noi
The dressed photon exchange amplitude, with 
$(p_{e^+}+p_{e^-})^2$, 
\beq
A_{\g} \,=\, \frac{e^2}{1+\Pigr(s)} \cdot\frac{Q_eQ_f}{s} \cdot
\left[ (1+\fvge)\gamu-\fage\gamu\gafi \right] \otimes
\left[ (1+\fvgf)\gimu-\fagf\gimu\gafi \right]           \, .
\eeq
contains $\Pigr$
as the $\g$ self-energy subtracted at $s=0$. Writing it in
the denominator  takes into account the resummation of the leading
log's from the light fermions, around the $Z$ given by
\beq
\Pigr_{ferm}(s) = -0.0593 - \frac{40\al}{18\pi}
\log\frac{s}{\mz} \pm 0.0007 \, + \,i\,
\frac{\al}{3}\,\sum_{f\neq t} \,Q_f^2 N_C^f \, .
\eeq
The form factors $F_{V,A}(s)$ arise
from the vertex correction diagrams together with the external
fermion self-energies. They vanish for real  photons:
$F_{V,A}^{\g e,f}(0) = 0$.
The typical size of the various corrections is  (real parts):
\bea
\Pigr(\mz) & = & -0.06  \nn \\
\fvge(\mz) & \simeq & \fage(\mz) \; \simeq 10^{-3} \,. \nn
\eea
For the region around the $Z$ peak, the photon vertex form factors
are negligibly small.

\newpage \noi
%\bigskip \noi
{\bf Dressed $Z$ amplitude and effective neutral current couplings:}

\smallskip \noi
More important is the weak dressing of the $Z$ exchange amplitude.
Without the box diagrams the corrections factorize and we obtain a
result quite close to the Born amplitude:
\bea
A_Z & = & \sqrt{2} \Gmu\mz (\rho_e\rho_f)^{1/2} \cdot     \\
 &   & \cdot \frac
{ [\gamu(I_3^e-2Q_e s_W^2\kappa_e)-I_3^e\gamu\gafi] \otimes
  [\gimu(I_3^f-2Q_f s_W^2\kappa_f)-I_3^f\gimu\gafi] }
  {s-\mz \,+\,i\,\frac{s}{\mz} \cdot M_Z \Gamma_Z }  .\nn
\eea
The weak corrections appear in terms of fermion-dependent
form factors $\rho$ and $\kappa$ in the coupling constants
and in the width in the denominator.

\smallskip
The s-dependence of the imaginary part is due to the s-dependence of
$\imag\Sigma^Z$; the linearization is completely sufficient in the
resonance region.
We postpone the discussion of the $Z$ width for the moment and continue
with the form factors.

\bigskip
The form factors $\rho$ and $\kappa$ in Eq.\ (144) have {\it universal}
parts     (i.e.\ independent of the fermion species) and
{\it non-universal} parts which explicitly depend on the type of the
external fermions.
The universal parts arise from the counterterms and the boson
self-energies, the non-universal parts  from the vertex corrections and
the fermion self-energies in the external lines:
\bea
\rho_{e,f} & = & 1 + (\dro)_{univ} + (\dro)_{non-univ} \, \\
\kappa_{e,f} & = & 1 +(\dkap)_{univ} + (\dkap)_{non-univ} \, . \nn
\eea
In their leading terms the universal contributions are  given by
\bea
 (\dro)_{univ} & = & \dro + \cdots      \\
 (\dkap)_{univ} & = & \ctanw \, \dro + \cdots   \nn
\eea
with $\dro$ from Eq.\ (82). For incorporating the
next order leading terms one has to perform the substitutions     .
\bea
 \rho_{e,f} = 1 +\dro + \cdots & \ra &
  \frac{1}{1-\drb} + \cdots  \nn \\
 \kappa_{e,f} = 1+\frac{c_W^2}{s_W}\dro +\cdots & \ra &
                1+\frac{c_W^2}{s_W}\drb +\cdots \nn
\eea
with $\drb$ from Eq.\ (110).

\medskip
The leading structure of the universal parts can
easily be understood from the bare amplitude  with the
counter term expansion
\bea
\frac{e_0^2}{4\swo\cwo} & = & \frac{e^2}{4 s_W^2 c_W^2} \left[
 1+2\frac{\delta e}{e} -\frac{c_W^2-s_W^2}{s_W^2}\left(\dmz-\dmw
 \right)  \right]  \nn \\
 & = & \sqrt{2}\Gmu\mz \left[1 + \dmz-\dmw  +
 \cdots   \right]          \nn
\eea
and
$$ \dmz-\dmw = \dro +\cdots $$ in the quadratic $m_t$-term.
Thereby, $\Gmu$ was introduced by means of Eq.\ (87) together with the
expression (98) for $\Dr$.    In a similar way one finds
from Eq.\ (86):
$$
\swo =  s_W^2 \left[1 + \ctanw\left(\dmz-\dmw\right) \right]
   =   s_W^2 \left[ 1+\ctanw \, \dro +\cdots \right]
$$
recovering $(\dkap)_{univ}$.

\bigskip     \noi
The factorized $Z$ amplitude  allows us to define
NC vertices at the $Z$ resonance
with effective coupling constants
$g_{V,A}^f$, synonymously to the use of $\rho_f, \kappa_f$:
\bea
 J_{\m}^{NC} & = & \left( \sqrt{2}\Gmu\mz \rho_f \right)^{1/2}
\left[ (I_3^f-2Q_fs_W^2\kappa_f)\gamu-I_3^f\gamu\gafi \right] \nn\\
  & = & \left( \sqrt{2}\Gmu\mz \right)^{1/2} \,
  [g_V^f \,\gamu -  g_A^f \,\gamu\gafi]  \, .
\eea
The complete expressions for the
 effective couplings read as follows:
\bea
 g_V^f & = &  \left[ v_f +2s_W c_W \, Q_f \frac{\Pgzrz}{1+\Pigr(\mz)}
             + \fvzf \right] \left( \frac
              {1-\Dr}{1+\Pizr(\mz)} \right)^{1/2}, \nn \\
 g_A^f & = &  \left[ a_f
             + \fazf \right] \left( \frac
              {1-\Dr}{1+\Pizr(\mz)} \right)^{1/2} \, .
\eea
The building blocks are the following finite combinations of 2-point
functions, besides $\Dr$ in Eq.\ (89), evaluated at $s=\mz$:
\bea
\Pir^Z(s) & = & \frac{\real\,\Sigma^Z(s)-\dmmz}{s-\mz}- \Pi^{\g}(0)
   +\,\frac{c_W^2-s_W^2}{s_W^2} \left(\dmz-\dmw
        - 2\,\frac{s_W}{c_W} \,\frac{\Sigma^{\g Z}(0)}{\mz}
       \right)   \nn \\
\hat{\Pi}^{\g Z}(s)
    & = & \frac{\sgz(s)-\sgz(0)}{s}\,-\,\frac{c_W}{s_W}
   \left(\dmz-\dmw\right) +\,2\,\frac{\sgz(0)}{\mz}
\eea
and the finite vector and axial vector form factors $F_{V,A}$
at $s=\mz$ from the vertex
corrections together with the external fermion wave function 
renormalizations
\bea
   &    &   \nn \\
     &  & i\,\frac{e}{2s_Wc_W} \left\{ \gamu\fvzf(s)
               - \gamu\gafi\fazf(s) +\,I_3^f \gamu(1-\gafi) \cdot
               \frac{c_W}{s_W} \,\frac{\sgz(0)}{\mz} \right\}\, , \nn\\
   &    &   \nn
\eea
after splitting off the singular part $\sim\sgz(0)$.
Due to the imaginary parts of the self energies and vertices,
the form factors and the effective couplings, respectively, are
complex quantities. The approximation, where the couplings are
taken as real, is called the ``improved Born approximation''.

\bigskip     \noi
{\it The $Zbb$ couplings:}

\medskip \noi
The separation of a universal part in the effective couplings is
sensible for two reasons: for the light fermions ($f\neq b,t$)
the non-universal contributions are small, and (practically)
independent of the unknown parameters $m_t,M_H$ which enter only the
universal part. This is, however, not true for the $b$-quark where
also the non-universal parts have a strong dependence on $m_t$
\cite{vertex}
resulting from the virtual top quark in the vertex corrections.
The difference between the $d$ and $b$
couplings can be parametrized in the following way
\beq
  \rho_b = \rho_d (1+\tau)^2, \;\;\;\;
  s^2_b = s^2_d (1+\tau)^{-1}
\eeq
with the quantity
$$
 \tau = \Delta\tau^{(1)}
      + \Delta\tau^{(2)}
      + \Delta\tau^{(\al_s)}
$$
calculated perturbatively, at the present level comprising:
the complete 1-loop order term \cite{vertex} with $x_t$
from Eq.\ (105): 
\beq
\Delta\tau^{(1)} = -2 x_t - \frac{\Gmu\mz}{6\pi^2\sqrt{2}}
 (c_W^2+1)\log\frac{m_t}{M_W} + \cdots ,
\eeq
 the leading
electroweak 2-loop contribution of $O(\Gmu^2 m_t^4)$
\cite{barbieri,dhl}
\beq
\Delta\tau^{(2)} = -2\, x_t^2 \, \tau^{(2)} \, ,
\eeq
where
 $\tau^{(2)}$ is a function of $M_H/m_t$
with
 $\tau^{(2)} = 9-\pi^2/3$ for $M_H \ll m_t$;
the QCD corrections to the leading term of $O(\al_s\Gmu m_t^2)$
\cite{jeg}
\beq
\Delta\tau^{(\al_s)} =  2\, x_t \cdot \frac{\al_s}{\pi}
 \cdot \frac{\pi^2}{3} \, ,
\eeq
and the $O(\al_s)$ correction to the $\log m_t/M_W$ term in (151),
with a numerically very small coefficient \cite{log}.
 
For $M_H > 2 m_t$ a good approximation
for the 2-loop coefficient $\tau^{(2)}$ is given by
the asymptotic expression
\cite{barbieri}
 with $r= (m_t/M_H)^2$:
\bea
 \tau^{(2)} & = & \frac{1}{144} \left[
    311 + 24\pi^2 +282\log r + 90 \log^2 r  \right. \nn \\
  &  & -4r(40+6\pi^2 +15\log r +18\log^2 r )  \nn \\
  &  & \left. +\frac{3r^2}{100}
      (24209-6000\pi^2-45420 \log r - 18000 \log^2 r) \right]
\eea

\bigskip \noi
{\it Effective mixing angles:}

\setlength{\unitlength}{1cm}
\begin{figure}[hbt]
\begin{center}
\begin{picture}(15,13)
%\put(0,0){\rule{.2cm}{.2cm}}
%\put(0,13){\rule{.2cm}{.2cm}}
%\put(15,0){\rule{.2cm}{.2cm}}
%\put(15,13){\rule{.2cm}{.2cm}}
\put(-4.0,-11.2){\includegraphics{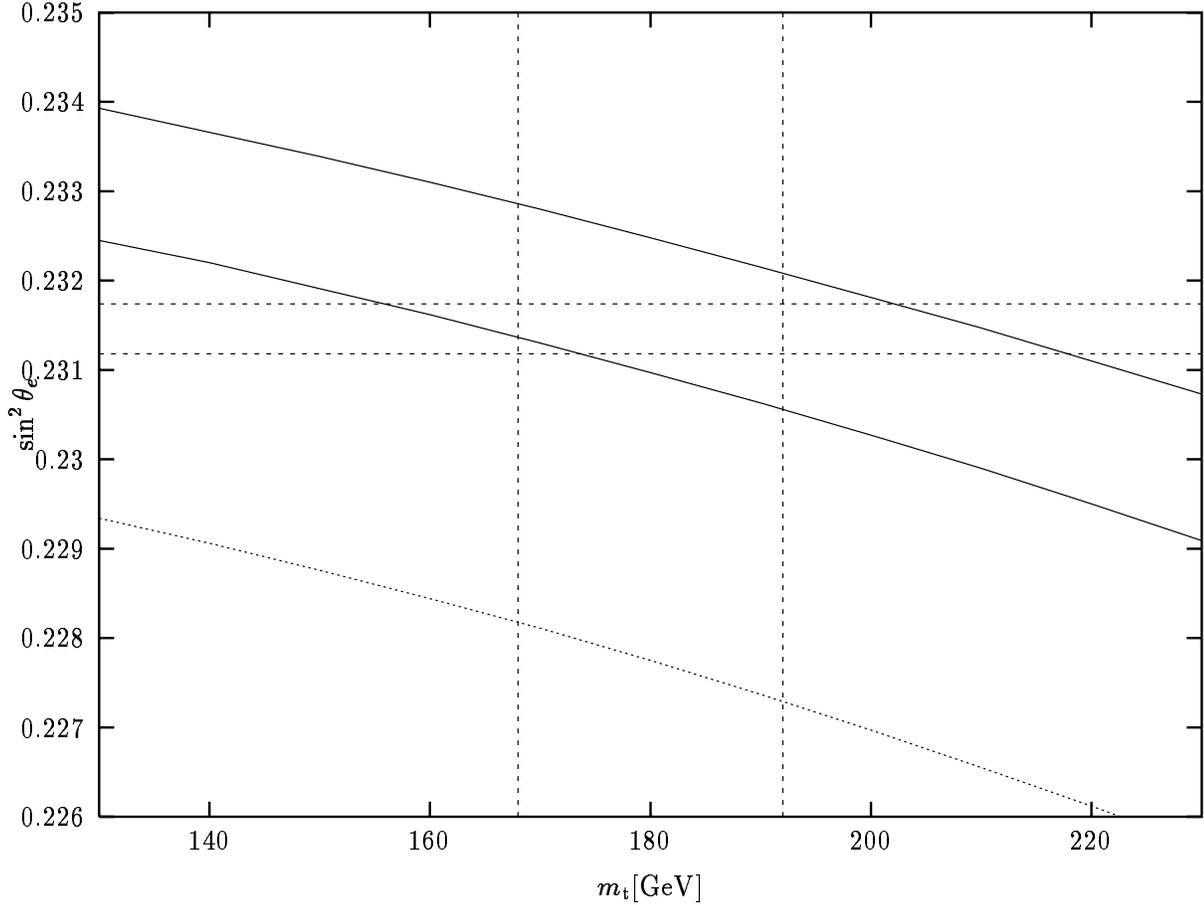}}
\end{picture}
\end{center}
\caption{$\sin^2\theta_e$ for $M_H=65$ GeV (lower line)
           and $M_H=1$ TeV (upper line).
           Also shown are the experimental data ($\pm 1\sigma$ ranges)
           for $m_t$ and $s_e^2$ (LEP/SLC average).
           The dotted curve includes only the fermionic loop effects via
           $\dal$ and $\Delta\rho$. }
\label{sine}
\end{figure}
\setlength{\unitlength}{0.7mm}

\medskip \noi
Due to the imaginary parts of the self energies and vertices,
the form factors and the effective couplings, respectively, are
complex quantities. We can define 
effective mixing angles
according to
\beq
 s_f^2 := 
 \sin^2\theta_f = s_W^2 \, \real \,\kappa_f =
  \frac{1}{4\mid Q_f\mid} \left(
              1-\frac{\real\, g_V^f}{\real\, g_A^f} \right) \, .
\eeq
 from the effective coupling constants in (148).
They are of particular interest since they determine the on-resonance
asymmetries, which will be discussed later in section 7.4.
Compared to $s_W^2$,
the on-resonance couplings are less sensitive to $m_t$ than the
$W$ mass.

\bigskip
In table 2 we put together the \sm predictions for
the leptonic (electron)  mixing angle $s_e^2$ for various values
of $m_t$ and $M_H$.
For the light quarks, the corresponding $s_q^2$ are very close
to the leptonic values (-0.0001 for $u$, -0.0002 for $d$ quarks).
Signficantly different is only $s_b^2$: +0.0014 for $m_t= 180$ GeV. 
Figure \ref{sine} displays the compatibilty of $s_e^2$ with the data. 

\bigskip
\begin{table}
\bc \btab{| l | c  c  c  c |}
\hline
  $m_t$  &  $M_H= 65$  &  100  &  300  & 1000  \\
\hline
     150   &     0.2319  &  0.2321 &   0.2327  &     0.2334  \\
     160   &     0.2316  &  0.2318 &   0.2324  &     0.2331  \\
     170   &     0.2313  &  0.2315 &   0.2321  &     0.2328  \\
     180   &     0.2310  &  0.2312 &   0.2318  &     0.2325  \\
     190   &     0.2306  &  0.2308 &   0.2314  &     0.2322  \\
     200   &     0.2303  &  0.2305 &   0.2311  &     0.2318  \\
\hline
\etab
\caption{The effective on-resonance mixing angle $s_e^2$ for electrons as
 predicted by the Standard Model
         for $M_Z=91.1884$ GeV and various top and Higgs masses (in GeV).
         $\al_s=0.123$.}
\ec
\end{table}

%\\[3cm]
%\bc      Effective mixing angles on-resonance
%                    for $M_Z=91.875$ GeV \\[3cm]
%         Normalization factors $\rho_f$ for the on-resonance
%         NC couplings   for $M_Z=91.187$ GeV
%\ec
%
%
\subsection{The Z line shape}
 The integrated cross section $\sigma(s)$
for $\epmf$ around the $Z$ resonance with unpolarized beams is
obtained from the formulae of the previous section in a
straight forward way, expressed in terms of the effective vector
and axial vector coupling constants.
It is, however, convenient to rewrite $\sigma(s)$
in terms of the $Z$ width and the partial widths $\G_e,\, \G_f$ in
order to have a more  model independent parametrization.
The following form \cite{berends,sisto}
 includes final state photon radiation
and QCD corrections in case of quark final states:
\footnote{Since initial state photon radiation is treated separately
the QED correction factor
 in Eq.\ (170) has to be removed in
$\G_e$.}
\bea
\sigma(s) & = & \frac{12\pi\G_e \G_f}
 {\mid s-\mz+iM_Z\G_Z(s)\mid ^2} \left\{
\frac{s}{\mz} + R_f\, \frac{s-\mz}{\mz}
              + I_f\, \frac{\G_Z}{M_Z}
               + \cdots \right\} \nn \\[0.2cm]
 &   & +\,\frac{4\pi\,\al(s)^2}{3s}\,Q_f^2 N_C^f \, K_{QCD} \,
       (1+\delta_{QED})
\eea
with
\beq
 \G_Z(s) = \G_Z \left\{ \frac{s}{\mz} + \eps\, \frac{s-\mz}{\mz}
           + \cdots \right\}
\eeq
and $N_C^f = 1$ for leptons and $=3$ for quarks.
The QCD correction in case $f=q$ is given in Eq.\ (172).
The terms $R_f,\, I_f,\, \eps$ are small
 quantities calculable in terms of
the basic parameters.
$R_f$ and $I_f$ describe
 the $\g$-$Z$ interference (improved Born approximation)
\bea
  R_f & = & \frac{2\,Q_e Q_f g_V^e g_V^f}
  {[(g_V^e)^2+(g_A^e)^2] [(g_V^f)^2+(g_A^f)^2]}
 \frac{4\pi\al(s)}{\sqrt{2}\Gmu\mz} \nn \\[0.2cm]
  I_f & = & \frac{2\,Q_e Q_f g_V^e g_V^f}
  {[(g_V^e)^2+(g_A^e)^2] [(g_V^f)^2+(g_A^f)^2]}
 \frac{4\pi\al(s)}{\sqrt{2}\Gmu\mz} \cdot \frac{s}{\mz}
 \, \imag\, \Pigr \, ,
\eea \\[0.2cm]
  and the last term is
the QED background from pure photon exchange
with
$$
  \al(s) = \frac{\al}{1+\real\, \Pigr_{ferm}(s)}
$$
and $\Pigr_{ferm}$ from Eq.\ (143).
The small correction
$$
\eps = \sum_f \eps_f, \;\;\;
 \eps_f \simeq \frac{6m_f^2}{\mz}\frac{\G_f}{\G_Z}
\frac{(g_A^f)^2}{(g_V^f)^2+(g_A^f)^2}
$$
is due to finite fermion mass effects in the final states.
 $I_f$ and $\eps$ have negligible influence on the line shape.

\smallskip
The $s$-dependent width gives rise to a dislocation of the
peak maximum by $\simeq -34$ GeV \cite{bbhvn,blrs}
 The first term in the
expansion (156) is the pure $Z$ resonance. It differs from a
Breit-Wigner shape by the $s$-dependence of the width:
\beq
\sigma_{res}(s) = \sigma_0 \,\frac{s \G_Z^2}
              {(s-\mz)^2 +s^2 \G_Z^2/\mz},\;\;\;\;
 \sigma_0 = \frac{12\pi}{\mz}\cdot\frac{\G_e\G_f}{\G_Z^2} \, .
\eeq
By means of the substitution \cite{blrs}
\beq
 s-\mz+i s\G_Z/M_Z = (1+i\g)(s-\hat{M}_Z^2+i\hat{M}_Z \hat{\G}_Z)
\eeq
with
\beq
 \hat{M}_Z = M_Z (1+\g^2)^{-1/2}, \;\;
 \hat{\G}_Z = \G_Z (1+\g^2)^{-1/2}, \;\;
 \g = \frac{\G_Z}{M_Z}
\eeq
a Breit-Wigner resonance shape is recovered:
 \beq
\sigma_{res}(s) = \sigma_0 \, \frac{s\gm}{(s-\mm)^2 + \mm\gm}
\eeq
Numerically one finds: $\hat{M}_Z-M_Z \simeq - 34$ MeV,
$\hat{\G}_Z - \G_Z\simeq -1$ MeV. $\sigma_0$ is not changed.
$\hat{M}_Z$ corresponds to the real part of the $S$-matrix pole
of the $Z$-resonance \cite{smatrix}.

\bigskip \noi
%\newpage \noi
{\it QED corrections:}

\medskip \noi
 The observed cross section is the result of convoluting
Eq.\ (156) with the initial state QED corrections consisting
of virtual photon and real photon bremsstrahlung
contributions:
\beq
\sigma_{obs}(s)  =
 \int_0^{k_{max}}\, dk \, H(k)\, \sigma(s(1-k)) \, .
\eeq
$k_{max}$ denotes a cut to the radiated energy. Kinematically
it is limited by $1-4m_f^2/s$ or
$1-4m_{\pi}^2/s$ for hadrons, respectively.
For the required accuracy, multi-photon radiation has to be
included. The radiator function $H(k)$ with soft-photon
resummation and the exact $O(\al^2)$ result \cite{bbvn}
 for initial state
QED corrections is given by \cite{berends}

\bigskip \noi
The QED corrections have two major impacts on the line shape:
\begin{itemize}
\item
a reduction of the peak height of the resonance cross section
by
 \beq
 \sigma_{obs}^{peak} \simeq \sigma_{res}^{peak}
  \left(\frac{\G_Z}{M_Z}\right) (1+\delta_1^{V+S})
 \, \simeq \, 0.74 \,\sigma_{res}^{peak} \, ,
 \eeq
\item
a shift in the peak position compared to the non-radiative
cross section by \cite{bbm}
 \beq \Delta \sqrt{s_{max}} \,= \, \frac{\beta\pi}{8}\G_Z
 \eeq
resulting in the relation between the peak position and the
nominal $Z$ mass:
 \bea
 \Delta \sqrt{s_{max}}&  \simeq & M_Z +\frac{\beta\pi}{8}\G_Z
      -\frac{\G_Z^2}{4M_Z}    \nn \\
  & \simeq & 89\, \mv  \, .
\eea
\end{itemize}
It is important to note that, to high accuracy, these effects
are practically universal, depending only on $M_Z$ and $\G_Z$
as parameters. This allows a model independent determination
of these parameters from the measured line shape.

\bigskip
 A final remark concerns the QED corrections resulting from
the interference between initial and final state radiation.
They are not included in the treatment above, but they can be
added in $O(\al)$ since they are small anyway. For not too tight
cuts, as it is the case for practical applications,
these interference corrections to the line shape are
 negligible and we do not list them here.

\bigskip
From line shape measurements one obtains the parameters
$M_Z,\, \G_Z,\, \sigma_0$ or the partial widths, respectively.
Whereas $M_Z$ is used as a precise input parameter, together
with $\al$ and $\Gmu$, the width and partial widths allow
comparisons with the predictions of the Standard Model to be
discussed next.
\subsection{$Z$ widths and partial widths}
The total
$Z$ width $\Gamma_Z$ can be calculated
as the sum over the partial decay widths
\beq
 \Gamma_Z = \sum_f \, \G_f + \cdots, \;\;\; \G_f =
 \Gamma  (Z\ra f\bar{f}) \, ,
\eeq
where the ellipses indicate other decay channels which, however,
are not significant.
 The fermionic partial
widths,
 when
expressed in terms of the effective coupling constants defined
in section 7.1, read:
\bea
\Gamma_f  & = &  \G_0 \,\sqrt{1-\frac{4m_f^2}{\mz}}
 \, \left[
 \mid g_V^f \mid ^2 \left(1+\frac{2m_f^2}{\mz}\right) +
 \mid g_A^f\mid ^2 \left(1-\frac{4m_f^2}{\mz}\right)
                           \right]
 \cdot   (1+\delta_{QED})  \nn \\[0.2cm]
  &   &
     \, +\, \Delta\G^f_{QCD}     \\[0.2cm]
  & \simeq & \G_0
 \, \left[
 \mid g_V^f\mid ^2  +
 \mid g_A^f\mid ^2 \left(1-\frac{6m_f^2}{\mz}\right)
                           \right]
 \cdot   (1+\delta_{QED})
     \, +\, \Delta\G^f_{QCD} \nn
\eea
with
\beq
\G_0 \, =\,
  N_C^f\,\frac{\sqrt{2}\Gmu M_Z^3}{12\pi} \, .
\eeq
The photonic QED correction
\beq
\delta_{QED} = Q_f^2\, \frac{3\al}{4\pi}
\eeq
 is very
small, maximum  0.17\% for charged leptons.

\smallskip \noi
The QCD correction for hadronic final states is given by
\beq
 \Delta\G^f_{QCD}\, =\, \G_0
  \left[ \mid g_V^f\mid ^2+\mid g_A^f\mid ^2 \right]
 \cdot K_{QCD}
\eeq
with \cite{qcd1}
\bea
K_{QCD}  & = & \frac{\al_s}{\pi} +1.41 \left(
  \frac{\al_s}{\pi}\right)^2 -12.8 \left(
  \frac{\al_s}{\pi}\right)^3 
 - \frac{Q_f^2}{4} \frac{\al\al_s}{\pi^2}
\eea
 for the light quarks with $m_q\simeq 0$.

\smallskip
For $b$ quarks
the QCD corrections are different due to  finite $b$ mass terms
and to top quark dependent 2-loop diagrams
 for the axial part
\cite{qcdb}:
\bea
 \Delta\G_{QCD}^b & = & \G_0 \left[
             \mid g_V^b\mid ^2+\mid g_A^b\mid ^2 \right]
                       \cdot K_{QCD} \nn \\[0.1cm]
    &  & +\G_0\,\mid g_V^b\mid ^2 \cdot \frac{12 m_b^2}{\mz}
    \left[ \alspi +\left(\alspi\right)^2 (6.07-2\ell)
          \right.   \nn \\[0.1cm]
    &   & \hspace*{3cm}
         \left.
        +\left(\alspi\right)^3 (2.38-24.29\ell+0.083\ell^2)
       \right] \nn \\[0.1cm]
    &  & +\G_0\,\mid g_A^b\mid ^2 \cdot \frac{6 m_b^2}{\mz}
    \left[ \alspi (2\ell-1)
     +\left(\alspi\right)^2 (17.96+14.14\ell-0.083\ell^2)
       \right] \nn \\[0.1cm]
    &  & +\G_0\,\mid g_A^b\mid ^2 \cdot \frac{1}{3}
      \left(\alspi\right)^2 I(\mz/4m_t^2)
\eea
with
$$
 \ell = \log \frac{\mz}{m_b^2}
$$
and
$$
  I(x) \simeq -9.250 + 1.037 x +0.632 x^2
  + 6\,\log(2\sqrt{x}) \, .
$$
The finite $b$-mass terms
contribute  $+ 2$ MeV to the partial $Z$ width
into $b$ quarks. Moreover,
the top mass dependent correction
 at the 2-loop level
 yields an additional, but negative, contribution.
For large $m_t$
this top-dependent term  cancels part of the positive and constant
correction resulting from $m_b\neq 0$ in $O(\al_s)$.

\medskip
Radiation of secondary fermions through photons
from the primary final state fermions
can yield another sizeable contribution to the partial $Z$ widths
which, however, is compensated by the corresponding virtual
contribution through the dressed photon propagator in the final state
vertex correction. For this compensation it is essential that
the analysis is inclusive enough, i.e.\ the cut to the invariant mass
of the secondary fermions is sufficiently large \cite{hoang}.

\bigskip
In table 3  the \sm predictions for the various partial widths
and the total width of the $Z$ boson are collected.
They include all the electroweak, QED and QCD corrections
discussed above.
Of particular interest are the following ratios of partial widths
\beq
R_{had} = \frac{\G_{had}}{\G_e}, \;\;\;
R_b = \frac{\G_b}{\G_{had}}, \;\;\;
R_c = \frac{\G_c}{\G_{had}} .
\eeq

\bigskip
\begin{table}
\bc
\btab{| r | r | c | c | c | c | c | c | c | c |}
\hline
 &  & & & & & & & & \\
 $m_t$ & $M_H$ & $\G_{\nu}$ & $\G_e$ & $\G_u$ & $\G_d$ &
 $\G_b$ & $ \G_{had}$ & $\G_{tot}$ & $R_{had}$ \\
 &  & & & & & & & & \\
\hline
\hline
     150  &    60   & 166.9 & 83.84 & 299.4 & 382.4 & 376.8 &
            1740.2 & 2492.2 & 20.76 \\
          &   300   & 166.8 & 83.74 & 298.7 & 381.7 & 376.1 &
            1736.8 & 2488.2 & 20.74\\
          &  1000   & 166.6 & 83.60 & 298.0 & 380.9 & 375.3 &
            1732.9 & 2483.3 & 20.73\\
\hline
     175  &    60   & 167.3 & 84.06 & 300.6 & 383.5 & 376.1 &
            1744.1 & 2497.9 & 20.75 \\
          &   300   & 167.2 & 83.95 & 299.9 & 382.7 & 375.4 &
            1740.6 & 2493.7 & 20.73 \\
          &  1000   & 166.9 & 83.81 & 299.1 & 381.9 & 374.6 &
            1736.6 & 2488.6 & 20.72 \\
\hline
     200  &    60   & 167.7 & 84.32 & 301.9 & 384.8 & 375.4 &
            1748.6 & 2504.5 & 20.74 \\
          &   300   & 167.6 & 84.20 & 301.1 & 384.0 & 374.7 &
            1744.9 & 2499.9 & 20.72 \\
          &  1000   & 167.3 & 84.05 & 300.3 & 383.1 & 373.8 &
            1740.7 & 2494.6 & 20.71\\
\hline
\etab
\caption{Partial and total $Z$ widths in MeV for various top
         and Higgs masses (in GeV).
          $\al_s=0.125$.
         Not listed are the values for
          $\G_{\tau}=0.9977\, \G_e$ and
          $\G_c$ which are very close to $\G_u$. }
\ec
\label{widths}
\end{table}

%\newpage
% . \\[9cm]
%{\bf Fig.\ 11:} {\it Total $Z$ width $\Gamma_Z$ in the \sm for $M_Z=
%           91.172$ GeV. \\ $M_H=42$ GeV ($\cdots$), 100 GeV ($---$),
%              1000 GeV ($-$ $-$ $-$). Not included is the
%              QCD uncertainty of $\simeq 12$ MeV.
%              1-$\sigma$ limits from LEP$^5$.} \\[10cm]
%{\bf Fig.\ 12:} {\it Partial $Z$ widths (in MeV) into $d$-quarks
%(upper curves) and into $b$-quarks (lower curves). $M_Z=91.172$ GeV.
%$M_H = $ 42 ($\cdots$), 100 ($---$), 1000 ($-$ $-$ $-$) GeV.
%The overall normalization is different for $d$ and $b$ due to the
%finite mass of the $b$-quark}

%\newpage\noi
%.
%newpage \noi  .\\[10cm]
%{\bf Fig.\ 7:} {\it The ratio $\G_b/\G_{had}$ as function of $m_t$}
% \\[2cm]

%
\subsection{Asymmetries}
\subsubsection{Left-right asymmetry}
The left-right \as is defined as the ratio
\beq
  \alr = \frac{\sigma_L-\sigma_R}{\sigma_L +\sigma_R}
\eeq
where $\sigma_{L(R)} $ denotes the integrated cross section for left
(right) handed electrons. $\alr$, in case of lepton universality, is
equal to the final state polarization in $\tau$-pair production:
\beq
     A_{pol}^{\tau} = \alr \, .
\eeq \\[0.2cm]
The on-resonance \as ($s=\mz$) in the improved Born approximation
 is given by
\beq
\alr(\mz) =  A_e
 \,+\,\Delta \alr^I \, +\, \Delta \alr^Q
\eeq
 where the combination
   \beq
    A_e = \frac{2g_V^e g_A^e}{(g_V^e)^2+(g_A^e)^2}
        = \frac{2(1 -4  \sefl)}
         {1+(1-4 \sefl)^2}
\eeq
depends only on the  effective mixing angle Eq.\ (155) for the
electron.
The small contributions from the
interference with the photon exchange
\bea
\Delta\alr^I
      & = & \frac{2 Q_e Q_f\, g_A^e g_V^f}
% {[(g_V^e)^2+(g_A^e)^2] [(g_V^f)^2+(g_A^f)^2]}
  {\gve \,\gvf}
 \frac{4\pi\al(\mz)}{\sqrt{2}\Gmu\mz} \cdot \frac{\G_Z}{M_Z}
 \cdot \imag\, \Pigr
\eea
and from the pure photon exchange part
\bea
\Delta\alr^Q
      & = & -\,
% {[(g_V^e)^2+(g_A^e)^2] [(g_V^f)^2+(g_A^f)^2]}
    \frac{A_e \, Q_e^2 Q_f^2}
   {\gve \, \gvf}
 \left(\frac{4\pi\al(\mz)}{\sqrt{2}\Gmu\mz}\right)^2
 \left( \frac{\G_Z}{M_Z}\right)^2  \, .
\eea
are listed in table 4 for the various final state fermions.
Except from lepton final states, they are negligibly small.
Mass  effects from final fermions practically cancel.
The same holds for QCD corrections in the case of quark final states,
final state QED corrections, and QED corrections from the interference
of initial-final state photon radiation. Initial state QED
corrections can be treated in complete analogy to Eq.\ (163)
applied to $\sigma_{L,R}(s)$. Their net effect in the \as
is also very small and practically independent of cuts
\cite{bh,alrqed}.
$\alr$ thus represents a unique laboratory for testing the
non-QED part of the electroweak theory.
Measurements  of $\alr$  are essentially measurements of $\sefl$ or of
the ratio $g_V^e/g_A^e$.

\smallskip
\begin{table}
\bc \btab{| c | c | c | c |}
\hline
          &   &   &   \\
 $f$   &  $ A_e $ & $\Delta\alr^I$ &
      $\Delta\alr^Q$  \\
          &   &   &   \\
\hline
          &   &   &   \\
 $\mu$ & 0.1511 &  0.0002 & -0.0009 \\
          &   &   &   \\
$\tau$ & 0.1511 & 0.0002 & -0.0009 \\
          &   &   &   \\
$ c$   & 0.1511 & 0.0005 & -0.0003 \\
          &   &   &   \\
$ b$   & 0.1511 & 0.0004 & -0.0001 \\
          &   &   &   \\
\hline
\etab
\caption{Contributions to the
         on-resonance left-right asymmetry for various final state
          fermions.
          $\sefl=0.2314$.}
\ec
\end{table}
\subsubsection{Forward-backward asymmetries}
The forward-backward \as is defined by
\beq
 \afb = \frac{\sigma_F -\sigma_B}
             {\sigma_F +\sigma_B}
\eeq
with
\beq
 \sigma_F = \,\int_{\theta>\pi/2} d\Omega \,
                \frac{d\sigma}{d\Omega} \, , \;\;\;
 \sigma_B = \,\int_{\theta<\pi/2} d\Omega \,
                \frac{d\sigma}{d\Omega} \, .
\eeq

\medskip \noi
For the on-resonance \as ($s=\mz$) we get in the improved Born
approximation:
\beq
A_{FB}(\mz) \,=\,\frac{3}{4}\cdot A_e\, A_f \left(
 1-4\m_f
 +6\mu_f\frac{(g_A^f)^2}{(g_V^f)^2+(g_A^f)^2} \right)
 \,+\,\Delta \afb^I \, +\, \Delta \afb^Q \, .
\eeq
 $A_{f}$ is defined as
\beq
A_f \,=\, \frac{2g_V^f g_A^f}{(g_V^f)^2+(g_A^f)^2}\,
 =\, \frac{2(1-4\mid Q_f\mid s_f^2)}
 {1+(1-4\mid Q_f\mid s_f^2)^2}
\eeq
with the short-hand notation for the effective mixing angle
in Eq. (155):
$$ s_f^2 \equiv \sin^2\theta_f \, .$$
The small contributions $\Delta\afb^{I,Q}$ result from the
the interference with the photon exchange
\bea
\Delta\afb^I
      & = & \frac{3}{4}\,\frac{2 Q_e Q_f\, g_A^e g_A^f}
% {[(g_V^e)^2+(g_A^e)^2] [(g_V^f)^2+(g_A^f)^2]}
  {\gve \,\gvf}
 \frac{4\pi\al(\mz)}{\sqrt{2}\Gmu\mz} \cdot \frac{\G_Z}{M_Z}
 \cdot \imag\, \Pigr
\eea
and from the pure photon exchange part:
\bea
\Delta\afb^Q
      & = & -\, \frac{3}{4}
% {[(g_V^e)^2+(g_A^e)^2] [(g_V^f)^2+(g_A^f)^2]}
 \, \frac{A_e A_f \, Q_e^2 Q_f^2}
   {\gve \, \gvf}
 \left(\frac{4\pi\al(\mz)}{\sqrt{2}\Gmu\mz}\right)^2
 \left( \frac{\G_Z}{M_Z}\right)^2  \, .
\eea

\medskip  \noi
The on-resonance \ass are essentially determined by the values of the
effective mixing angles for  $e$ and $f$ entering the product
$A_e A_f$.  Through $s^2_{e,f}$ also the  dependence of the \ass on
the basic \sm parameters $m_t, M_H$ is fixed.
 The small corrections
from finite mass effects, interference and photon exchange  can be
considered practically independent of the details of the model.
For demonstrational purpose  we list in
table 5  the various terms in the
 on-resonance \ass according to Eq.\ (183) for a common value
of the effective mixing angle $s^2_e = s^2_f =0.2314$.

\smallskip
\begin{table}
\bc \btab{| c | c | c | c | c |}
\hline
       &   &   &   &   \\
 $f$   &  $\frac{3}{4}A_e A_f$ & mass correction & $\Delta\afb^I$ &
      $\Delta\afb^Q$  \\
       &   &   &   &   \\
\hline
       &   &   &   &   \\
 $\mu$ & 0.0171 &  $< 10^{-6}$ & 0.0018 & -0.0001 \\
       &   &   &   &   \\
$\tau$ & 0.0171 & $ 1.3\cdot 10^{-5}$ & 0.0018 & -0.0001 \\
       &   &   &   &   \\
$ c$   & 0.0758 & $ 2.5\cdot 10^{-5} $ & 0.0011 & -0.0002 \\
       &   &   &   &   \\
$ b$   & 0.1061 & $ 1.5\cdot 10^{-5} $ & 0.0004 & $ -5\cdot 10^{-5}$ \\
       &   &   &   &   \\
\hline
\etab
\caption{On-resonance forward-backward asymmetries for $s^2_f
         =0.2314$.}
\ec
\end{table}

\bigskip \noi
{\it Final state QED corrections:}

\medskip \noi
According to the representation of $\afb$ as the ratio
 of the antisymmetric to the symmetric part of the cross
 section, the effects can be summmarized as follows:

\medskip
\noi
If  no cuts are applied, only the symmetric part
$$ \sigma \,=\, \sigma_F + \sigma_B $$
gets a correction:
\beq
  \sigma \ra \sigma\cdot \left(
   1+\frac{3\al}{4\pi}\,Q_f^2 \right)
\eeq
whereas \cite{bardin1,bardin2}
\beq
 \delta (\sigma_F -\sigma_B) \,=\, 0\, .
\eeq
This results in a correction to the \as
\beq
 \afb\,\ra\, \afb\cdot \left(
  1- \frac{3\al}{4\pi}\,Q_f^2 \right)
\eeq
which is a very small negative contribution  ($< 0.17$ \% relative
to $\afb$).

\bigskip \noi
{\it QCD corrections:}

\medskip \noi
Quite in analogy, for the
QCD single gluon emission \cite{laerman,djouadi} the following
correction to the
\as for quark final states with $m_q \ra 0$ arises:
\beq
 \afb\,\ra\, \afb\cdot \left(
  1- \frac{\al_s}{\pi} \right) \, .
\eeq \\[0.2cm]
For massive quarks, the QCD final state corrections can be included by
multiplying the purely electroweak \as by a factor
\beq 1 - \frac{\al_s}{\pi}\, \Delta_q \, .  \eeq
The coefficient $\Delta_q$ is, to a very good approximation (1\%) for
the known quarks given by \cite{djouadinew}
\beq
\Delta_q \, =\, 1 - \frac{16}{3}\,\frac{m_q}{M_Z} + \cdots
\eeq
which yields
\beq
 \Delta_q \, =\, \left\{ \begin{array}{ll}
   1  &  \mbox{ for } u,d \mbox{-quarks}  \\
 1- 0.02 & \mbox{ for } s \mbox{-quarks}  \\
 1-0.07  & \mbox{ for } c \mbox{-quarks} \\
 1-0.21  & \mbox{ for } b \mbox{-quarks}
 \end{array}  \right.
\eeq
with $m_s = 500$ MeV, $m_c = 1.5$ GeV, $m_b = 4.5$ GeV.
The exact formulae are given in the report ``Heavy Quarks''
 \cite{heavy}.

\bigskip     \noi
{\it Initial state QED corrections:}

\medskip \noi
As we know from the integrated cross section,
the initial state corrections give rise to a significant
reduction of the peak height which is due to the
rapid variation of $\sigma(s)$ with the energy.
Since the \as $\afb (s)$ is a steeply increasing function
around the $Z$ the energy loss from initial-state radiation
$s \ra s'<s$ leads to a reduction in the \as as well:
  $$\afb(s') < \afb (s) \, .$$
Quantitatively, the $O(\al)$ correction to $\afb$ for muons
 close to the peak
$  \delta\afb \simeq  -0.02  $
is of the order of the on-resonance
\as itself. Therefore it is obvious that
also the higher order QED contributions
have to be taken into account carefully.

\bigskip
We can express the initial state QED corrections to $\afb$ in
a compact form, quite in analogy to the convolution integral
for the integrated cross section $\sigma(s)$
 in Eq.\ (163):
\beq
\afb(s) \, =\, \frac{1}{\sigma(s)}\,\int_{0}^{k_{max}} dk\;
 \frac{4(1-k)}{(2-k)^2}\,\tilde{H}(k)\,\sigma_{FB}(s')\, ,
 \;\;\;\; k_{max} \leq 1-\frac{4m_f^2}{s} \, .
\eeq
The basic ingredients are the expression for the non-radiative
antisymmetric cross section are
    $$\sigma_{FB}(s) =      \sigma_F(s) -\sigma_B(s)
$$
and the radiator function $\tilde{H}(k)$.
The quantity
$$ s' = (1-k)\, s
    = (p_f+p_{\bar{f}})^2$$ is the invariant mass of
the outgoing fermion pair.
The effect of the change in the scattering angle by the boost
from the $f\bar{f}-$cms to the laboratory frame
is taken into account by the kinematical factor in front of
 $\tilde{H}$ in the
convolution integral.

\bigskip
$\tilde{H}$ is different from the radiator function $H$ for the
symmetric cross section in Eq.\ (163) in the hard photon terms.
 According to the present
status of the calculation, $\tilde{H}$ contains the exact
$O(\al)$ contribution \cite{bardin1,bardin2,bardin3},
the $\oal$ contributions in the leading-log approximation
\cite{ringberg}, and the resummation of soft photons to all
orders \cite{afb}.

\bigskip
The behaviour of $\afb$ under initial state QED corrections
 is  qualitatively similar to that of the
integrated cross section where the higher order QED contributions
bring the prediction closer to the lowest order result compared to the
$O(\al)$ corrections.

\bigskip
The QED corrections from the interference of
initial-final  state radiation
 are very  small ($\mid \delta\afb\mid <0.001$)
 if no tight cuts
to the photon phase space are applied.
More restrictive cuts make the interference contributions to
$\afb$ important exceeding the level of 0.01 (for muons) when the
photon is restricted to energies below 1 GeV \cite{afb}.
 The complete set of QED corrections is
available in (semi-) analytic form,
 exact in $O(\al)$  and with leading higher order terms, also for
situations with
cuts, covering: energy or invariant mass cuts,
accollinearity cuts, acceptance cuts
\cite{bardin3,cuts,zfitter,topaz}
showing agreement
 within 0.2\%.
 \subsection{Uncertainties of the \sm predictions}
In order to establish in a significant manner
possibly                    small effects from unknown
physics we have to know the uncertainties of our theoretical predictions
which have to be confronted with the experiments.

\medskip \noi
The sources of uncertainties in theoretical predictions are the following:
\begin{itemize}
\item the experimental errors of the parameters used as an input. With the
choice $\al$, $\Gmu$, and $M_Z$ from LEP we can keep these errors as small as
possible. The errors from this source are then determined by $\delta M_Z$
since the errors of $\al$ and $\Gmu$ are negligibly small. For
any of the     mixing angles with
$s_W^2,\sms, s^2_f$
\beq
 \frac{\delta s^2}{s^2} = \frac{2\,c^2}{c^2-s^2}\,
 \frac{\delta M_Z}{M_Z}
\eeq
one finds
$$ \delta s^2 \, \simeq 2\cdot 10^{-5} \, .$$
\item the uncertainties from quark loop contributions to the \rc.
Here, we have to distinguish two cases: the uncertainties from the light
quark contributions to $\Delta \al$ and the uncertainties from the
heavy quark contributions to $\Delta \rho$. In both cases the
uncertainties are due to strong interaction effects, which are not sufficiently
under control theoretically. The problems are due to:\\
(i) the QCD parameters. The scale of $\al_s$  and the
definition and scale of quark masses to be used in the calculation of
a particular quantity are quite ambiguous in many cases.\\
(ii) the bad convergence and/or breakdown of perturbative QCD. In
particular at low $q^2$ and in the resonance regions theoretically poorly
known nonperturbative effects are non-negligible.

\medskip
The theoretical problems with
the hadronic contributions of the 5 known light quarks
to $\Delta \al$ can be circumvented by
using the experimental $\epm$-annihilation cross-section $
\sigma_{tot}(\epm \rightarrow \gamma^* \rightarrow hadrons)$.
   The error \cite{eidelman}
$$
\delta(\dal) = \pm 0.0007
$$
is dominated by the large experimental errors
in $\sigma_{tot}(\epm \rightarrow \gamma^* \rightarrow hadrons)$
 and can be improved
 only by more precise measurements of
hadron production in $\epm$-annihilation at energies well below $M_Z$.
The present uncertainty leads to an error
 in the $W$-mass prediction
$$
\frac{\delta M_W}{M_W} = \frac{s_W^2}{c_W^2-s_W^2}
 \frac{\delta(\Dr)}{2(1-\Dr)} $$ of
$\delta M_W$ = 13 MeV
and $\delta \sin^2 \theta$
= 0.00023 in the prediction of the various  weak mixing parameters
$s_W^2, \sms, s^2_f$. This matches with the present (and even more
the future) experimental precision in the electroweak mixing angle.

\item
The uncertainties from the QCD contributions,
 besides the 3 MeV in the
hadronic $Z$ width from $\delta \al_s=0.006$,
 can essentially be traced back to
those in the top quark loops for the $\rho$-parameter.
They  can be combined into the following errors
\cite{kniehl95}, which have improved due to the recently available
3-loop result:
$$
 \delta(\dro) \simeq 1.5\cdot 10^{-4},   \;
 \delta s^2_{\ell} \simeq 0.0001
$$
for $m_t = 174$ GeV, and slightly larger for heavier top.

\item the uncertainties from omission of higher order effects.
The size of unknown higher order contributions can be estimated
by different treatments of non-leading terms
of higher order in the implementation of radiative corrections in
electroweak observables (`options')
and by investigations of the scheme dependence.
Explicit comparisons between the results of 5 different computer codes  
based on  on-shell and $\ms$ calculations
for the $Z$ resonance observables are documented in the ``Electroweak
Working Group Report'' \cite{ewgr} in ref.\ \cite{yb95}
(see also  \cite{bardin}).
The typical size of the genuine electroweak uncertainties
is of the order 0.1\%.
The following table 6 shows the uncertainty in a selected set of
precision observables. In particular for the very precise $s^2_e$
the theoretical uncertainty is still remarkable.
Improvements of the accuracy displayed in table 6
 require systematic electroweak and QCD-electroweak
 2-loop calculations.
As an example for the importance of electroweak non-leading
2-loop effects, an explicit calculation of these terms has been
performed for $\rho$ (the overall normalization) in neutrino
scattering \cite{padova}: they are sizeable and comparable to
the $O(\Gmu^2 m_t^4)$ term. Hence, one should take the registered       
uncertainties also for the $Z$ region very seriously.

\begin{table}[htbp]\centering
\vspace{2cm}
\begin{tabular}{|c|c|c|}
\hline \hline
Observable $O$ & $\Delta_c O$  & $\Delta_g O$ \\
\hline
            & & \\
$M_W\,$(GeV)          & $4.5\tmth$ & $1.6\tmt$\\
$\G_e\,$(MeV)          & $1.3\tmt$ & $3.1\tmt$\\
$\G_Z\,$(MeV)          & $0.2$     & $1.4$\\
$ s^2_e$             & $5.5\tmfv$ & $1.4\tmf$\\
$ s^2_b$             & $5.0\tmfv$ & $1.5\tmf$\\
$R_{had}$                 & $4.0\tmth$& $9.0\tmth$\\
$R_b$                 & $6.5\tmfv$ & $1.7\tmf$ \\
$R_c$                 & $2.0\tmfv$& $4.5\tmfv$ \\
$\sigma^{had}_0\,$(nb)    & $7.0\tmth$ & $8.5\tmth$\\
$\afb^l$             & $9.3\tmfv$ & $2.2\tmf$\\
$\afb^b$             & $3.0\tmf$ & $7.4\tmf$ \\
$\afb^c$             & $2.3\tmf$ & $5.7\tmf$ \\
$\alr$                & $4.2\tmf$ & $8.7\tmf$\\
\hline \hline
\end{tabular}
\caption[]
{Largest half-differences among central values $(\Delta_c)$ and among
maximal and minimal predictions $(\Delta_g)$ for $m_t = 175\,\gv$,
$60\,\gv < M_H < 1\,\tv$ and $\al_s(\mz) = 0.125$
(from ref.\ \cite{ewgr}) }
\end{table}
\clearpage
\end{itemize}
\section{Standard model and precision  data}
\subsection{Standard model predictions versus data}
In table 7
the \sm predictions for $Z$ pole observables and the $W$ mass  are
put together. The first error corresponds to
the variation of $m_t$ in the observed range (1) and $ 60 < M_H < 1000$ GeV.
The second error is the hadronic
uncertainty from $\al_s=0.123\pm 0.006$, as measured
by QCD observables at the $Z$ \cite{alfas}.
 The recent combined LEP results \cite{lep} on the $Z$ resonance
parameters, under the assumption of lepton universality,
are also shown in table 1, together with $s^2_e$ from
the left-right asymmetry at the SLC \cite{sld}.
The quantities $\rho_{\ell}$, $s^2_{\ell}$ are the leptonic
NC coupling normalization and mixing angle, assumed to be universal.

The value for the leptonic mixing angle from the left-right asymmetry
$A_{LR}$ has come closer to the LEP result, but due to its smaller
error the deviation from the cumulative LEP average
is still about  $3\sigma$.

%\newpage
\begin{table}
  \bc
 \btab{| l | l | r | }
\hline
\hline
 observable & exp. (1995) & \sm prediction \\
\hline
\hline
$M_Z$ (GeV) & $91.1884\pm0.0022$ &  input \\
\hline
$\Gamma_Z$ (GeV) & $2.4963\pm 0.0032$ & $2.4976 \pm 0.0077\pm 0.0033$ \\
%\hline
%$\Gamma_{had}$ (GeV) & $1.740\pm 0.008$ &
% $1.736\pm 0.008 \pm 0.007$ \\
%\hline
%$\Gamma_e$ (MeV) & $83.2\pm 0.4$ & $83.7\pm 0.4 $ \\
\hline
$\sigma_0^{had}$ (nb) & $41.4882\pm 0.078$ & $41.457\pm0.011\pm0.076$ \\
\hline
 $\G_{had}/\G_e$ & $20.788\pm 0.032 $ & $20.771\pm 0.019\pm 0.038$ \\
%\hline
%$\Gamma_e$ (MeV) & $83.82\pm 0.27$ & $83.7\pm 0.4 $ \\
\hline
$\Gamma_{inv}$ (MeV) & $499.9\pm 2.5$ & $501.6\pm 1.1$ \\
\hline
$\G_b/\G_{had}=R_b$  & $0.2219\pm 0.0017$ & $0.2155\pm 0.0004$ \\
\hline
$\G_c/\G_{had}=R_c$  & $0.1540\pm 0.0074$ & $0.1723\pm 0.0002$ \\
  \hline
$A_b$            & $0.841\pm 0.053$  & $0.9346 \pm 0.0006$ \\
\hline
$\rho_{\ell}$ & $1.0044\pm 0.0016$ & $1.0050\pm 0.0023$ \\
\hline
$s^2_{\ell}$ (LEP) & $0.23186\pm 0.00034$ & $0.2317\pm 0.0012$ \\
\hline
$s^2_e (A_{LR})$ & $0.23049\pm 0.00050$ & $0.2317\pm 0.0012$   \\
 LEP$+$SLC   &  $0.23143\pm 0.00028$    &                    \\
\hline
$M_W$ (GeV) & $80.26 \pm 0.16$ & $80.36\pm 0.18$  \\
\hline
\hline
\etab
            \caption{Precision observables: experimental results
%            \cite{lep,sld,wmass}
%            (from refs.\ 1,2,3)
             and standard model         
             predictions. }  
\ec
\clearpage
\end{table}

\begin{figure}[htb]
\vspace{-1cm}
\centerline{
\epsfig{figure=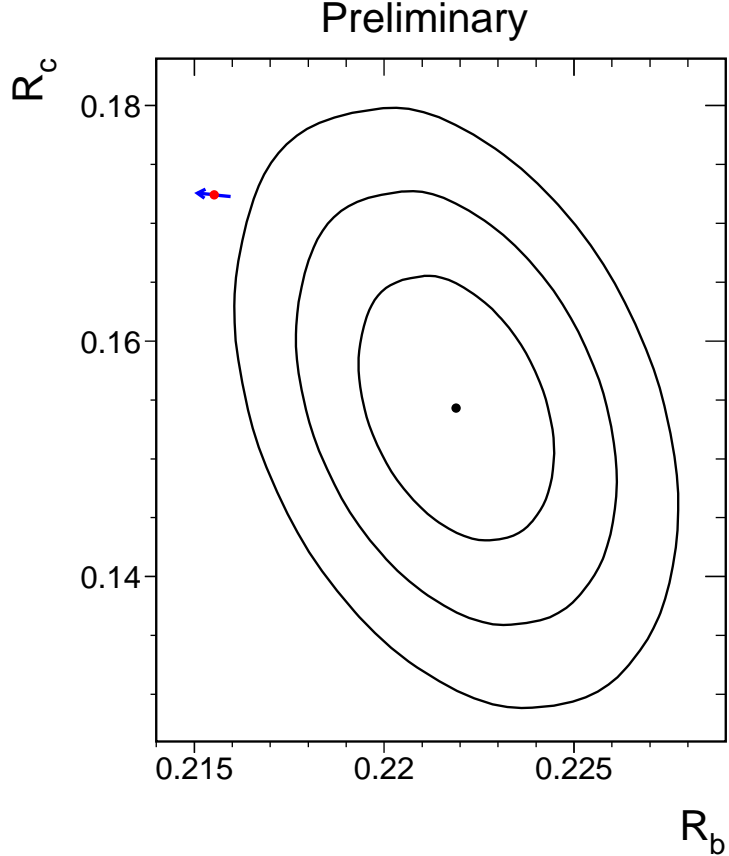,height=15cm,angle=0}}
%\vspace{-1.5cm}
\caption{Contours in the $R_b$-$R_c$ plane derived from LEP data,
         corresponding to 68\%, 95\% and 99.7\% confidence level
         assuming Gaussian systematic errors. The short line with the
         arrow is the \sm prediction for $m_t=180\pm 12$ GeV. The arrow
         points in the direction of increasing values of $m_t$
         (from ref.\ \protect\cite {lep}). } 
\label{lep1}
\end{figure}

\smallskip
Significant deviations from the \sm predictions are observed in the
ratios  $R_b = \Gamma_b/\Gamma_{had}$ and
 $R_c = \Gamma_c/\Gamma_{had}$.
The theoretical predictions are practically independent of $M_H$ and
$\al_s$ and only sensitive to $m_t$.
 The experimental values,
together with the top mass  from the Tevatron, are compatible
with the \sm at a confidence level of less than 1\% (see
Figure \ref{lep1}), 
enough to claim a deviation from the Standard Model.
The other precision observables are in perfect agreement with the
Standard Model. Note that
the experimental value for $\rho_{\ell}$ exhibits the presence of
genuine electroweak corrections by nearly 3 standard deviations.
The importance of bosonic corrections is visible for $s_e^2$,
Figure 2.

%\clearpage

\subsection{Standard model fits}
Assuming the validity of the \sm a global fit to all electroweak
results from LEP, SLD, $p\bar{p}$ and $\nu N$
constrains the parameters $m_t,\al_s$ as follows:
\cite{lep}
\beq
    m_t = 178\pm 8^{+17}_{-20}\, \gv, \;\;\     
    \al_s = 0.123 \pm 0.004 \pm 0.002
\eeq
with $M_H= 300$ GeV for the central value.
The second error is from the variation of $M_H$
between 60 GeV and 1 TeV.
The fit results include the
uncertainties of the \sm calculations.
 
The $W$ mass prediction in table 7 is obtained  by Eq.\ (116) from
 $M_Z,\Gmu,\al$ and  $M_H,m_t$.
The indirect determination of the       
$W$ mass  from LEP/SLD data,
$$ M_W = 80.359\pm 0.055^{+0.013}_{-0.024} \, \gv \, , $$
is in best agreement with the direct measurement (see table 7). 
Moreover, the value obtained for
$\al_s$ at $M_Z$ coincides with the one
 measured from others than electroweak
observables at the $Z$ peak \cite{alfas}.

\subsection{Neutrino - electron scattering }
The cross section for $\mu$-neutrino  electron scattering
and the electroweak mixing angle measured
by the CHARM II Collaboration  \cite{charm}
agree with the standard model values:
\bea
 \sigma(\nu e)/E_{\nu} &=& 16.51\pm 0.93)\cdot 10^{-42} cm^2 GeV^{-1}
 \nn \\
 (\mbox{SM} & : & 17.23\cdot 10^{-42} \, )     \nn \\
 \sin^2\! \theta_{\! \nu\! e} &=&  0.2324\pm 0.0083 \, .
\eea
The mixing angle is determined from the ratio
$$
 R_{\nu e} = \frac{\sigma(\nu_{\mu} e)}{\sigma(\bar{\nu}_{\mu} e)}
 = \frac{1+\snue+\snue^2}{1-\snue+\snue^2}
$$
It  coincides with the result
on $s^2_{\ell}$ from the $Z$,
table 7, as expected by the theory. The major
loop contributions in the difference,
the different scales and the neutrino
charge radius, largely cancel each other by numerical coincidence
\cite{cradius}.

%    subsection on Higgs
\subsection{The Higgs boson}
The minimal model with a single scalar doublet is the simplest way
to implement the electroweak symmetry breaking. The experimental
result that the $\rho$-parameter is very close to unity is a
natural feature of models with doublets and singlets.
In the standard model, the mass $M_H$ of the Higgs boson
appears as the only additional parameter beyond the vector boson
and fermion masses. $M_H$ cannot be predicted but has to taken from
experiment. The present lower limit (95\% C.L.) from the search at
LEP  is 65.2 GeV \cite{grivaz}.
 
\begin{figure}[htb]
%\vspace{-1cm}
\centerline{
\epsfig{figure=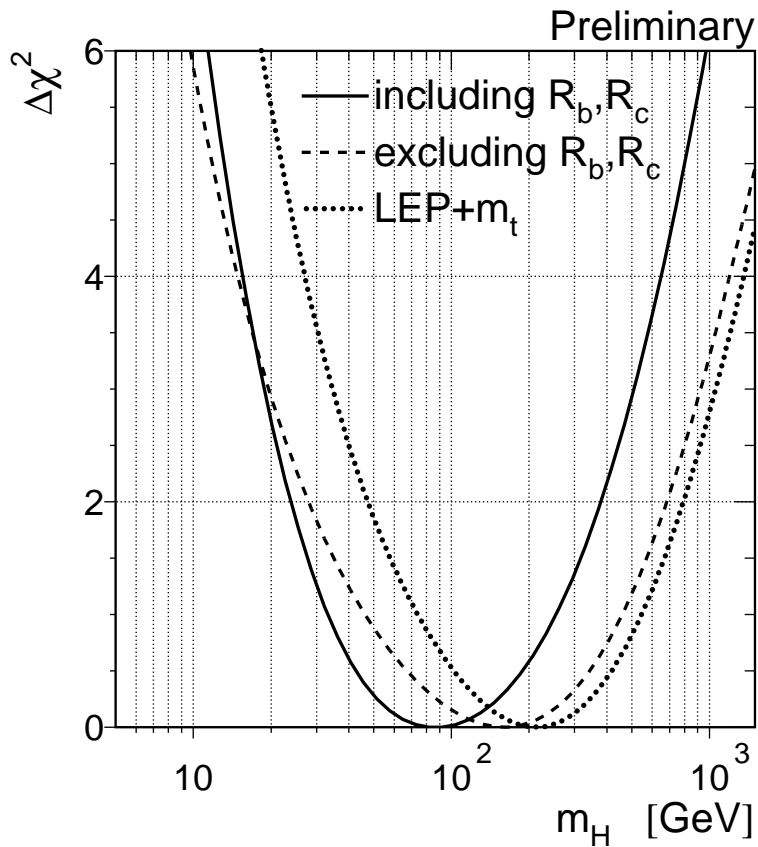,height=12cm,angle=0}}
\vspace{-1.cm}
\caption{\it $\Delta\chi^{2}=\chi^2-\chi^2_{min}$ vs $m_H$ curves.
  Continuous line: based on all LEP, SLD, $p\bar{p}$ and $\nu$N data;
  dashed line: as before, but excluding the LEP+SLD measurements of
  $R_b$ and $R_c$; dotted line: LEP data including measurements of
  $R_b$ and $R_c$. In all cases, the direct measurement of $m_t$ at
  the TEVATRON is included. (From \protect\cite{lep}) }
\label{mhfit}
\end{figure}

\medskip
Indirect determinations of the Higgs mass can be obtained from the 
precsion data.
The main Higgs dependence of the electroweak predictions is only
logarithmic in the Higgs mass. Hence, the sensitivity of the data
to $M_H$ is not very pronounced. Using the Tevatron value for $m_t$ as
an additional experimental constraint, the electroweak fit to all data
yields $M_H < 650$ GeV with approximately 95\% C.L.
\cite{lep}, as shown in Figure \ref{mhfit} (see also \cite{mhfit} for
similar results).
 These indirect mass bounds
depend sensitively on small changes in the input data, and their
reliability suffers at present from averaging data points which
fluctuate by several standard deviations. As a general feature,
it appears that the data prefer light Higgs bosons.

\begin{figure}[htb]
\vspace{-1cm}
\centerline{
\epsfig{figure=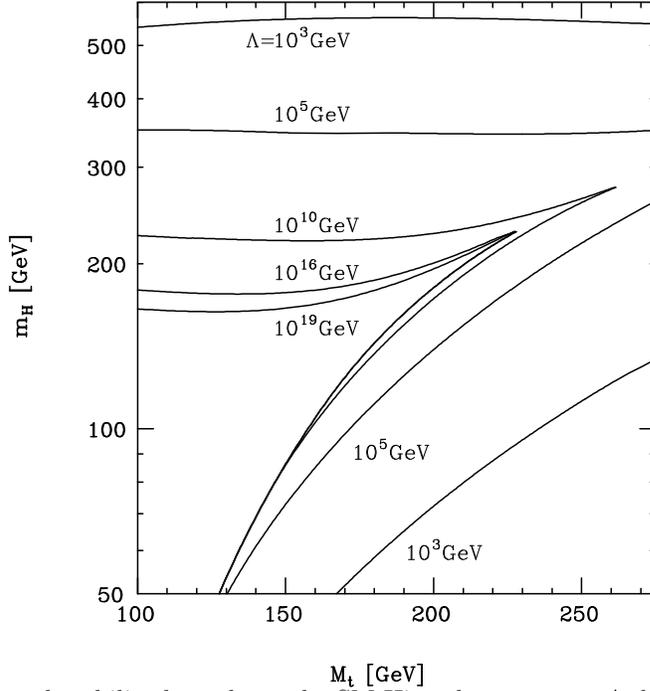,height=15cm,angle=90}}
\vspace{-1.5cm}
\caption{Strong interaction and stability bounds on the SM Higgs 
  boson mass. $\Lambda$ denotes the energy scale where the particles
  become strongly interacting
  (from \protect\cite{carena}). }
\label{landau}
\end{figure}

There are also  theoretical constraints on the Higgs mass
from vacuum stability and from absence of a Landau pole \cite{lindner} 
as illustrated in Figure \ref{landau},
and from lattice calculations \cite{lattice}.

 A recent calculation
of the decay width for $H\ra W^+W^-,ZZ$  in the large $M_H$ limit
in 2-loop order \cite{ghinculov} has shown that the 2-loop
contribution exceeds the 1-loop term in size (same sign) for
 $M_H > 930$ GeV. The requirement of applicability of
perturbation theory therefore puts a stringent upper limit on the
Higgs mass.
 
\smallskip
Higgs boson searches at LEP2
 require precise predictions for the Higgs
production and decay signatures together with detailed background
studies. For a recent report see \cite{carena}.

%\clearpage

%%%%%%%%%%%%%%%%%%%%%%%%%%%%%%%%%%%%%%%%%%%%%%%%%%%%
%%%%%%%%%%%%%%%%%%%%%%%%end of section %%%%%%%%%%%%%%%%%%%%%%%%%%%%%%%
\section{Beyond the minimal model}
We want to conclude  with an outlook on
renormalizable generalizations of the minimal model and
their effect on electroweak observables.
Extended models can be classified in terms of the following
categories:

\begin{itemize}
\item[(i)]
extensions within the minimal gauge group \su with $\rho_{tree}=1$
\item[(ii)]
 extensions within \su with $\rho_{tree}\ne 1$
\item[(iii)]
extensions with larger gauge groups SU(2)$\times$U(1)$\times$G
and respective extra gauge bosons.
\end{itemize}
Extensions of the class (i) are, for example,
 models with
additional (sequential) fermion doublets, more Higgs doublets,
and the minimal supersymmetric version of the Standard Model.

\subsection{Generalization of self energy corrections}
If ``new physics'' would be present in form of new particles
which couple to the gauge bosons but not directly to the external
fermions in a 4-fermion process, only the self energies are
affected.
In order to have a description which is as far as possible independent
of the special type of extra heavy particles, it is convenient
to introduce a
parametrization of the radiative corrections from the vector
boson self-energies in terms of the static $\rho$-parameter
\bea
  \dro (0)               & = &
% \frac{\Sigma^{33}(0)}{\mz} -
% \frac{\Sigma^{WW}(0)}{\mw}       \nn \\
% &   & \nn \\
% & = &
  \frac{\Sigma^{ZZ}(0)}{\mz} -
  \frac{\Sigma^{WW}(0)}{\mw} -2\frac{s_W}{c_W}
  \frac{\sgz (0)}{\mz}
\eea
and the combinations
\bea
\Delta_1               & = & \frac{1}{s_W} \Pi^{3\g}(\mz)
                -\Pi^{33}(\mz) \nn \\
\Delta_2               & = &
 \Pi^{33}(\mz) - \Pi^{WW}(\mw)  \nn \\[0.2cm]
\Delta\al              & = &
 \Pi^{\g\g}(0) - \Pi^{\g\g}(\mz)  \, .
\eea
The quantities in Eq.\ (199) are the isospin components of the
self-energies
\bea
\Sigma^{\g Z} & = &-\frac{1}{s_W}\left(
 \Sigma^{3\g} - s_W^2\Sigma^{\g\g} \right)  \nn \\
\Sigma^{ZZ} & = & \frac{1}{s_W^2} \left(
 \Sigma^{33}-2s_W\Sigma^{3\g}+s_W^2\Sigma^{\g\g}\right)
\eea
 in the expansions
\bea     \real\,
\Sigma^{ij}(k^2) = \Sigma^{ij}(0) + k^2 \Pi^{ij}(k^2) \, . \\
 \nn
\eea

\noi
The $\Delta$-notation above has been introduced in \cite{delta}.
Several other conventions are used in the literature, for example:
\begin{itemize}
\item
The $S,T,U$ parameters of \cite{pt}
are related to (199) by
\beq
 S = \frac{4s_W^2}{\al}\Delta_1, \;\;
 T = \frac{1}{\al} \Delta\rho(0), \;\;
 U = \frac{4s_W^2}{\al} \Delta_2 ,
\eeq
\item
the $\eps$-parameters of \cite{epsilon}
by
\beq
  \eps_1 = \dro, \;\;
  \eps_2 = - \Delta_2,\;\;
  \eps_3 = \Delta_1 ,
\eeq
\item
the $h$-parameters of \cite{kennlang} by
\beq
 h_V = \frac{1}{\al} \dro(0), \;\;
 h_{AZ} = \frac{4\pi}{\sqrt{2}\Gmu\mw}\,\Delta_1, \;\;
 h_{AW} = h_{AZ} +
  \frac{4\pi}{\sqrt{2}\Gmu\mw}\,\Delta_2,
\eeq
\item
and the parameters of \cite{lps}
by
\beq
\Delta_{\rho}(0) =
  \frac{1}{4\sqrt{2}\Gmu}\,\dro, \;\;
\Delta_3 =
 - \frac{1}{4\sqrt{2}\Gmu c_W^2}\,\Delta_1, \;\;
\Delta_{\pm} = c_W^2 \Delta_3
 - \frac{1}{4\sqrt{2}\Gmu}\,\Delta_2 .\;\;
\eeq
\end{itemize}
Further literature can be found in \cite{pt1}. 
The combinations (196) of self energies contribute in a universal
way to the electroweak parameters
(the residual corrections not from self-energies are dropped since
 they are identical to the \sm ones):
\begin{enumerate}
\item
the $M_W-M_Z$ correlation in terms of $\Dr$:
\beq
 \Delta r = \Delta\al - \frac{c_W^2}{s_W^2}\dro(0)
 -\frac{c_W^2-s_W^2}{s_W^2}\Delta_2 + 2\Delta_1
\eeq
\item
the normalization
of the NC couplings at $\mz$
\beq
\dro_f = \dro(0) + \Delta_Z
\eeq
where
the extra quantity
$$
  \Delta_Z = \mz\, \frac{d\,\Pi^{ZZ}}{dk^2}(\mz)
$$
in (207) is from the residue of the $Z$ propagator at the peak.
Heavy particles decouple from $\Delta_Z$.
\item
the effective mixing angles
\beq
     s_f^2 = (1+\Delta\kappa') \, \tilde{s}^2, \;\;\;\;
   \tilde{s}^2 = \frac{1}{2} \left( 1 -
\sqrt{1 - \frac{4\pi\al(\mz)}{\sqrt{2}\Gmu\mz}} \right) \, ,
\eeq
with
\beq
\Delta\kappa' = -\frac{c_W^2}{c_W^2-s_W^2} \dro(0)
      + \frac{\Delta_1}{c_W^2-s_W^2} \, .
\eeq
\end{enumerate}

\bigskip  \noi
The finite combinations of self energies (198) and (199)
are of practical interest since they can be extracted from
precision data in a fairly model independent way.
 An experimental observable particular sensitive to
$\Delta_1$ is the weak charge $Q_W$ which determines the
atomic parity violation in Cesium \cite{mrosner}
\beq
 Q_W = -73.20\pm 0.13 - 0.82\dro(0) - 102 \Delta_1
\eeq
being almost independent of $\dro(0)$.

\bigskip  \noi
The theoretical interest in the $\Delta$'s is based on their
selective sensitivity to different kinds of new physics.

\begin{itemize}
\item
$\dal$ gets contributions only from light charged particles
whereas heavy objects decouple.
\item
$\dro(0)$ is a measure of the violation of the custodial $SU(2)$
symmetry. It is sensitive to particles with large mass splittings
in multiplets. As an example, we have already  encountered
fermion doublets with different masses.
Another example are the Higgs bosons of a 2-Higgs doublet model
\cite{toussaint,bertolini,dgk,dghk}
with masses $M_{H^+}, M_{h}, M_{H}, M_{A}$ and mixing
angles $\beta,\al$
for the charged $H^{\pm}$ and the neutral $h^0, H^0, A^0$
Higgs bosons,
yielding
\bea
\dro(0) & = & \frac{\Gmu}{8\pi^2\sqrt{2}}  \left[
 \sin^2(\al-\beta) F(M^2_{H^+},M_A^2,M_H^2) \right. \nn \\
  &   &  \left.
+\cos^2(\al-\beta) F(M^2_{H^+},M_A^2,M_h^2) \right]
\eea
with
$$ F(x,y,z) = x +
  \frac{yz}{y-z} \log\frac{y}{z}
 -\frac{xy}{x-y} \log\frac{x}{y}
 -\frac{xz}{x-z} \log\frac{x}{z} \,  .
$$
For either $M_{H^+} \gg M_{neutral}$ or vice versa one finds
a positive contribution
\beq
\dro(0) \simeq \frac{\Gmu M_{H^+}^2}{8\pi^2\sqrt{2}}
  \;\;\;  \mbox{or} \;\;\;
               \frac{\Gmu M_{neutral}^2}{8\pi^2\sqrt{2}}
 \,  > 0 \, .
\eeq
Also a negative contribution
$$ \dro(0) < 0 \;\;\;  \mbox{for} \;\;\;
 M_{h,H}  < M_{H^+}  <  M_A  \;\;\; \mbox{and} \;\;\;
 M_{A}  < M_{H^+}  <  M_{h,H}
$$
is possible in the unconstrained 2-doublet model.
\item
$\Delta_1$ is sensitive to chiral symmetry breaking by masses.
In particular, a doublet of mass degenerate heavy fermions
yields a contribution
\beq
\Delta_1 =  N_C^f\, \frac{\Gmu\mw}{12\pi^2\sqrt{2}} \, ,
\eeq
whereas the contribution of degenerate heavy fermions to
$\dro(0)$ is zero.
Hence, $\Delta_1$ can directly count the number $N_{deg}$
of mass degenerate fermion doublets:
$$ \Delta_1^{f} = 4.5\cdot 10^{-4}\cdot N_{deg}. $$
$\Delta_1$ also gets sizeable contributions from models with a
large number of additional fermions like in technicolor models.
For example, $\Delta_1 \simeq 0.017$ for $N_{TC}=4$  and one
family of technifermions \cite{pt,technicolor}.
\item
A further quantity $\eps_b$ has been introduced \cite{abc1} in order    
to parametrize specific non-universal
left handed contributions to the $Zbb$ vertex via
\beq
 g_A^b =g_A^d(1+\eps_b), \;\;\;
 g_V^b/g_A^b = (1-\frac{4}{3}s^2_d+\eps_b)\, (1+\eps_b)^{-1} \, .
\eeq
%There is a wide literature \cite{pt} in this field  with various
%conventions.
\end{itemize} 
Phenomenologically, the $\eps_i$ are parameters which
can be determined experimentally from the electroweak precision data.
An updated analysis \cite{caravaglios}
on the basis of the recent electroweak results
yields for $\eps_b$ the value
$$
   \eps_b = 9.9 \pm 4.5 \;\;\; (\mbox{SM}:\;\; -6.6)    
$$
The large difference to the standard model value is an another way
of visualizing the deviation between the measured and predicted
number for the ratio $R_b$ (table 7).

\subsection{Models with
 $\rho_{tree}\neq 1 $}
One of the basic relations of the minimal \sm is the tree
level correlation between the vector boson masses
and the electroweak mixing angle
 $$ \rho_{tree} = \frac{\mw}{\mz \cow} = 1 \, .
$$
Many extensions of the minimal model, like those discussed in the
previous section, preserve this feature.

\medskip
The formulation of the electroweak theory
               in terms of a local gauge theory requires at least
a single scalar doublet for breaking the electroweak symmetry
\su $\ra $ U(1)$_{em}$.
In contrast to the fermion and vector boson part, very little is known
empirically
about the scalar sector.
Without the assumption of minimality, quite a lot of options are at
our disposal,
including more complicated
multiplets  of Higgs fields.
In general models the tree level $\rho$-parameter
$\rho_{tree} = \rho_0$
is determined by
$$
 \rho_0 = \frac{\sum_i v_i^2 [I_i(I_i+1)-I_{3i}^2] }
 {2 \sum_i v_i^2 I_{3i}^2}
$$
where $v_i, I_{3i}$ are the vacuum expectation values
and third isospin component of the neutral component of the
 $i$-th Higgs multiplet in the representation
with isospin $I_i$.
The presence of at least a triplet of Higgs fields gives rise
to $\rho_0 \neq 1$.
As a consequence,
the  tree level relations between
 the electroweak parameters
have to be generalized according to
\beq
\sin^2 \theta_W \, \ra \,
\stt = 1 - \frac{M_W^2}{\rho_0\mz}
\eeq
and
\beq
\frac{\Gmu}{\sqrt{2}} = \frac{e^2}{8\stt M_W^2} =
\frac{e^2}{8\stt\cth^2\rho_0\mz}
\eeq
Writing $\rho_0 = (1-\dro_0)^{-1}$,
 we obtain for the mixing angle:
\beq
\stt = 1 -\frac{M_W^2}{\mz} + \frac{M_W^2}{\mz}\,\dro
    \equiv s_W^2 + c_W^2 \dro_0 \, ,
\eeq
for the overall normalization factor
in the NC vertex:
\beq
\frac{e}{2\sth\cth} = \left(\sqrt{2} \Gmu\mz\rho_0\right)^{1/2} \, ,
\eeq
 and for the $M_W-M_Z$ interdependence:
\beq
  M_W^2 \left(1-\frac{M_W^2}{\rho_0\mz}\right) =
  \frac{e^2}{4\sqrt{2}\Gmu} \, ,
\eeq
in complete analogy to what we have found
 from the top quark loops.

\bigskip \noi
At the level of radiative corrections, a small $\dro_0$ may be
included by
\beq
\Dr \ra \ \Dr -\frac{c_W^2}{s_W^2}\dro_0
\eeq
for the $M_W$-$M_Z$  correlation, and
\beq
 \rho_f \ra \rho_f + \dro_0, \;\;\;
 s_f^2 \ra s_f^2 + c_W^2 \dro_0
\eeq
for the normalization and the effective mixing angles of the
$Zff$ couplings.

\medskip
 A complete discussion of radiative corrections requires not only
the calculation of the extra loop diagrams from the non-standard
Higgs sector but also an extension of the
renormalization procedure \cite{pass,nardi}.
 Since  $M_W,M_Z$ and $\siw$ (or
$\rho_0$, eqivalently) are now independent parameters,
one extra renormalization condition  is required.
A natural condition would be to define the mixing angle for
electrons $s_e^2$ in terms of the ratio
of the dressed coupling constants at the $Z$ peak
$$ \frac{g_V^e}{g_A^e}   =: 1- 4s_e^2 $$
which is measureable in terms of the left-right or the
forward-backward asymmetries.
This fixes
the counter term for $s_e^2$ by
\beq
\frac{\delta s_e^2}{s_e^2} =
\frac{c_e}{s_e} \frac{\real\,\sgz(\mz)}{\mz} +  
\frac{c_e}{s_e} \frac{\sgz(0)}{\mz} + \Delta\kappa_e
\eeq
with the finite part
$\Delta\kappa_e$
 of the electron-$Z$ vertex correction.
The counter terms for the other parameters $\al,M_Z$
 are treated as usual.
With this input, we obtain a renormalized $\rho$-parameter
and the corresponding counter term for the bare $\rho$-parameter
$\rho_0^{b} = \rho +\delta\rho $ as follows:
\bea
      \rho  & = & \frac{\mw}{\mz c_e^2}  \, ,\nn \\[0.2cm]
\frac{\delta\rho}{\rho} & = & \dmw -\dmz
       +\frac{\delta s_e^2}{c_e^2} \, .
\eea
Other derived quantities are:
\begin{itemize}
\item
The relation between $M_W$ and $\Gmu$:
\beq
 \mw = \frac{\pi\al}{\sqrt{2}\Gmu s_e^2}\cdot
       \frac{1}{1-\Dr}
\eeq
with
\beq
\Dr =
   \frac{\sw(0)-\dmmw}{\mw} +
  \Pi^{\g}(0)
 -\frac{\delta s_e^2}{s_e^2}
+ 2\frac{c_e}{s_e}
                \frac{\sgz(0)}{\mz} +\delta_{VB}      \, .
\eeq
\item
The normalization of the $Zff$ couplings at 1-loop:
\bea   &   &
 \frac{e^2}{4s_e^2 c_e^2} \left[ 1 + \Pi^{\g}(0)
- \frac{c_e^2-s_e^2}{c_e^2} \frac{\delta s_e^2}{s_e^2}
+ 2\frac{c_e^2-s_e^2}{c_e s_e}
                \frac{\sgz(0)}{\mz}
 + \Delta\rho_f \right]   \\[0.2cm]
 & = & \sqrt{2}\Gmu\mz\rho \left[ 1
 -\frac{\sw(0)-\dmmw}{\mw} +\frac{\delta s_e^2}{c_e^2}
- 2\frac{s_e}{c_e}
                \frac{\sgz(0)}{\mz}
 -\delta_{VB}
 + \Delta\rho_f \right]            \nn
\eea
  where $\Delta\rho_f$ denotes the finite part of the $Zff$
vertex correction.
\item
The effective mixing angles of the $Zff$ couplings:
$$
 s_f^2 \, =\, s_e^2 \,(1-\Delta\kappa_e +\Delta\kappa_f)  .
$$
\end{itemize}
These relations predict the $Z$ boson couplings, $M_W$  and
$\rho$ in terms of the data points
$\al, \Gmu, M_Z, s_e^2$.
 By this procedure,
the $m_t^2$-dependence of the self energy corrections
to theoretical predictions is absorbed into the renormalized
$\rho$-parameter,
leaving a $\sim \log m_t/M_Z$ term as an observable effect.
For the $Zbb$-vertex, an additional $m_t^2$ dependence is found
in the non-universal vertex corrections $\Delta\rho_b$
and $\Delta\kappa_b$. This makes observables containing
this vertex the most sensitive top indicators in the class of
models with $\rho_{tree} \neq 1$.

\bigskip
In the minimal Standard Model, the quantity
 equivalent to (223) can be
calculated in terms of the data points $\al,\Gmu,M_Z$ and
the parameters $m_t,M_H$.
With the experimental constraints from $M_W$ and the $Z$ boson observables
one obtains
\beq
\rho_{SM} = 1.0103\pm 0.0015 \, .
\eeq
In the extended model we can get a value
for $\rho$ from directly using the data on $\mw$
and $s_e^2 = 0.23186 \pm0.00034$  yielding
\beq
 \rho = 1.0085\pm 0.0040 \, .
\eeq
The difference $\rho -\rho_{SM}$ can be interpreted as a
measure for a deviating tree level structure. The data
imply that it is compatible with zero.

\begin{figure}[htb]
\vspace{-3cm}
\centerline{
\epsfig{figure=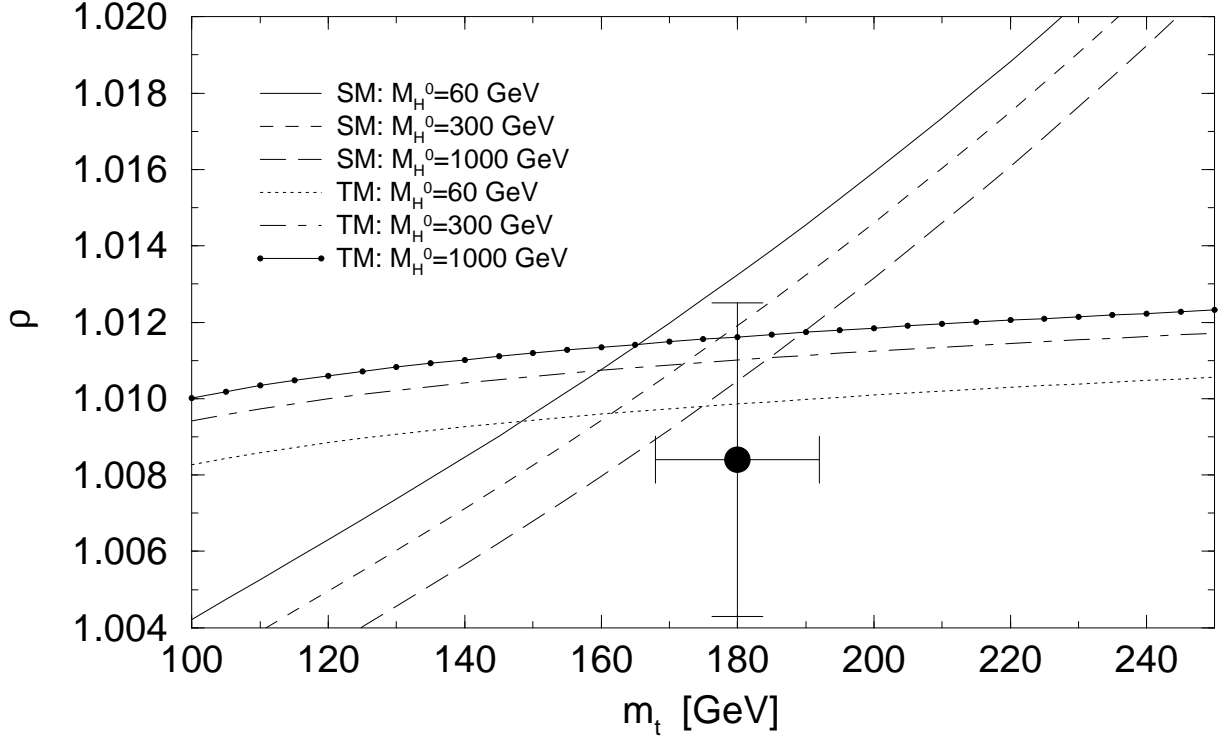,height=15cm,angle=0}}
\vspace{-1cm}
\caption{The $\rho$ parameter, Eq.\ (223),  in the \sm (SM) and in 
        a model with an extra Higgs triplet (TM). The masses of the
        non-standard Higgs bosons of TM are put to 300 GeV. 
        $\rho(TM)$ is calculated from Eq.\ (229) with the LEP input
        on $s_e^2$.}
\label{rho}
\end{figure}

\smallskip \noindent
In a specific model one can calculate the value for $\rho$ from
\beq
 \rho = \frac{\pi\al}{\sqrt{2}\Gmu\mz s_e^2 c_e^2}\cdot
 \frac{1}{1-\Dr}
\eeq
in terms of the input data $\al,\Gmu,M_Z, s_e^2$ together with
$m_t$ and the
parameters of the Higgs sector. Such a complete calculation has been performed
for a model with an extra Higgs triplet, involving one extra neutral and
a pair of charged Higgs bosons \cite{blank}. Figure \ref{rho} shows
that the model predictions coincide in the aerea of the experimental data
points on $m_t$ and $\rho$ from Eq.\ (228).

%\clearpage

%
\subsection{Extra $Z$ bosons}
The existence of additional vector bosons is predicted by
GUT models based on groups bigger than SU(5), like $E_6$
and $SO(10)$, by models with symmetry breaking in terms of a
strongly interacting sector, and composite scenarios.
Typical examples of extended gauge symmetries are the
$SU(2)\times U(1)\times U(1)_{\chi,\psi,\eta}$ models following
from $E_6$ unification, or LR-symmetric models.
In the following we consider only models with an extra $U(1)$.

\bigskip
The mixing between the mathematical states $Z_0$ of the minimal gauge
group and $Z'_0$ of an extra hypercharge form the physical mass
eigenstates $Z,\, Z'$,
where the lighter $Z$ is identified with the resonance at LEP.
The mass eigenstates are obtained by a rotation
\bea
    Z  & = & \cosm\, Z_0 +\sinm\, Z'_0  \nn \\
    Z' & = & -\sinm\, Z_0 + \cosm\, Z'_0
\eea
with a mixing angle $\theta_M$ related to the mass eigenvalues
by
 \beq
\tan^2 \theta_M = \frac{\mzoo-\mz}{\mzop-\mzoo},\;\;\;
 \mzoo = \cosmm\, \mz + \sinmm\, \mzop \, .
\eeq
$\mzoo$ denotes the nominal mass of $Z_0$.
In constrained models with the Higgs fields in doublets and
singlets only, the usual \sm relation holds
$$ \siw = 1-\frac{\mw}{\mzoo} $$
between the masses and the mixing angle in the Lagrangian
\beq
{\cal L}_{NC} = \frac{g_2}{\cosw}\, J_{Z_0}^{\m}\,Z_0^{\m}
 \, +\, g'\,
            J_{Z'_0}^{\m}\,Z_0^{'\,\m}
\eeq
with
$$
  J_{Z_0}^{\m} = J_L^{\m} - \siw\, J_{em}^{\m} \, .
$$
It is convenient to introduce the quantity
\beq
 s_W^2 =  1-\frac{\mw}{\mz}, \;\;\; c_W^2=1-s_W^2
\eeq
with the physical mass of the lower eigenstate.
For small mixing angles $\theta_M$ we have the following relation:
 \beq
 \siw = s_W^2 + c_W^2 \droz
\eeq
with
\beq
 \droz = \sinmm \left(\frac{\mzop}{\mz}-1\right) \, .
\eeq \\

\noi
The $W$ mass is obtained from
$$
 \mw = \frac{\pi\al}{\sqrt{2}\Gmu \siw (1-\Dr)}
$$
after the substitution (237):
\beq
 \mw = \frac{\mz}{2}\left( 1 + \sqrt{1
     -\frac{\pi\al}{\sqrt{2}\Gmu\mz\rho_{Z'}\,(1-\Dr)}} \: \right)
\eeq
with
$  \rho_{Z'} = (1-\droz)^{-1} $.
Formally, $\rho_{Z'}$ appears as a non-standard tree level
$\rho$-parameter.
In all present practical applications the radiative correction
$\Dr$ was approximated by the standard model correction.

\bigskip \noi
 The mass
mixing has
two implications for the NC couplings of the $Z$ boson:
\begin{itemize}
\item
$\droz$ contributes to the overall normalization by a factor
 $$ \rho_{Z'}^{1/2} \simeq 1 + \frac{1}{2}\droz $$
and to the mixing angle by a shift
 $$ s_W^2 \ra s_W^2 + c_W^2 \droz \, .$$
Both effects are universal, parametrized by $M_{Z'}$ and the mixing
angle $\theta_M$ in a model independent way,
\item
A non-universal contribution is present as the
 second term in the vertex
\bea
  (Zff) & = & \cos\theta_M (Z_0ff) + \sin\theta_M (Z'_0ff)  \nn \\
      & \simeq & (Z_0ff) + \theta_M (Z'_0ff)  \, . \nn
\eea
It depends on the classification of the fermions under the
extra hypercharge and is strongly model dependent.
\end{itemize}
Complete 1-loop calculations are not available as yet.
The present standard approach consists in the implementation
of the standard model corrections to the $Z_0$ parts of the
coupling constants in terms of the
form factors $\rho_f$ for the normalization and $\kappa_f$
for the effective mixing angles
$$ s_W^2 \ra s_f^2 = \kappa_f s_W^2. $$
In this approach the effective
 $Zff$ vector and axial vector couplings read:
\bea
v_Z^f & = & \left[ \sqrt{2}\Gmu\mz \rho_f (1+\droz)\right]^{1/2}
 \left[ I_3^f -2Q_f(\kappa_f s_W^2 +c_W^2\droz) \right]
 \nn \\
   &   &  +\, \sinm \,v_{Z'_0}^f \, , \nn \\[0.2cm]
a_Z^f & = & \left[ \sqrt{2}\Gmu\mz \rho_f\right]^{1/2}\, I_3^f
           \,+\, \sinm\, a_{Z'_0}^f \, .
\eea
The quantities  $a_{Z'_0}^f \, v_{Z'_0}^f $ denote the extra $U(1)$
 couplings
between the fermion $f$ and the $Z'_0$.

\bigskip
From an analysis of the electroweak precision data
the mixing angle is constrained typically to
$\mid \theta_M\mid < 0.01$, not very much dependent on the
specification of the model \cite{altarelli,aguila}.

\smallskip
Quite recently, models with an extra $Z'$
 have received new
attention in order to explain the observed deviations from
the standard model in $R_b$ and $R_c$ by a `hadrophilic' coupling
to quarks only \cite{hadrophil}. 
\subsection{New physics in $R_b$?}
If the observed difference between the measured and calculated
values of $R_b$ is explained by a  non-standard contribution
$\Delta\G_b$ to the partial width $\G(Z\ra b\bar{b})$, then also
other hadronic quantities like $\G_Z, R_{had}, \dots$ are increased
unless the value of $\al_s$ is reduced simultaneously. Including
the new physics  $\Delta\G_b$ as an extra free parameter in the fit
yields the values \cite{lep}:
 $$\al_s=0.102\pm 0.008, \;\;
 \Delta\G_b =11.7\pm3.8\pm1.4 \mv \, .$$
The top mass is affected only marginally, shifting the central
value by $+3$ GeV, but the impact on $\al_s$ is remarkable.
\subsection{The minimal supersymmetric standard model (MSSM):}
The MSSM deserves a special discussion
as the most predictive framework beyond the minimal model.
Its structure allows a similarly complete calculation of
the electroweak precision observables
as in the standard model in terms of one Higgs mass
(usually taken as $M_A$) and $\tan\beta= v_2/v_1$,
together with the set of
SUSY soft breaking parameters fixing the chargino/neutralino and
scalar fermion sectors.
It has been known since quite some time
\cite{higgs}
that light non-standard
Higgs bosons as well as light stop and charginos
% all around 50 GeV or little higher,
predict larger values for the ratio $R_b$ and thus diminish the
observed difference  \cite{susy1,susy3,susy4,susy5}.
Complete 1-loop calculations are meanwhile available for
$\Delta r$ \cite{susydelr} and for the $Z$ boson observables
\cite{susy3,susy4,susy5}.

\begin{figure}[htb]
\vspace{-1cm}
\centerline{
\epsfig{figure=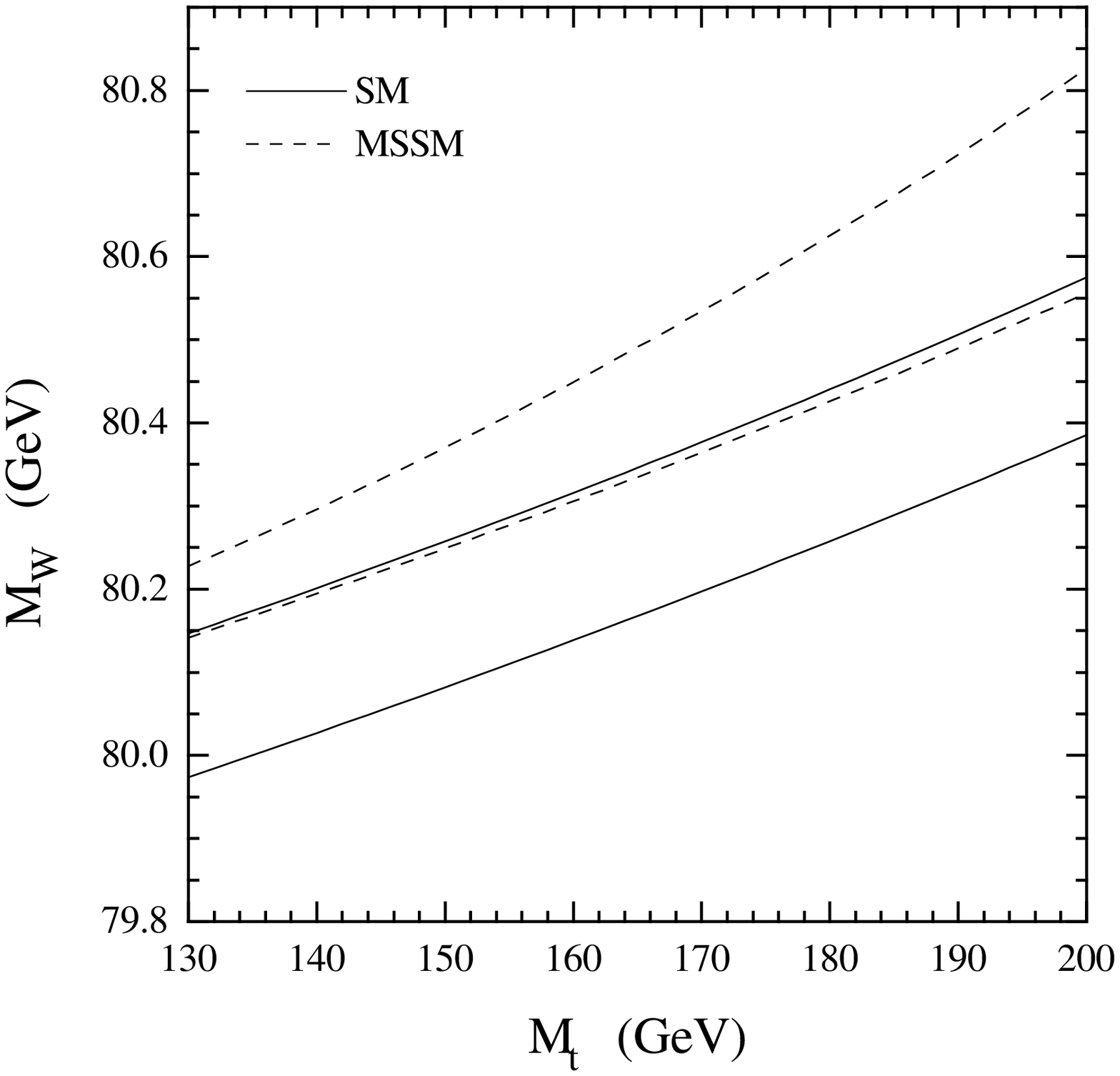,height=15cm,angle=0}}
\vspace{2cm}
\caption{The $W$ mass range in the standard model (-----) and the
         MSSM (- - -). Bounds are from the non-observation of Higgs
         bosons and SUSY particles at LEP2.} 
\label{susymw}
\end{figure}

Figure \ref{susymw}
displays the range of predictions for $M_W$ in the minimal model
and in the MSSM. Thereby it is assumed that no direct discovery has been
at LEP2. As one can see, precise determinations of $M_W$ and $m_t$
can become decisive for the separation between the  models.

\begin{figure}[htb]
\vspace{-1cm}
\centerline{
\epsfig{figure=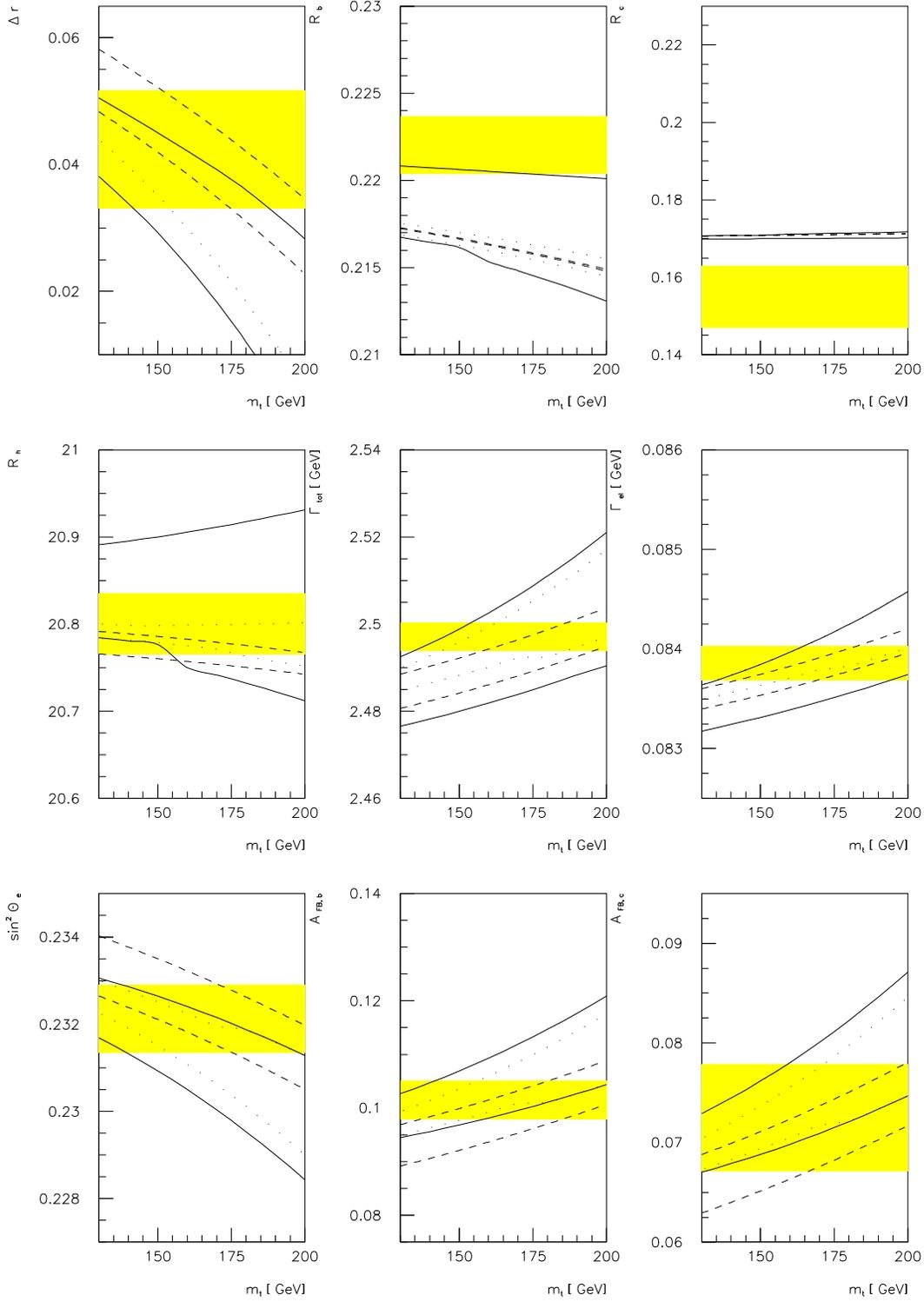,height=20cm,angle=0}}
%\vspace{-1.5cm}
\caption{Range of precision observables in the standard model (- - -)
         and in the MSSM (---), and present experimental data (dark area).
          The MSSM parameters are restricted by the mass bounds from direct
          searches at LEP I and Tevatron, the dotted lines indicate the 
          bounds to be expected from LEP II.}
\label{zfig5}
\end{figure}

\begin{figure}[htb]
\vspace{-1cm}
\centerline{
\epsfig{figure=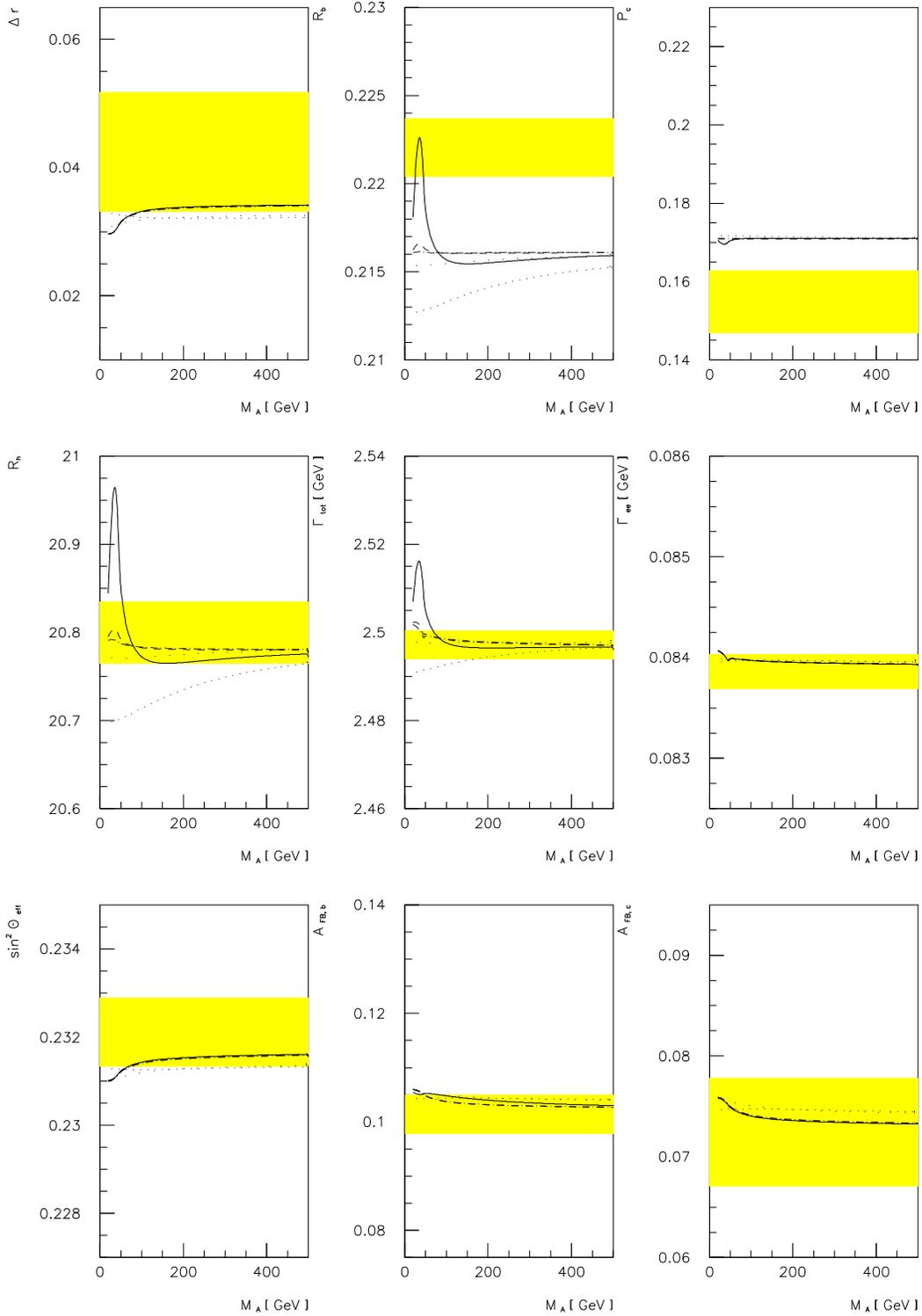,height=20cm,angle=0}}
%\vspace{-1.5cm}
\caption{Precision observables as function of the pseudoscalar
        Higgs mass $M_A$ for $\tan\beta = 0.7 (\cdot \cdot \cdot), \,
        1.5 (\cdot\;\;\cdot\;\;\cdot), \, 8 (- \cdot -\cdot -), \,
        20$(- - -),  70 (-----). $m_t=174$ GeV, $\al_s=0.123$. 
        $m_{\tilde{l}}=800$ GeV, $m_{\tilde{q}}=500$ GeV,
        $\mu=100$ GeV, $M_2= 300$ GeV.} 
\label{zfig1}
\end{figure}

The range of predictions for $\Dr$ and the $Z$ boson observables in 
the MSSM is visualized in Figure \ref{zfig5}
 (between the solid lines)
 together with the standard model predictions (between the dashed lines)
and with the present experimental data (dark area).
$\tan\beta$ is thereby varied between 1 and 70, the other parameters are
restricted according to the mass bounds from the direct search for
non-standard particles at LEP I and the Tevatron.
From a superficial inspection, one might get the impression that the
MSSM, due to its extended set of parameters, is more flexible to
accomodate also the critical observable $R_b$. A more detailed analysis
shows, however, that those parameter values yielding a ``good'' $R_b$
are incompatible with other data points. An example is given in
Figure \ref{zfig1}: a light $A$ boson together with a large $\tan\beta$
can cure $R_b$, but violates the other hadronic quantities and the
efefctive leptonic mixing angle. Whereas the hadronic quantities can be
repaired (at least partially) by lowering the value of $\al_s$, the 
mixing angle and $A_{FB}^b$ remain off for small Higgs masses.
Thus, even in the MSSM it is not possible to simultaneously find
agreement with all the individual  precision data. The results of a 
global fit are discussed below.   

\clearpage

\clearpage

\clearpage

%The range of predictions for the $Z$ observables is shown in Figure 5.

\noi
The main results in view of the recent precision data are: \hfill
\\
$\bullet$ $R_c$ can hardly be moved towards the measured range.
 \hfill \\ 
$\bullet$ $R_b$ can come closer to the measured value, in particular
for light $\tilde{t}_R$ and light charginos. \hfill \\
 $\bullet$  $\al_s$ turns out to be smaller than in the 
 minimal model because of  the
reasons explained in the beginning of this section.
\hfill \\ 
$\bullet$ There are strong constraints from the other precision
observables which forbid parameter configurations shifting $R_b$
into the observed $1\sigma$ range.

\begin{figure}[hbt]
\vspace{1cm}
\centerline{
%\begin{picture}(100,110)(0,1)
%\mbox{\epsfxsize8.0cm\epsffile[0 20 595 794]{mssm1.eps}}
\mbox{\epsfxsize8.0cm\epsffile{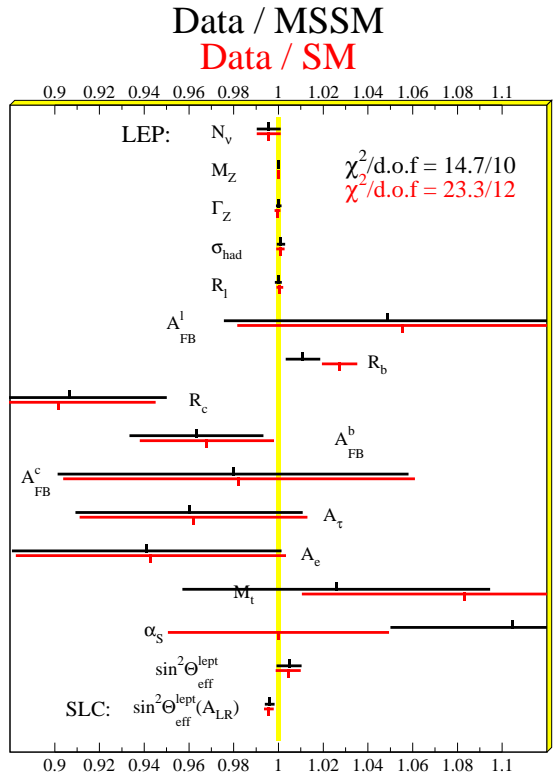}}} 
%\end{picture}
\caption{Experimental data normalized to the best fit results in
         the SM and MSSM. }
\label{mssm}
\end{figure}

For obtaining the optimized SUSY parameter set, therefore, a global
fit to all the electroweak precision data (including the top
mass measurements)
 has to be performed,
as done in refs.\ \cite{susy4,deboer}. Figure \ref{mssm} displays the
experimental data normalized to the best fit results  
in the SM and MSSM, with 
the data from this conference \cite{deboer}.
For the SM, $\al_s$ identified with the experimental
number, therefore the corresponding result in Figure 6 is
centered at 1. The most relevant conclusions are: \hfill \\
(i) The difference between the experimental and theoretical value
of $R_b$ is diminished by a factor $\simeq 1/2$, \hfill \\
(ii) the central value for the strong coupling is $\al_s=0.110$
and thus is very close to the value obtained from deep inelastic
scattering, \hfill \\
(iii) the other observables are practically unchanged, \hfill \\
(iv) the $\chi^2$ of the fit is slightly better than in the minimal
model.

\setlength{\unitlength}{0.7mm}

\section{Conclusions}
The experimental data for testing the electroweak theory have
achieved an impressive accuracy.
For the interpretation of the precision
experiments 
radiative corrections, or quantum effects, play
a crucial role.
The calculation of radiative corrections is theoretically well
established, and many contributions have become available
over the past few years
to improve and stabilize the Standard Model predictions.
After taking the measured $Z$ mass, besides $\al$ and $\Gmu$, for
completion of the input, each other precision observable
provides a test of the electroweak theory.
The theoretical predictions of the \sm depend  on the mass of the recently
discovered top quark and 
of the as yet experimentally unknown Higgs boson
through the virtual presence of these particles in the loops.
As a consequence, precision data can be used to pin down the allowed
range of the mass parameters, yielding $m_t$ in beautiful agreement
with the directly measured value.

\medskip
Theoretical uncertainties in the \sm predictions have their origin essentially
in the uncertainties of the hadronic vacuum polarization of the
photon and from
the unknown higher order contributions.
In order  to reach a theoretical
accuracy at the level 0.1\% or below, new experimental data on
$\dal$ and more complete electroweak 2-loop calculations are required.
The observed deviations of several $\sigma$'s in $R_b,R_c,\alr$
reduce the quality of the \sm fits significantly, but the
indirect  determination of $m_t$ is remarkably stable.
Still impressive is the perfect agreement between
theory and experiment for the whole set of the other
precision observables. Supersymmetry can improve the situtation due to an 
enhancement of $R_b$ by new particles in the range of 100
GeV or even below, but it is not possible to accomodate $R_c$.
Within the MSSM analysis, the value for $\al_s$ is close to the one
from deep-inelastic scattering.   

\section*{Acknowledgements}
I want to thank the organizers of the Hellenic School for the
invitation and for the very pleasant stay at Corfu. Many thanks also to
A. Dabelstein and G. Weiglein for their support  in the preparation of
these lecture notes.   
%%%%%%%%%%%%%%%%%%%%%%%%%%%%%%%%%%   end of text    %%%%%%%%%%%%%%%

\bigskip \noi
%{\bf References}:
%\vglue 0.2cm
%


\begin{thebibliography}{9}
%
\bibitem{gsw}
S.L. Glashow, \np {\bf B22} (1961) 579;\\
S. Weinberg, \prl {\bf  19} (1967)  1264; \\
A. Salam, in: {\it Proceedings of the  8th Nobel Symposium},
p.\ 367, ed. N. Svartholm, Almqvist and Wiksell,
Stockholm 1968
\bibitem{gim}
S.L. Glashow, I. Iliopoulos, L. Maiani, \pr
{\bf D2} (1970) 1285;
\bibitem{cabibbo}
N. Cabibbo, \prl {\elevenbf 10} (1963) 531; \\
M. Kobayashi, K. Maskawa, {\elevenit Prog.\ Theor.\ Phys.\ }
{\elevenbf 49} (1973) 652
\bibitem{color}
H.Y. Han, Y. Nambu, \pr {\elevenbf 139} (1965) 1006; \\
C. Bouchiat, I. Iliopoulos,  Ph. Meyer, \pl
{\elevenbf B138} (1972)  652
\bibitem{thooft}
G. 't Hooft, \np {\elevenbf B33} (1971) 173;
             \np {\elevenbf B35} (1971) 167
\bibitem{lep}
The LEP Collaborations ALEPH, DELPHI,L3, OPAL and the LEP Electroweak
Working Group, CERN-PPE/95-172; \\
A. Olshevsky,
             talk at the {\elevenit
             International Europhysics Conference on High Energy Physics},
             Brussels 1995 (to appear in the Proceedings); \\
P. Renton, talk at the {\elevenit 17th International Symposium on
           Lepton-Photon Interactions}, Beijing 1995, Oxford OUN-95-20
           (to appear in the Proceedings)
\bibitem{wmass}  UA2 Collaboration, J. Alitti et al.,
                 \plb {\bf B276} (1992) 354; \\
                 CDF Collaboration, F. Abe et al.,
                 \prd {\bf D43} (1991) 2070; \\
             D0 Collaboration,  C.K. Jung, talk at the 27th
             International Conference on High Energy Physics,
             Glasgow 1994; \\
             CDF Collaboration, F. Abe et al., FERMILAB-PUB-95/033-E;
                                               FERMILAB-PUB-95/035-E
                                               (1995)
\bibitem{top}
CDF Collaboration, F. Abe et al., \prl {\elevenbf 74} (1995) 2626; \\
D0 Collaboration, S. Abachi et al., \prl {\elevenbf 74} (1995) 2632 
\bibitem{fp}
L.D. Faddeev, V.N. Popov, \pl {\elevenbf B25} (1967) 29
\bibitem{bhs} M. B\"ohm, W. Hollik, H. Spiesberger,
{\elevenit Fortschr.\ Phys.\ } {\elevenbf 34} (1986) 687
\bibitem{ds}
A. Denner, T. Sack, \np {\elevenbf B347} (1990) 203
\bibitem{pa}
G. Passarino, in:
 {\elevenit Proceeedings of the
 LP-HEP 91 Conference}, Geneva 1991,
eds.\ S. Hegarty, K. Potter, E. Quercigh,
World Scientific, Singapore 1992
\bibitem{pass}
G. Passarino,  \np {\elevenbf B361} (1991) 351;\\
G. Passarino, in  {\it Proceedings of the XXVIth
Rencontre de Moriond:
 '91 Electroweak Interactions and Unified
 Theories}, ed. Tran Thanh Van
\bibitem{minsub}
G. 't Hooft, \np {\elevenbf B61} (1973) 455;
             \np {\elevenbf B62} (1973) 444
\bibitem{ross}
D.A. Ross, J.C. Taylor, \np {\elevenbf B51} (1973) 25
\bibitem{pav}
G. Passarino, M. Veltman, \np {\bf B160} (1979) 151
\bibitem{con}
M. Consoli, \np {\bf B160} (1979) 208
\bibitem{sirmar}
A. Sirlin, \pr {\bf D22} (1980) 971; \\
W. J. Marciano, A. Sirlin, \pr {\bf D22} (1980) 2695;\\
A. Sirlin, W. J. Marciano, \np {\bf B189} (1981) 442
\bibitem{dubna}
D.Yu.\ Bardin, P.Ch.\ Christova, O.M. Fedorenko,
\np {\bf B175} (1980) 435; \np {\bf B197} (1982) 1; \\
D.Yu.\ Bardin, M.S. Bilenky, G.V. Mithselmakher, T. Riemann,
M. Sachwitz,
\zp {\bf C44} (1989) 493
\bibitem{fjeg}
J. Fleischer, F. Jegerlehner, \pr {\bf D23} (1981) 2001
\bibitem{aoki}
K.I. Aoki, Z. Hioki, R. Kawabe, M. Konuma, T. Muta,
{\it Suppl.\ Prog.\ Theor.\ Phys.\ } {\bf 73} (1982) 1; \\
Z. Hioki, \prl {\bf 65} (1990) 683, E:1692;
 \zp {\bf C49} (1991) 287
\bibitem{maiani}
M. Consoli, S. LoPresti, L. Maiani,
 \np {\bf B223} (1983) 474
\bibitem{dz}
D.Yu.\ Bardin, M.S. Bilenky, G.V. Mithselmakher, T. Riemann,
M. Sachwitz,
\zp {\bf C44} (1989) 493
\bibitem{hollik}
W. Hollik, {\it Fortschr.\ Phys. } {\bf 38} (1990) 165
\bibitem{LEP}
M. Consoli, W. Hollik, F. Jegerlehner, in:
{\it
$Z$ Physics at LEP 1},
  eds.\ G. Altarelli, R. Kleiss
and C. Verzegnassi,
 CERN 89-08   (1989)
\bibitem{pittau}
G. Passarino, R. Pittau, \pl {\bf B228} (1989) 89;\\
V.A. Novikov, L.B. Okun, M.I. Vysotsky, CERN-TH.6538/92 (1992)
\bibitem{green}
M. Veltman, \pl {\bf B91} (1980) 95; \\
 M. Green, M. Veltman, \np
{\bf B169} (1980) 137, E: \np {\bf B175} (1980) 547; \\
F. Antonelli, M. Consoli, G. Corbo, \pl {\bf B91} (1980) 90;\\
F. Antonelli, M. Consoli, G. Corbo, O. Pellegrino,
\np {\bf B183} (1981) 195
\bibitem{pavelt}
G. Passarino, M. Veltman, \pl {\bf B237} (1990) 537
\bibitem{msbar}
W.J. Marciano, A. Sirlin, \prl {\bf 46} (1981) 163; \\
A. Sirlin, \pl {\bf B232} (1989) 123
\bibitem{msbar1}
G. Degrassi, S. Fanchiotti, A. Sirlin,
\np {\bf B351} (1991) 49
\bibitem{msbar2}
G. Degrassi, A. Sirlin,
\np {\bf B352} (1991) 342
\bibitem{star}
D.C. Kennedy, B.W. Lynn, \np {\bf B322} (1989) 1
\bibitem{star1}
M. Kuroda, G. Moultaka, D. Schildknecht, \np {\bf B350} (1991) 25
\bibitem{ward}
J.C. Ward,  \pr {\bf 78} (1950) 1824
\bibitem{thooftvelt}
G. 't Hooft, M. Veltman, \np {\bf B135} (1979) 365
\bibitem{dimreg}
C. Bollini, J. Giambiagi {\it Nuovo Cim.} {\bf 12B} (1972) 20;\\
J. Ashmore, {\it Nuovo Cim.\ Lett.\ } {\bf 4} (1972) 289; \\
G. 't Hooft, M. Veltman, \np {\bf B44} (1972) 189
\bibitem{breitmaison}
P. Breitenlohner, D. Maison, {\it Comm.\ Math.\ Phys.\ }
{\bf 52} (1977) 11, 39, 55
\bibitem{rho}
M. Veltman, \np  {\bf B123} (1977) 89;\\
M.S. Chanowitz, M.A. Furman, I. Hinchliffe, \pl {\bf B78} (1978) 285
\bibitem{muon}
R.E Behrends, R.J. Finkelnstein, A. Sirlin,
 \pr {\bf 101} (1956) 866; \\
T. Kinoshita, A. Sirlin, \pr {\bf 113} (1959) 1652
\bibitem{eidelman} S. Eidelman, F. Jegerlehner,
                   \zp {\bf C67} (1995) 585
\bibitem{burkhardt} H. Burkhardt, B. Pietrzyk,
                    \plb {\bf B356} (1995) 398
\bibitem{vacpol}
F. Jegerlehner, {\elevenit Progress in Particle and Nuclear Physics}
                {\bf 27} (1991) 1, updated from:
H. Burkhardt, F. Jegerlehner, G. Penso, C. Verzegnassi,
\zp {\bf C43} (1989) 497;
\bibitem{swartz} M.L. Swartz, preprint SLAC-PUB-95-7001 (1995)
\bibitem{martin} A.D. Martin, D. Zeppenfeld,
                 \plb {\bf B345} (1995) 558
\bibitem{screening}
M. Veltman, {\it Acta Phys.\ Polon.\ } {\bf B8} (1977) 475.
\bibitem{marciano}
W.J. Marciano, \pr {\bf D20} (1979) 274
\bibitem{chj}
M. Consoli, W. Hollik, F. Jegerlehner, \pl {\bf B227} (1989) 167.
\bibitem{hoog}
J.J. van der Bij, F. Hoogeveen, \np {\bf B283} (1987) 477
\bibitem{barbieri}
R. Barbieri, M. Beccaria, P. Ciafaloni, G. Curci, A. Vicere,
 \plb {\bf B288} (1992) 95; \np {\bf B409} (1993) 105; \\
J. Fleischer, F. Jegerlehner, O.V. Tarasov, \plb {\bf B319} (1993) 249
\bibitem{adjouadi} A. Djouadi, C. Verzegnassi, \plb {\bf B195} (1987) 265
\bibitem{tarasov} L. Avdeev, J. Fleischer, S. M. Mikhailov, O. Tarasov,
                  \plb {\bf B336} (1994) 560;
                  E: \plb {\bf B349} (1995) 597; \\
                  K.G. Chetyrkin, J.H. K\"uhn, M. Steinhauser,
                  \plb {\bf B351} (1995) 331
\bibitem{qcd}
A. Djouadi, {\it Nuovo Cim.} {\bf 100A} (1988) 357; \\
D. Yu.\ Bardin, A.V. Chizhov, Dubna preprint E2-89-525 (1989); \\
B.A. Kniehl, \np {\bf B347} (1990) 86; \\
F. Halzen, B.A. Kniehl, \np {\bf B353} (1991) 567; \\
A. Djouadi, P. Gambino, \prd {\bf D49} (1994) 3499
\bibitem{dispersion1}
B.A. Kniehl, J.H. K\"uhn, R.G. Stuart, \plb {\bf B214} (1988) 621; \\
B.A. Kniehl, A. Sirlin, \np {\bf B371} (1992) 141;
                        \prd {\bf D47} (1993) 883; \\
 S. Fanchiotti, B.A. Kniehl, A. Sirlin, \prd {\bf D48} (1993) 307
\bibitem{cks} K.G. Chetyrkin, J.H. K\"uhn, M. Steinhauser,
              preprint KA-TTP95-13 (1995)
\bibitem{nonleading}
A. Sirlin, \pr {\bf D29} (1984) 89
\bibitem{beg}
S. Fanchiotti, A. Sirlin, New York University preprint
NYU-Th-91/02/04 (1991), in:
{\it  Beg Memorial Volume}, eds.\ A. Ali and P. Hoodbhoy,
World Scientific, Singapore 1991
\bibitem{neutrino}
G.L. Fogli and D. Haidt, \zp
{\elevenbf C40} (1988) 379;\\
CDHS Collaboration, H. Abramowicz et al.,
 \prl {\elevenbf B57} (1986) 298;
A. Blondel et al, \zp {\elevenbf C45} (1990) 361;\\
CHARM Collaboration, J.V. Allaby et al.,
 \pl {\elevenbf B177} (1987) 446; \\
\zp {\bf C36} (1987) 611;\\
CHARM-II Collaboration, D. Geiregat et al.,
 \pl {\elevenbf B247} (1990) 131;\\
\pl {\elevenbf B259} (1991) 499; \\
CCFR Collaboration, C.G. Arroyo et al., \prl {\elevenbf 72} (1994) 3452; \\
D. Harris (CCFR Collaboration), talk at the {\it
                International Europhysics Conference on High Energy Physics},
                Brussels 1995 (to appear in the Proceedings) 
\bibitem{llewellyn}
C.H. Llewellyn Smith, {\it \np}{\bf B228} (1983) 205
\bibitem{vertex}
A.A. Akhundov, D.Yu.\ Bardin, T. Riemann, \np {\bf B276} (1986) 1; \\
W. Beenakker, W. Hollik, \zp {\bf C40} (1988) 141; \\
J. Bernabeu, A. Pich, A. Santamaria, \pl {\bf B200} (1988) 569;
               \np {\bf B363} (1991) 326
\bibitem{luo}
P. Langacker, M. Luo, \pr {\bf D44} (1991) 817
\bibitem{susy}
U. Amaldi, W. de Boer, H. F\"urstenau, \pl {\bf B260} (1991) 447; \\
J. Ellis, S. Kelley, D.V. Nanopoulos, \pl {\bf B260} (1991) 131;
 \pl {\bf B287} (1992) 725; \\
G.G. Ross, R.G. Roberts, \np {\bf B377} (1992) 571
\bibitem{dhl} A. Denner, W. Hollik, B. Lampe,
             {\elevenit \zp}{\elevenbf C60} (1993) 93
\bibitem{jeg} J. Fleischer, O.V. Tarasov, F. Jegerlehner,
              P. R\c{a}czka,
              {\it \pl}{\bf B293} (1992) 437; \\ 
               G. Buchalla, A.J. Buras,
               \np {\bf B398} (1993) 285; \\
              G. Degrassi, \np {\bf B407} (1993) 271; \\
              K.G. Chetyrkin, A. Kwiatkowski, M. Steinhauser,
              {\elevenit Mod.\ Phys. Lett.\  }{\bf A8} (1993) 2785
\bibitem{log} A. Kwiatkowski, M. Steinhauser, \plb {\bf B344} (1995) 359; \\
              S. Peris, A. Santamaria, CERN-TH-95-21 (1995)
\bibitem{berends} F.A. Berends et al.,
 in:
 {\it Z Physics at LEP 1}, CERN 89-08 (1989), eds.\
 G. Altarelli, R. Kleiss, C. Verzegnassi, Vol.\ I, p.\  89;\\
W. Beenakker, F.A. Berends, S.C. van der Marck,
 \zp{\bf C46} (1990) 687
\bibitem{sisto}
A. Borelli, M. Consoli, L. Maiani, R. Sisto, \np{\bf B333} (1990) 357
\bibitem{bbhvn} G. Burgers, F.A. Berends, W. Hollik,
                W.L. van Neerven, \pl{\bf B203} (1988) 177
\bibitem{blrs}  D. Yu.\ Bardin, A. Leike, T. Riemann, M. Sachwitz,
                \pl{\bf B206} (1988) 539
\bibitem{smatrix}
       G. Valencia, S. Willenbrock, \pl{\bf B 259} (1991) 373;\\
       R.G. Stuart, \pl{\bf B272} (1991) 353
\bibitem{bbvn} G. Burgers, F.A. Berends, W.L.van Neerven,
              \np{\bf B297} (1988) 429;
              E: \np{\bf B304} (1988) 921
\bibitem{bbm} W. Beenakker, F.A. Berends, S.C. van der Marck,
             \zp{\bf C46} (1990) 687
\bibitem{qcd1}
K.G. Chetyrkin, A.L. Kataev, F.V. Tkachov, \pl{\bf B85} (1979) 277; \\
M. Dine, J. Sapirstein, \prl{\bf 43} (1979) 668;\\
W. Celmaster, R. Gonsalves, \prl{\bf 44} (1980) 560;\\
S.G. Gorishny, A.L. Kataev, S.A. Larin, \pl{\bf B259} (1991) 144; \\
L.R. Surguladze, M.A. Samuel, \prl{\bf 66} (1991) 560; \\
A. Kataev, \plb {\elevenbf B287} (1992 209
\bibitem{qcdb}
T.H. Chang, K.J.F Gaemers, W.L. van Neerven, \np {\bf B202} (1982) 407;\\
J.H. K\"uhn, B.A. Kniehl, \plb {\bf B224} (1990) 229;\\
 \np {\bf B329} (1990) 547;\\
  K.G. Chetyrkin, J.H. K\"uhn, \plb {\bf B248} (1992) 359;\\
  K.G. Chetyrkin, J.H. K\"uhn, A. Kwiatkowski,
             \plb {\bf B282} (1992) 221;\\
  K.G. Chetyrkin, A. Kwiatkowski, \plb {\bf B305} (1993) 285;
     Karlsruhe preprint TTP93-24 (1993);\\
  K.G. Chetyrkin,
     Karlsruhe preprint TTP93-5 (1993);\\
  K.G. Chetyrkin, J.H. K\"uhn, A. Kwiatkowski, in [91], p.\ 175;\\
  S. Larin, T. van Ritbergen, J.A.M. Vermaseren, ibidem, p.\ 265;
     \plb {\bf B320} (1994) 159 
\bibitem{hoang} A. Hoang, J.H. K\"uhn, T. Teubner, \np {\bf B455}
               (1995) 3; {\bf B452} (1995) 173
\bibitem{bh} M. B\"ohm, W. Hollik
     {\it \np}{\bf B204} (1982) 45; {\it \zp}{\bf C 23} (1984) 31
\bibitem{alrqed} S. Jadach, J.H. K\"uhn, R.G. Stuart,
                 Z. W\c{a}s, {\it \zp}{\bf C38} (1988) 609; \\
      J.H. K\"uhn, R.G. Stuart, {\it \pl}{B200} (1988) 360
\bibitem{bardin1} D. Bardin, M.S. Bilenky, O.M. Fedorenko,
                 T. Riemann, Dubna preprint JINR-E2-88-324 (1988)
\bibitem{bardin2} D. Bardin, M.S. Bilenky, A. Chizhov, A. Sazonov,
                 Yu.\ Sedykh, T. Riemann, M. Sachwitz,
                 {\it \pl}{\bf B229} (1989) 405
\bibitem{laerman}
J. Jersak, E. Laerman, P.M. Zerwas, \pr{\bf D25} (1980) 1218
\bibitem{djouadi}
A. Djouadi, \zp{\bf C39} (1988) 561
\bibitem{djouadinew}
  A. Djouadi, B. Lampe, P. Zerwas, \zp {\elevenbf C67} (1995) 123
\bibitem{afb} M. B\"ohm, W. Hollik et al.,
 in:
 {\it Z Physics at LEP 1}, CERN 89-08 (1989), eds.\
 G. Altarelli, R. Kleiss, C. Verzegnassi, Vol.\ I, p.\  203;\\
 W. Beenakker, F.A. Berends, S.C. van der Marck,
 \pl{\bf B252} (1990) 299
\bibitem{heavy} J.H. K\"uhn, P.Zerwas et al., in:
 {\it Z Physics at LEP 1}, CERN 89-08 (1989), eds.\
 G. Altarelli, R. Kleiss, C. Verzegnassi, Vol.\ I, p.\ 267
\bibitem{bardin3} D. Bardin, M.S. Bilenky, A. Chizhov, A. Sazonov,
                 O. Fedorenko, T. Riemann, M. Sachwitz,
                 {\it \np}{\bf B351} (1991) 1
\bibitem{ringberg} W. Beenakker, F.A. Berends, W.L. van Neerven,
 in: Proceedings of the 1989 Ringberg Workshop
 {\it Radiative Corrections for $\epm$ Collisions}, p.\ 3,
 ed.\ J.H. K\"uhn, Springer, Berlin - Heidelberg - New York 1989
\bibitem{cuts} D. Bardin, L. Vertogradov,
               Yu.\ Sedykh, T. Riemann, CERN-TH.5434/89 (1989)
\bibitem{zfitter} D. Bardin et al., CERN-TH.6443/92 (1992)
\bibitem{topaz} G. Montagna, F. Piccinini, O. Nicrosini,
               G. Passarino, R. Pittau,
            Pavia-Torino preprint FNT/T-92/02,
            DFTT/G-93-1 (1993)
\bibitem{kniehl95} B.A. Kniehl, in [91], p.\ 299
\bibitem{ewgr} D. Bardin et al., in [91], p.\ 7
\bibitem{yb95}  {\it Reports of
             the Working Group on Precision Calculations
             for the $Z$ Resonance}, CERN 95-03 (1995), eds.\
             D. Bardin, W. Hollik, G. Passarino
\bibitem{bardin} D. Bardin, 
                talk at the {\elevenit
                International Europhysics Conference on High Energy Physics},
                Brussels 1995 (to appear in the Proceedings)
\bibitem{padova} G. Degrassi, S. Fanchiotti, F. Feruglio, P. Gambino,
      A. Vicini, in [91], p.\ 163
\bibitem{alfas}
     S. Bethke, in: {\it Proceedings of the  Tennessee International
                Symposium on
                Radiative Corrections}, Gatlinburg 1994,
                Ed.\ B.F.L. Ward, World Scientific 1995
\bibitem{sld} SLD Collaboration, K. Abe et al., \prl {\bf 73} (1994) 25;
          M. Woods (SLD Collaboration), 
                talk at the {\elevenit
                International Europhysics Conference on High Energy Physics},
                Brussels 1995 (to appear in the Proceedings)
\bibitem{charm} CHARM II Collaboration, P. Vilain et al.,
               \plb {\bf B335} (1994) 246; {\bf B345} (1995) 115
\bibitem{cradius}
     S. Sarantakos, A. Sirlin, \np {\bf B217} (1983) 84;
     D.Yu. Bardin V.A. Dokuchaeva, \np {\bf B246} (1984) 221;
     M. B\"ohm, W. Hollik, H. Spiesberger,
      \zp {\bf C27} (1985) 523
\bibitem{grivaz} J.-F. Grivaz,
                talk at the {\elevenit
                International Europhysics Conference on High Energy Physics},
                Brussels 1995 (to appear in the Proceedings), LAL-95-83
\bibitem{mhfit} J. Ellis, G.L. Fogli, E. Lisi, CERN-TH-95-202,
                hep-ph/9507424; \\
                P. Chankowski, S. Pokorski, hep-ph/9509207
\bibitem{lindner} M. Lindner, M. Sher, H. Zaglauer, 
                 \plb {\bf B228} (1989) 139
\bibitem{lattice} Kuti et al., \prl {\bf 61} (1988) 678;
        Hasenfratz et al., \np {\bf B317} (1989) 81;
        M. L\"uscher, P. Weisz, \np {\bf B318} (1989) 705
\bibitem{ghinculov}  A. Ghinculov, \np {\bf B455} (1995) 21
\bibitem{carena} M. Carena. P. Zerwas (conveners) et al., report on
     {\elevenit Higgs Physics}, hep-ph/9602250, to appear in the {\it
     Proceedings of the LEP2 Workshop}, eds. G. Altarelli, T, Sj\"ostrand,
     F. Zwirner
\bibitem{delta}
G. Burgers, F. Jegerlehner, in:
{\it $Z$ Physics at LEP 1}, eds.\ G. Altarelli, R. Kleiss
and C. Verzegnassi, CERN 89-08 (1989)
\bibitem{pt}
M.E. Peskin, T. Takeuchi, \prl  {\bf 65} (1990) 964
\bibitem{epsilon}
G. Altarelli, R. Barbieri, \pl {\bf B253} (1991) 161; \\
G. Altarelli, R. Barbieri, S. Jadach, \np {\bf B269} (1992) 3;
E:  \np {\bf B276} (1992) 444
\bibitem{kennlang}
D.C. Kennedy, P. Langacker, \prl {\bf 65} (1990) 2967
\bibitem{mrosner}
W.J. Marciano, J.L. Rosner, \prl {\bf 65} (1990) 2963
\bibitem{lps}
B.W. Lynn, M.E. Peskin, R.G. Stuart, in: {\it Physics with LEP},
 eds.\ J. Ellis and R. Peccei,
CERN 86-02 (1986)
\bibitem{pt1}
R. Barbieri,  M.Frigeni, F. Caravaglios, \plb {\bf B279} (1992) 169;\\
V.A. Novikov, L.B. Okun, M.I. Vysotsky, CERN-TH.6943/93 (1993); \\
M. Bilenky, K. Kolodziej, M. Kuroda, D. Schildknecht,
\plb {\bf B319} (1993) 319; \\
S. Dittmaier, D. Schildknecht, M. Kuroda, \np {\bf B448} (1995) 3
\bibitem{technicolor}
B. Holdom, J. Terning, \pl {\bf B247} (1990) 88; \\
M. Golden, L. Randall, \np {\bf B361} (1991) 3; \\
C. Roiesnel, T.N. Truong, \pl {\bf B256} (1991) 439
\bibitem{abc1} G. Altarelli, R. Barbieri, F. Caravaglios,
           \np {\elevenbf B405} (1993) 3; CERN-TH.6895/93 (1993)
\bibitem{caravaglios} F. Caravaglios, talk at the {\elevenit
                International Europhysics Conference on High Energy Physics},
                Brussels 1995 (to appear in the Proceedings)
\bibitem{toussaint}
D. Toussaint, \pr {\bf D18} (1978) 1626; \\
J.M Frere, J. Vermaseren, \zp {\bf C19} (1983) 63
\bibitem{bertolini}
S. Bertolini, \np {\bf B272} (1986) 77; \\
W. Hollik, \zp {\bf C32} (1986) 291; \zp {\bf C37} (1988) 569
\bibitem{dgk}
A. Denner, R. Guth, J.H. K\"uhn, \pl {\bf B240} (1990) 438
\bibitem{dghk}
A. Denner, R. Guth, W. Hollik, J.H. K\"uhn, \zp {\bf C51} (1991) 695
\bibitem{nardi}
B.W. Lynn, E. Nardi, \np {\bf B381} (1992) 467
\bibitem{blank}
T. Blank, Diploma thesis, Univ.\ Karlsruhe 1995
\bibitem{altarelli}
G. Altarelli et al.,
 \np {\bf B342} (1990) 15; \pl {\bf B245} (1990) 669; \\
M.C. Gonzalez-Garcia, J.W.F. Valle, \pl {\bf B259} (1991) 365; \\
J. Layssac, F.M. Renard, C. Verzegnassi,
\zp {\bf C53} (1992) 97
\bibitem{aguila}
F. del Aguila, W. Hollik, J.M. Moreno, M. Quiros,
 \np {\bf B372} (1992) 3
%
\bibitem{hadrophil} P. Chiapetta, J. Layssac, F.M. Renard,
                C. Verzegnassi, PM/96-05, hep-ph/9602306; \\
             G. Altarelli, N. Di Bartolomeo,F. Feruglio,
             R, Gatto, M. Mangano, CERN-TH/96-20,
             hep-ph/9601324 
\bibitem{higgs} A. Denner, R. Guth, W. Hollik, J.H. K\"uhn,
                \zp {\bf C51} (1991) 695; \\
                J. Rosiek, \plb {\bf B252} (1990) 135; \\
                F. Cornet, W. Hollik, W. M\"osle,
                 \np {\bf B428} (1994) 61; \\
                M. Boulware, D. Finnell, \prd {\bf D44} (1991) 2054
\bibitem{susy1}  G. Altarelli, R. Barbieri, F. Caravaglios,
                CERN-TH.7536/94 (1994); \\
                C.S. Lee, B.Q. Hu, J.H. Yang, Z.Y. Fang,
                {\elevenit J. Phys.\ } {\bf G19} (1993) 13; \\
                Q. Hu, J.M. Yang, C.S. Li, {\elevenit Comm.\ Theor.\ Phys.\ }
                {\bf 20} (1993) 213; \\
                J.D. Wells, C. Kolda, G.L. Kane, \plb {\bf B338} (1994) 219;\\
                G.L. Kane, R.G. Stuart, J.D. Wells,
                \plb {\bf B354} (1995) 350
\bibitem{susydelr} P. Chankowski, A. Dabelstein, W. Hollik, W. M\"osle,
                   S. Pokorski, J. Rosiek, \np {\bf B417} (1994) 101; \\
                   D. Garcia,  J. Sol\`a, {\elevenit Mod.\ Phys.\
                   Lett.\  }{\bf A9} (1994) 211
\bibitem{susy3} D. Garcia, R. Jim\'enez, J. Sol\`a,
                \plb {\bf B347} (1995) 309; {\bf B347} (1995) 321; \\
                D. Garcia, J. Sol\`a, \plb {\bf B357} (1995) 349
\bibitem{susy4} P. Chankowski, S. Pokorski, preprint
                IFT-UW-95/5 (1995); \\
                P. Chankowski,
                talk at the {\elevenit
                International Europhysics Conference on High Energy Physics},
                Brussels 1995 (to appear in the Proceedings)
\bibitem{susy5} A. Dabelstein, W. Hollik, W. M\"osle, preprint
                KA-TP-5-1995 (1995),  {\it Proceedings of
                the Ringberg Workshop ``Perspectives for  Electroweak
                Interactions in $\epm$ Collisions''},
                 Ringberg Castle, February 1995,
                Ed.\ B.A. Kniehl, World Scientific 1995 (p.\ 345)
\bibitem{deboer} W. De Boer, S. Meyer, A. Dabelstein, W. Hollik,
                W. M\"osle, U. Schwickerath (to be published); \\
                W. Hollik,
                 talk at the {\elevenit
                International Europhysics Conference on High Energy Physics},
                Brussels 1995 (to appear in the Proceedings)
\end{thebibliography}
\end{document}